\documentclass[twocolumn,tighten]{aastex63}
\usepackage{graphicx}
\usepackage{gensymb}
\usepackage{longtable}
\usepackage{amsmath}

\begin{document}

\title{Cassiopeia A's Reverse Shock and its Effects on the Expanding SN Ejecta\footnote{Based in part on 
observations with the NASA/ESA Hubble Space Telescope, obtained at the Space Telescope Science Institute,
which is operated by the Association of Universities for Research in Astronomy, Inc.} }

\author[0000-0003-3829-2056]{Robert A.\ Fesen}
\affil{6127 Wilder Lab, Department of Physics and Astronomy, Dartmouth
       College, Hanover, NH 03755 USA}

\author[0000-0002-0763-3885]{Dan Milisavljevic}
\affil{Department of Physics and Astronomy, Purdue University, 
       525 Northwestern Avenue, West Lafayette, IN 47907 USA}

\author[0000-0002-7507-8115]{Daniel Patnaude}
\affil{Smithsonian Astrophysical Observatory, MS-3, 60 Garden Street, Cambridge, MA 02138 USA}

\author[0000-0002-7868-1622]{Roger A. Chevalier}
\affil{Department of Astronomy, University of Virginia, P.O. Box 400325 Charlottesville, VA 22904-4325, USA}

\author[0000-0002-7868-1622]{John C.\ Raymond}
\affil{Center for Astrophysics, 60 Garden St., Cambridge, MA 02176, USA}

\author[0000-0002-4024-6967]{McKinley Brumback}
\affil{6127 Wilder Lab, Department of Physics and Astronomy, Dartmouth College, Hanover, NH 03755 USA}
\affil{Department of Physics, Middlebury College, Middlebury, VT 05753, USA}

\author[00000-0002-4471-9960]{Kathryn E.\ Weil}
\affil{6127 Wilder Lab, Department of Physics and Astronomy, Dartmouth
       College, Hanover, NH 03755 USA}
\affil{Department of Physics and Astronomy, Purdue University, 525 Northwestern Avenue, West Lafayette, IN 47907 USA}

\begin{abstract}

Using optical and near-infrared images of the Cassiopeia A (Cas A) supernova remnant covering the time period 1951 to 2022, together with optical spectra of selected filaments, we present an investigation of Cas~A's reverse shock velocity and the effects it has on the remnant's metal-rich ejecta. We find the sequence of optical ejecta brightening and the appearance of new optical ejecta indicating the advancement of the remnant's reverse shock in the remnant's main shell has velocities typically between 1000 and 2000 km s$^{-1}$, which is $\sim 1000$ km s$^{-1}$ less than recent measurements made in X-rays. We further find the reverse shock appears to move much more slowly and is nearly even stationary in the sky frame along the remnant's western limb. However, we do not find the reverse shock to move inward at velocities as large as $\sim 2000$ km s$^{-1}$ as has been reported. Optical ejecta in Cas A's main emission shell have proper motions indicating outward tangential motions $\simeq 3500 - 6000$ km s$^{-1}$, with the smaller values preferentially along the remnant's southern regions which we speculate may be partially the cause of the remnant's faint and more slowly evolving southern sections. Following interaction with the reverse shock, ejecta knots exhibit extended mass ablated trails $0.2'' - 0.5''$ in length leading to extended emission indicating reverse shock induced decelerated velocities as large as $\simeq$1000 km s$^{-1}$. Such ablated material is most prominently seen in higher ionization line emissions, whereas denser parts of ejecta knots show surprisingly little deceleration.

\end{abstract}

\bigskip

\keywords{ISM: individual (Cassiopeia A) - ISM: kinematics and dynamics }

\section{Introduction}

Supernova (SN) explosions generate high-velocity ejecta which initially act in
unison like a spherical piston creating an expanding blast wave which
propagates outward into the surrounding medium \citep{Shklovskii1973}.  
This outwardly moving shock subsequently undergoes deceleration 
through its interaction with either the progenitor's pre-SN 
circumstellar medium (CSM) or local ambient interstellar medium (ISM),
forming behind it a high pressure shell of swept-up material.  

When the relatively low temperature ($\sim$10$^{2-3}$ K) and low pressure SN ejecta makes contact with the much hotter ($\sim$10$^{8-9}$ K) and higher pressure post-shocked circumstellar medium behind the forward shock, a secondary or ``reverse'' shock forms.  
With a velocity less than that of the forward shock, models show that this reverse shock, despite its name, still moves outward from the remnant's center of expansion with respect to fixed sky coordinates until a significant fraction of a SN's mass has been shocked \citep{McKee74,Gull1975,McKee95,True1999}.

The interaction of this reverse shock with a remnant's
low density ejecta leads to strong X-ray and nonthermal emissions \citep{Yany2020}, 
while driving a relatively slow, radiative shock into clumps or ``knots'' of denser 
ejecta. These reverse shock heated ejecta knots are 
subsequently observed as UV, optical, and
infrared emission features \citep{McKee74,Chevalier77,Chevalier1982a}.
Following reverse shock passage, such compact ejecta knots can experience compression, 
distortion, and mass ablation \citep{Nag88,Mueller91,WC01,Kif03,Hammer2010}.

Young Galactic SN remnants (SNRs) offer the potential
of examining the progression of a SN's reverse shock in detail.
In this regard, Cassiopeia~A (Cas~A) is especially useful as it is
the youngest Galactic core-collapse supernova (CCSN) remnant (current age $\simeq$350 yr\footnote{ \citet{Soria13} suggest a transient 4th mag star reported by
Cassini around 1671 may have been a sighting of the Cas A SN. }), is
among the closest young SNR (3.4$\pm 0.2$ kpc; \citealt{Min59,Reed95,Rest08,Alarie14, Neumann2024}), 
and the only Galactic CCSN remnant with a secure SN classification (SN IIb) derived from light echo spectroscopy using multiple  lines of sight \citep{Krause08,Rest08,Rest11,Besel12}.  
Cas ~A is likely the remains of a 15-25 solar mass red supergiant progenitor that lost much of its hydrogen envelope \citep{Fesen87,Fesen88,Fesen01} possibly due to a binary companion \citep{Chevalier2003,Young06,Krause08,Weil2020}. 

Observationally, the Cas~A remnant consists of bright shell of optical, infrared, radio, and X-ray
emissions roughly 5$\arcmin$ diameter ($\simeq$ 5 pc at 3.4 kpc).
Multi-wavelength data show its reverse shock is currently
encountering O-Si-S-Ar-Ca rich ejecta \citep{ck78,ck79,Hughes00,Hwang00,LH03,Delaney10},
having long-ago progressed passed the thin residual H, He, and N-rich 
photospheric layers present at the time of SN outburst as evidenced in early
light-echo spectra \citep{Krause08,Rest11} and by its N-rich but H-poor outer ejecta knots
\citep{Fesen87,Fesen01,Fesen06b}. 

Cas~A's reverse shocked, high-density ejecta are most visible
via optical and infrared emissions while lower density ejecta are more
prominent in X-rays 
\citep{Hughes00,Hwang00,Gotthelf01,LH03,Helder08,Patnaude14}.
Kinematic studies have shown Cas~A's optical ejecta to be arranged into a cellular-like structure consisting of a dozen or so ejecta ``rings'' which surround hotter, X-ray bright, Fe-rich ejecta 
\citep{Delaney10,MF13,Mil15}. 
These rings represent cross-sections of large cavities or ``bubbles'' of ejecta. These ring
structures
arise from the interaction of a roughly spherical reverse shock front with these bubbles of ejecta.
Rings are composed of numerous individual ejecta knots of varying densities and elemental 
compositions and have typical angular dimensions of $0.2'' - 1.0''$ 
\citep{Fesen01etal, Fesen06a,Delaney10,Mil15}.

\begin{figure*}[t]
\begin{center}
\includegraphics[angle=0,width=8.5cm]{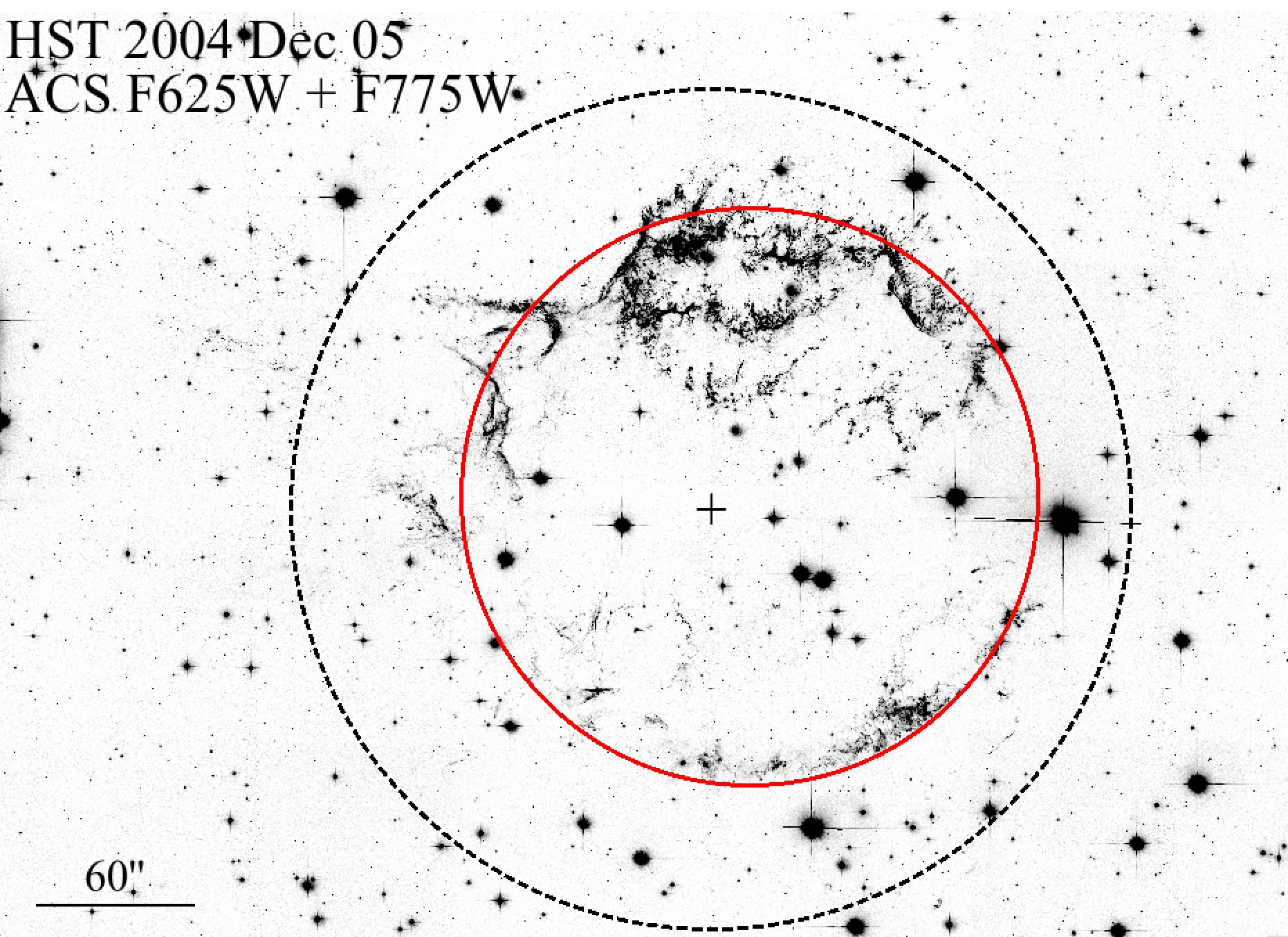} 
\includegraphics[angle=0,width=8.5cm]{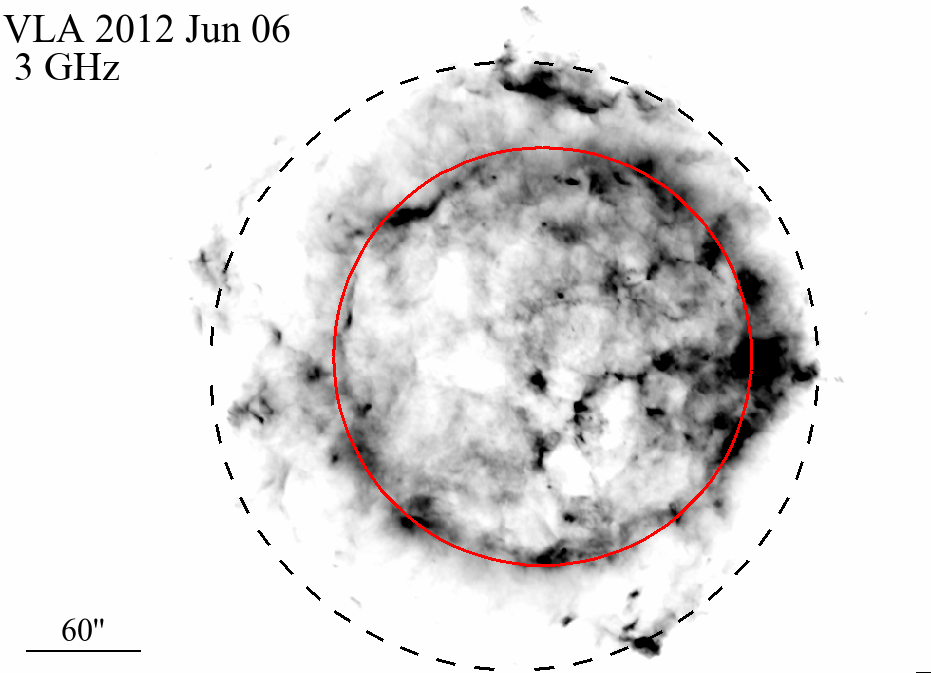}  \\
\includegraphics[angle=0,width=8.5cm]{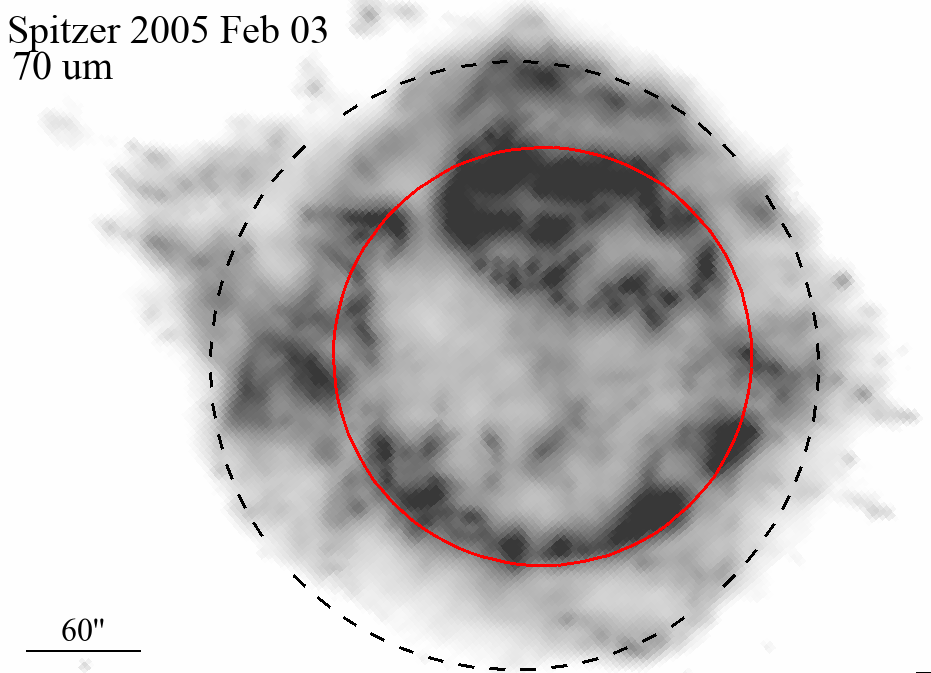} 
\includegraphics[angle=0,width=8.5cm]{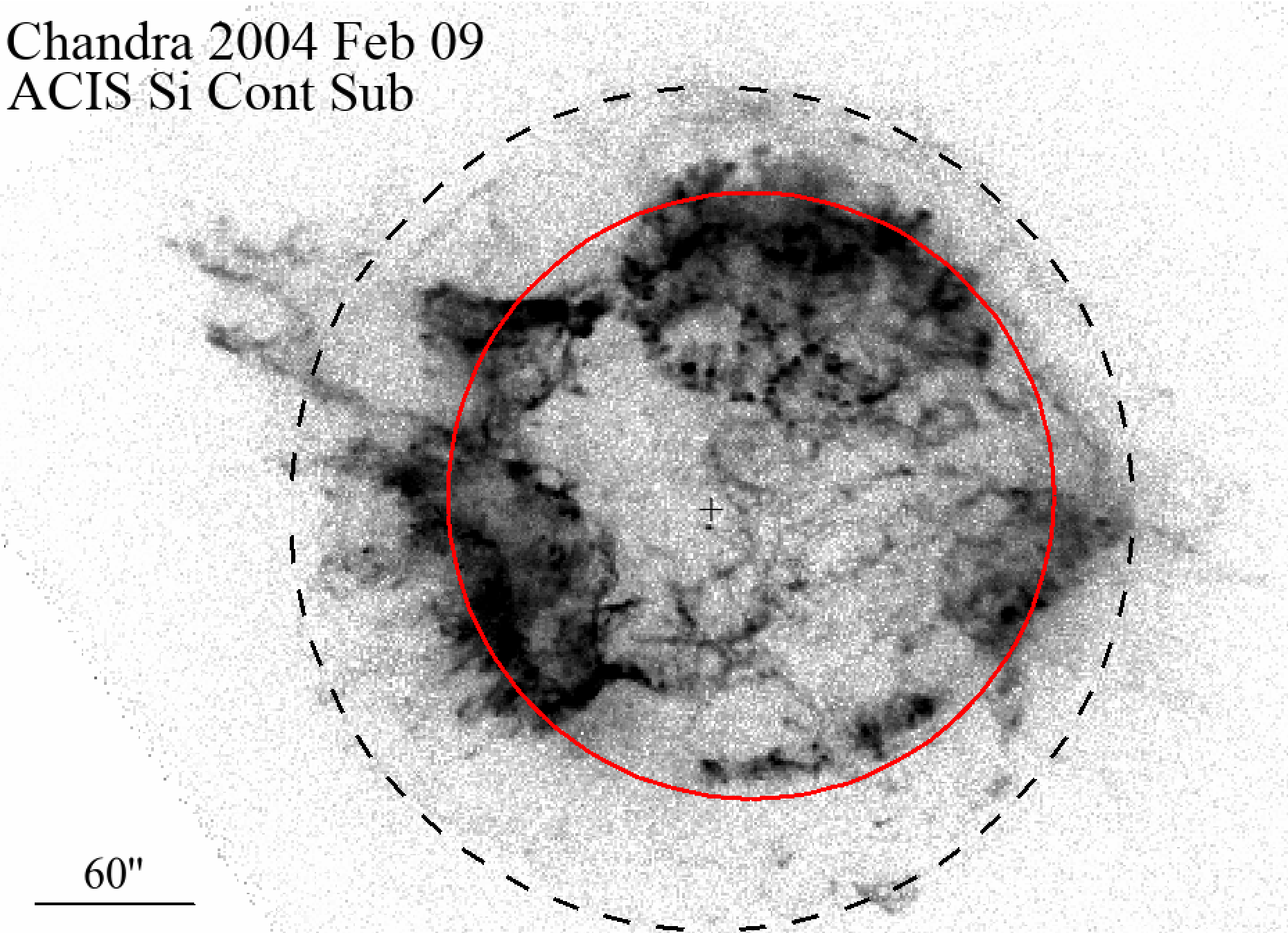} 
\caption{Comparison of Cas~A's large-scale optical ({\sl HST}), radio (VLA), infrared (Spitzer) and  X-ray ({\sl Chandra}) emissions.  The dashed black circles are 160$''$ in radius and centered on the remnant's expansion center (\citealt{Thor01};  
$\alpha(\rm J2000)$ = $23^{\rm h} ~ 23^{\rm m} ~ 27.77^{\rm s}$, 
$\delta({\rm J2000})$ = $58\degr ~ 48' ~ 49.4''$) marked as a small cross in the optical and X-ray images. 
The dashed circles show the extent of the remnant's outermost radio and infrared emissions
and the location of the forward shock, best seen in the X-rays.
The solid red circles are 110$''$ in radius and shifted
5$''$ north and 13$''$ west to match the expansion center defined by the remnant's bright radio emission shell (\citealt{Anderson1995}; $\alpha(\rm J2000)$ = $23^{\rm h} ~ 23^{\rm m} ~ 25.854^{\rm s}$, 
$\delta({\rm J2000})$ = $58\degr ~ 48' ~ 54.3''$). North is up, East to the left.
\label{circles}
}
\end{center}
\end{figure*}

\begin{table}
\centering
\caption{Estimates Cas A's Forward Shock Velocity\footnote{Based on X-ray data and assuming d = 3.4 kpc} \label{Tab1} }
\label{FRshock}
\begin{tabular}{lcc}
\hline
\hline
Reference   & $\mu$              &    V$_{\rm FS}$ \\
            & ($''$ yr$^{-1}$)   &  (km s$^{-1}$) \\
\hline
 \cite{Vink98}      & $\sim$0.32      &  $\simeq$ +5200     \\
 \cite{Kor98}       & $\sim$0.32      &  $\simeq$ +5200     \\
 \cite{Will02}      & \nodata         &  $\sim$ +4000     \\
 \cite{Delaney03}   & $\sim$0.32      &  +2500 to +6500    \\
 \cite{Patnaude09}  & 0.26 - 0.32     &  +4200 to +5200     \\
 \cite{Hwang12}     & 0.30 - 0.31     &  +4850 to +5050     \\
 \cite{Vink2022}    & $\simeq$0.36    &  +5000 to +7500      \\
 \cite{Sakai2024}   & \nodata    &  +5000 to +6500     \\
\hline
\end{tabular}
\end{table}

\begin{deluxetable*}{llc}
\tablecolumns{5}
\tablecaption{Model Estimates of Cas A's Reverse Shock Velocity\tablenotemark{a}  \label{Tab2}}
\tablewidth{0pt}
\tablehead{\colhead{Reference} & \colhead{Method ~~~~~~~~~~~} &  \colhead{V$_{\rm RS}$}  \\
 \colhead{}   & \colhead{}  & \colhead{(km s$^{-1}$)}  }
\startdata
\cite{LH03}        & X-ray emission models          & $\sim$ +1000  \\
 \cite{Schure08}    & hydrodynamical models          &  +990 to +3160    \\
 \cite{Hwang12}     & hydrodynamical models          & +2500           \\
 \cite{Mice16}      & dust destruction models        & +1600           \\ 
 \cite{Orlando16}   & hydrodynamical models          & +1500          \\
 \cite{Orlando21}  &  MHD + neutrino models          & +3000              \\
 \cite{Orlando22}  & MHD + asymmetric CSM shell     & +3000 to $-2000$  \\
\enddata
\tablenotetext{a}{Positive and negative values indicate 
motions outward or inward toward the remnant's interior, respectively.  }
\label{model_speeds}
\end{deluxetable*}

\begin{deluxetable*}{llccccl}
\tablecolumns{7}
\tablecaption{Direct Measurements of Cas A's Reverse Shock Velocity\tablenotemark{a} \label{Tab3} }
\tablewidth{0pt}
\tablehead{\colhead{Reference} & \colhead{Measured} & \colhead{$\Delta$t} & \colhead{Azimuth}  & 
 \multicolumn{2}{c}{\underline{~~~~~~~~Observed Expansion~~~~~~~}} &
\colhead{V$_{\rm RS}$\tablenotemark{b}}  \\
 \colhead{}   & \colhead{Feature}  &  \colhead{(yrs)} &  \colhead{(degrees)}  & 
 \colhead{(\% yr$^{-1}$)} & \colhead{($\mu$ yr$^{-1}$)} &    \colhead{(km s$^{-1}$)}  }
\startdata
 \cite{Kor98}     & radio ring    & 11.0     & $0 - 360$   & 0.07\% $-$ 0.14\%   & $0.108''\pm0.025''$   & $+1750 \pm 400$ \\
 \cite{Vink98}    & X-ray ring    & 17.0     & $0 - 360$   & 0.200\% $\pm$0.006\% & $0.220''\pm0.006''$  & $+3500\pm110$  \\
  \cite{Delaney03} & radio ring   &  9.2     & $115 - 250$ & 0.07\% $\pm$0.03\%  & $0.07'' \pm0.03''$    &  $+1155\pm500$ \\
 \cite{Morse04}   & optical knots &  2.0     & 225, 315    & \nodata             & $0.16'' - 0.20''$     &  +2600 to +3200  \\
 \cite{Helder08}  & western X-rays &  4.0    &  $\approx$290  & \nodata          & $-0.06''\pm 0.009''$  & ~~$-970 \pm 140$  \\
 \cite{Vink2022}  & X-ray ring    &  19      & $0 - 360$    &0.134\% $\pm$0.084\% & $0.153''\pm 0.009''$  & $+2460\pm1540$ \\ 
 ~~~~~~ " ~~~"    &   ~~~~~  "~~~ &  19      & $\sim$270    &$-$0.1023\% $\pm$0.0012\%  & $-0.1167'' \pm 0.0001''$ &  $-1880\pm22$ \\
 \cite{Wu2024}    & X-ray ring    &  19      & $ 315 - 5 $  &  \nodata           & $0.245'' \pm0.013''$   & $+3950\pm210$          \\      
   ~ " ~~~~~~~"  & X-ray ring    &  19      & $15 - 135$   &  \nodata           & $0.180'' \pm0.016''$   & $+2900\pm260$            \\
\enddata
\tablenotetext{a}{Positive and negative values indicate 
motions outward or inward, respectively, toward the remnant's interior.  }
\tablenotetext{\rm b}{Listed velocity is transverse velocity only and assumes a distance 3.4 kpc. }
\label{rs_speeds}
\end{deluxetable*}


\begin{deluxetable*}{lclcll}[ht]
\centering
\tablecolumns{6}
\tablewidth{0pc}
\tablecaption{Summary Log of Observations Used for this Analysis\tablenotemark{a} \label{4} }
\tablehead{
\colhead{Instrument} & \colhead{Date} &
\colhead{Obs. ID} & \colhead{Exposure Time}   & \colhead{Wave Band/Plate Info\tablenotemark{a}} & Observer/PI   }
\startdata
\underline{\bf{Ground Based} ~~~~~~~}         &                &              &    &             \\
Palomar Hale 5m & 1951-09-09 & PH520B        & 1800 s       & blue: 103aO+GG1         & Baade      \\
Palomar Hale 5m & 1951-10-31 & PH553B        & 7200 s       & ~red: 103aE+RG2         & Baade  \\
Palomar Hale 5m & 1951-11-03 & PH563B        & 7200 s       & ~red: 103aE+RG2         & Baade      \\
Palomar Hale 5m & 1954-11-25 & PH232M        & 5400 s       & blue: 103aJ+GG11        & Minkowski   \\
Palomar Hale 5m & 1958-08-11 & PH3033S       & 5400 s       & ~red: 103aF+RG2         & Shane     \\
Palomar Hale 5m & 1968-09-26 & PH5254vB      & 5400 s       & ~red: 103aF+RG2         & van den Bergh \\
Palomar Hale 5m & 1972-09-10 & PH6249vB      & 7200 s       & ~red: 103aF+RG2         & van den Bergh \\
Palomar Hale 5m & 1976-07-02 & PH7252vB      & 7200 s       & ~red: 098+RG645         & van den Bergh \\
Palomar Hale 5m & 1989-09-28 & PH8202vB      & 7200 s       & ~red: 098+RG645         & van den Bergh \\
Palomar Hale 5m & 1989-09-28 & PH8204vB      & 7200 s       & [S II]+098 (FWHM=166 A)  & van den Bergh \\
Palomar Hale 5m & 1989-09-29 & PH8206vB      & 7200 s       & ~red: 098-04+RG645      & van den Bergh \\
MDM Hiltner 2.4m        & 1996-10-06 & ~~~\nodata    &  900 s       & 5600--8800 \AA: R band  & Fesen         \\
MDM Hiltner 2.4m        & 1998-09-18 & ~~~\nodata    & 2400 s       & 5600--8800 \AA: R band  & Fesen  \\
MDM Hiltner 2.4m        & 1999-10-15 & ~~~\nodata    & 2400 s       & 5600--8800 \AA: R band  & Thorstensen  \\
MDM Hiltner 2.4m        & 2021-10-11 & ~~~\nodata    &  200 s       & 6000--7500 \AA: F675W   & Fesen  \\
\underline{\bf{Space Based} ~~~~~~~~~}         &                &              &    &                \\
{\it HST} WFPC2 & 1999-06-13 & U52B0*     & 1500 s       & 6000--7500 \AA: F675W   & Fesen  \\
{\it HST} WFPC2 & 2000-01-23 & U59T0*     & 1000 s       & 6000--7500 \AA: F675W   & Fesen  \\
{\it HST} WFPC2 & 2002-01-09 & U6D10*     & 1000 s       & 6000--7500 \AA: F675W   & Fesen  \\
{\it HST} ACS/WFC & 2004-12-05 & J8ZM0*   & 2400 s       & 5500--7100 \AA: F625W   & Fesen  \\
{\it HST} ACS/WFC & 2004-12-05 & J8ZM0*   & 2000 s       & 6900--8600 \AA: F775W   & Fesen  \\
{\it HST} WFPC2   & 2008-02-05 & UA5K0*   & 2400 s       & 6000--7500 \AA: F675W   & Patnaude  \\
{\it HST} WFPC2   & 2008-06-02 & UA5K0*   & 2400 s       & 6000--7500 \AA: F675W   & Patnaude  \\
{\it HST} WFC3/IR & 2010-10-28 & IBID*   & 22.1 ks       & 9000--10700 \AA: F098M  & Fesen  \\
{\it HST} WFC3/IR & 2011-11-18 & IBQH*   & 22.1 ks       & 9000--10700 \AA: F098M  & Fesen  \\
{\it HST} WFC3/UVIS & 2018-01-10 &IDG8*    &  2000 s      & 5500--7100 \AA: F625W   & Fesen \\
{\it HST} WFC3/UVIS & 2018-03-05 &IDG8*    &  2000 s      & 5500--7100 \AA: F625W   & Fesen \\
{\it HST} WFC3/UVIS & 2019-01-25 &IDWN*    &  4380 s      & 5500--7100 \AA: F625W   & Fesen \\
 JWST NIRCam & 2022-11-05 & jw01947   & 3350 s       & 1.54--1.71 $\mu$m: F162M & Milisavljevic \\
{\it HST} ACS/WFC & 2022-11-27 & JEW81*    & 2360 s       & 5500--7100 \AA: F625W   & Fesen  \\
{\it HST} ACS/WFC & 2022-11-27 & JEW81*    & 2360 s       & 6900--8600 \AA: F775W   & Fesen  \\
\enddata
\tablenotetext{a}{For the Palomar observations, listed are the specific photographic emulsion and filter used.}
\label{data_sources}
\end{deluxetable*}

Cas~A's forward shock is seen as a nearly
continuous surrounding ring of thin nonthermal, continuum emitting X-ray wisps and filaments 
in {\sl Chandra} X-ray images first noted by \citet{Gotthelf01}. Proper motion measurements
of the outward propagation of these filaments has led to velocity estimates of $\sim$ 4500 to 6500 km s$^{-1}$ assuming a distance of 3.4 kpc (see Table~\ref{FRshock}).
However, recent X-ray analysis suggests an average value around 5800 km s$^{-1}$ and velocities as high as 7500 km s$^{-1}$ \citep{Vink2022}.

Unlike the situation of Cas~A's forward shock,
determining Cas~A's reverse shock velocity is not straightforward. This has lead to a range of values derived from both
evolutionary expansion models (Table~\ref{model_speeds}) or via
direct expansion measurements (Table~\ref{rs_speeds}). 
At present, there is no firm
consensus as to the
precise velocity of Cas~A's reverse shock
except to say values generally lie between 1000 and 4000 km s$^{-1}$
and that it likely varies significantly at different points around the remnant's rim.
Indeed, \citet{Vink2022} find that it actually does not expand
outward in a region along the west rim  but instead actually
moves inward at a velocity of nearly 2000 km s$^{-1}$ (Table 3).

The changing location and hence motion 
of the reverse shock over time can be estimated through
noting a sharp rise in the radio or X-ray emission with increasing radius by comparing
radial 1D intensity profiles or by stretching different 2D images
taken at different epochs.

Cas~A's reverse shocked ejecta is marked by a bright optical, infrared,
X-ray, and radio shell
some $\sim95'' - 115''$ in radius
displaced roughly $15''$  to the northwest from 
the remnant's center of expansion (CoE)
determined from the
motions of ejecta knots
\citet{Thor01}.  This displacement is comparable in the optical \citep{Reed95},
radio \citep{Anderson1995,Arias2018}, and X-rays 
\citep{Kor98, Gotthelf01,   Delaney03, Vink2022}.

This northwest displacement of the reverse shocked ejecta shell from remnant center is shown Figure~\ref{circles}. Dashed black circles are shown centered on the remnant's CoE  with red circles indicating the location of the remnant's bright emission shell,  best seen in the radio and infrared. Significant changes in Cas~A's emissions across a wide range of wavelengths are largely confined to this ring, consistent with the appearance of freshly reverse shock heated ejecta.

The reverse shock's velocity may also be estimated through  measurements of the advancement of ejecta brightening in the following reverse shock passage.  However, this requires multi-epoch observations of sufficient spatial and temporal resolution in order to witness and measure the sequential brightening of shock heated ejecta following ejecta -- reverse shock interaction.
At a distance of 3.4 kpc, a reverse shock velocity of 3000 km s$^{-1}$ results in a proper motion less than $0.2''$ yr$^{-1}$, a spatial scale difficult to detect or measure optically from the ground or in the radio and X-rays. Therefore, extended data sets spanning several years are required.

Whereas in the radio and X-ray the time period over which a Cas~A ejecta knot becomes sufficiently bright after initial interaction with the reverse shock is variable \citep{Smith2009, Patnaude09, Patnaude14}, it is less than around one year for optical knots to turn-on \citep{Morse04}. 
Unfortunately, measuring the motion of the reverse shock in the optical is often affected by the ejecta's highly clumpy nature, leading to regions showing no appreciable emission changes over several years or to 
significant gaps in the sequence of ejecta brightening. 

\citet{Kor98} and  \citet{Vink98} estimated average reverse shock
velocities of 1750 km s$^{-1}$ and
3500 km s$^{-1}$ respectively across the remnant, while
\cite{Delaney03} estimated a much lower reverse shock velocity at 1155 km s$^{-1}$
through measurements of the motion of the remnant's bright radio ring over the azimuth range of $115^{\circ} - 250^{\circ}$. They viewed this value as consistent with
their hydrodynamic, power-law expansion simulation which gave a velocity of 1270 km s$^{-1}$ based on Sedov-Taylor models   \citep{True1999}. On the other hand,
\citet{Morse04} found much higher reverse shock velocities 
along the remnant's northwestern rim  based on proper motion measurements of optical {\it Hubble Space Telescope} 
({\sl HST}) images covering a time span of just two years, between 2000 and 2002. They estimated  the reverse shock velocity in thie region to be
$\sim 3000$ km s$^{-1}$.

Subsequent studies also arrived at very different velocities.
\cite{Schure08}
estimated a broad range of reverse velocities, 990 -- 3160 km s$^{-1}$, through
hydrodynamical simulations of the progenitor's circumstellar medium shaped by
a red giant wind followed by a brief Wolf-Rayet fast wind phase.
\cite{Hwang12} used a power law density envelope expansion simulation of
\cite{Matzner99} to model reverse shock propagation in Cas~A and found reverse
shock velocities in the range of 2400 -- 2600 km s$^{-1}$.  

In contrast, \cite{Helder08} reported a much lower reverse shock velocity along parts of the 
western half of Cas~A where they found the reverse shock to be nearly stationary or
inward moving at over 1000 km s$^{-1}$ meaning an 
expanding ejecta knot moving outward at 5500 km s$^{-1}$ 
would experience an effective reverse shock
velocity around 6500 km
s$^{-1}$.  \citet{Sato2018} estimated even higher inward moving
reverse shock velocities in the frame of the ejecta of 5100 to 8700 km s$^{-1}$
based on proper motion measurements of inward moving nonthermal X-ray emission
features seen in the interior of Cas~A on {\sl Chandra} images.

Recently \citet{Vink2022}, using
{\sl Chandra} X-ray data in the 4.2-6.0 keV continuum band covering a period of 19 years starting in January 2000,
divided the remnant X-ray emission into segments with opening angles of
20$^{\circ}$ and then measured the expansion of the remnant's bright X-ray emission shell 
projected onto the plane of the sky. They 
found that while the
reverse shock moves outward along Cas~A's eastern limb at velocities between 2000 and 4000 km s$^{-1}$, it actually moves inward toward the
remnant's center along a portion of the western limb with a velocity of 1880 km s$^{-1}$
with a unusually small uncertainty of 22 km s$^{-1}$. 

If this were true, then  
the remnant's ejecta in this region 
expanding outwardly at a velocity
of $\sim 5000 - 6000$ km s$^{-1}$ 
would collide with an inward moving reverse shock at a relative velocity
as large as $\sim$8000 km s$^{-1}$. \citet{Vink2022} viewed such a high velocity, inward moving reverse shock helpful in explaining the presence of X-ray synchrotron emitting filaments seen along the remnant's western limb. Interestingly, recent
dynamical modeling which included a thin circumstellar shell, densest along the remnant's
western limb, leads to better matches for X-ray estimated forward and reverse 
shock velocity variations around Cas~A's rim \citep{Orlando22}.

Finally, the most recent observational estimates are those of 
\citet{Wu2024} who, using the same 19 years of {\sl Chandra} data  as \cite{Vink2022} reported similar velocity estimates to those of \cite{Vink2022}.
They
determined 
the reverse shock is moving outward in the frame of the
observer at a velocity of 
$3950\pm 210$ km s$^{-1}$ 
in the NW (PA = 315 to 5)
similar to \citet{Vink2022}'s estimate of $4000 - 4500$
km s$^{-1}$, 
and $2900 \pm 260$ km s$^{-1}$ in the 
SE (PA = 15 to 135), comparable to $\sim 3000$
km s$^{-1}$ found by 
\citet{Vink2022}. 

Here we present an optical investigation of the velocity and location 
of the remnant's reverse shock front through the use of ground based, HST and JWST images of Cas~A. Although the data set span over seven decades (1951 to 2022),
because significant changes of the remnant's optical emission features can occur on the timescale of
a decade or less, our reverse shock velocity estimates are largely based on proper motions data typically spanning less than ten years. Combining transverse reverse shock proper motion measurements with ground-based radial velocity measurements \citep{MF13}, yield space velocity estimates on Cas~A's recent reverse shock. In addition, we also present {\sl HST}
images and ground-based spectral data on the structural and kinematic
effects that Cas A's reverse shock has on its dense optical ejecta knots.

The outline of the paper is as follows: 
The optical imaging and spectral data used in this study are described in 
$\S2$, with an overview of the remnant's optical emission along with descriptions on the limitations and on our methodology in studying Cas~A's reverse shock given in $\S3$. We present in $\S4$ our estimated reverse shock velocities at various locations in the remnant based on some archival ground based images but mostly using {\sl HST} images.
Structural and dynamical effects of the remnant's reverse shock on the dense optical ejecta as seen through both images and spectra are presented in $\S5$. Discussions of our key findings along with comparisons with previous results regarding Cas~A's reverse shock are presented in $\S6$. Conclusions are given in $\S7$.


\section{Data Sets} \label{sec:obs}
\subsection{Optical Images}
 
This study draws upon a unique series of optical and near infrared images taken from ground based and space-based observatories of
Cas~A spanning from 1951 to 2022 thereby covering 
roughly 70 years or roughly 20\% of the remnant's currently estimated 350 yr lifetime.  
The earliest deep optical images were made using a variety of photographic emulsions on glass plates using the Hale
5m telescope at Palomar Observatory between September 1951 and September 1989.
Information regarding the filter and photographic emulsion used in these
observations can be found in Table~\ref{data_sources}\footnote{A gap in the
Palomar 5m photographic record of Cas~A between 1958 and 1965 was due to the
nearly simultaneous retirements of Walter Baade and Rudolph Minkowski from the
staff of Mount Wilson and Palomar Observatories \citep{vdb76b}}. The first
Palomar image was taken in September 1951 by  W.\ Baade just 
two weeks after F. Graham Smith announced the first accurate
position of Cas~A based on radio measurements \citep{Smith1951}.


The Palomar 5m images represent the best ground-based photographs taken
of Cas~A during the period from 1951 to 1989.  
For consistency in comparison with {\sl HST} images,
we  analyzed mainly Palomar images taken with red filters, as the remnant is
substantially brighter at red optical wavelengths 
($6000$ \AA  ~$-$ 7500 \AA).  

These photographic plates were first
cleaned of surface dust and then digitized using the high-
speed digitizing scanning machine DASCH  \citep{Simcoe2006,Laycock2010} 
at the Harvard–Smithsonian
Center for Astrophysics. WCS image coordinates were then
applied using USNO UCAC3 catalog stars \citep{Zach2010},
along with several additional stars
located within and around the remnant\footnote{The best of the 
digitized Palomar plate images in fits format are available in an on-line  data set. See Appendix B.}. For additional details, see
\citet{Patnaude14}.

A few additional ground-based images used in 
this study were taken with the MDM Observatory's
2.4m Hiltner telescope 
at Kitt Peak. These images obtained in the years
1996, 1998, and 1999, were taken using an R band filter on 
in direct imaging mode with a Loral $2048 \times 2048$ CCD
with a scale of 0.341$''$ pixel$^{-1}$. Total exposure
times varied from 900  to 2400 s. A more recent sub-arcsecond 
red filter image of the remnant was taken with the MDM 2.4m 
in October 2021 using a filter identical to the {\sl HST} WFPC2 F675W filter.

Much higher resolution images of Cas~A were obtained using the 
{\sl HST} employing the WFPC2, ACS, and
WFC3 cameras between June 1999 and November 2022  using a variety of broad and
narrow passband optical and infrared filters. 
These data were obtained through a series of Guest Observer
programs (GO 7406, 8281, 9238, 9890, 10286, 12300, 12674, 14801,
15337, 15515, and 17210). Much of these data have already appeared in 
several papers regarding different aspects of the remnant's optical properties
\citep{Fesen01,Thor01,Morse04,Fesen06a,Fesen06b,HF2008,Fesen11,Patnaude14,Fesen16,Koo2018,Kerz2019,Weil2020}. 

Broadband filter {\sl HST} images cover the full  $\pm$6500 km
s$^{-1}$ radial velocity range from the chiefly red emission lines arising from Cas~A's main
shell of ejecta.  Below we briefly describe the various image data sets used in
this study and direct the reader to the papers listed above for more detailed
descriptions.  Table 1 lists only the
specific {\sl HST} observations of Cas~A used in this study. 

Multi-passband {\sl HST} images of Cas~A were obtained at several epochs using the
Wide Field Planetary Camera 2 (WFPC2) starting in 1999 and ending in 2008.  The
first set of images obtained in June 1999 only covered the remnant's
northwestern limb.  Later images covering nearly all of the remnant's bright
main shell were taken in January 2000 and January 2002.  Images were taken
using the F450W, F675W, and F850LP filters with exposures times ranging from $4
\times 400$ s to $4 \times 700$ s.  The primary optical and NIR emission lines
traced by these three filters are [\ion{O}{3}] $\lambda\lambda$4959, 5007,
[\ion{S}{2}] $\lambda\lambda$6716, 6731 plus [\ion{O}{2}] $\lambda\lambda$7320,
7330, and [\ion{S}{3}] $\lambda\lambda$9069, 9531, respectively.  Detailed
descriptions of these data can be found in \citet{Fesen01} and \citet{Morse04}. 
Additional WFPC2 images of portions of the remnant's
northwestern and southwest limbs were obtained on 5 February and 3 June 2008
but only in the F450W and F657W filters.

Somewhat higher-resolution images of Cas~A were obtained using the Wide
Field Channel (WFC) of the Advanced Camera for Surveys (ACS;
\citealt{Ford98,Pavlovsky04}) on 13 November 2003, 4--6 March 2004 and 4--5
December 2004.  The ACS/WFC consists of two $4096 \times 4096$ CCDs at an
average pixel size of $0\farcs05$ providing a $202'' \times 202''$ field of
view.  Images were taken in the four Sloan Digital Sky Survey (SDSS) filters
F450W, F625W, F775W, and F850LP (g,r,i, and z) at six target pointings covering
the entire remnant, including all known outlying ejecta knots.  Total
integration times in each filter were 2000 s, 2400 s, 2000 s, and 2000~s,
respectively. In order to compare with the WFPC2 F675W images, we combined the
F625W and F775W images. To remove cosmic ray hits, cover the $2\farcs5$
interchip gap, and minimize saturation effects of bright stars, a 2-point
ACS-WFC-dither-line (ACS Instrument Handbook; \citealt{Boffi03}) was used with
two exposures taken at each dither point for the six pointings in each of the
four filters.

Near-infrared images isolating mainly the remnant's [\ion{S}{3}]
$\lambda\lambda$9069,9531 and [\ion{S}{2}] $\lambda\lambda$10287--10371  line
emissions using the WFC3's F980M filter were obtained on 
29 October 2010 and 18 November 2011 using the Wide
Field Camera 3 (WFC3). For both observations, 12 sets of dithered 50 s
exposures were taken using the IR Channel HgCdTe detector and the F098M filter.
The IR Channel on WFC3 consists of a 1k $\times$ 1k HgCdTe detector yielding a
total field of view of $123 \times 136$ arcsecond with a pixel size of
$0\farcs13$.  Although these images were primarily taken to explore the
remnant's NE and SW outer ``jets'' of high-velocity ejecta, they did cover over
90\% of Cas~A's main shell thus making them useful for ejecta proper motion
studies.

An additional set of optical images using the WFC3/UVIS camera
on {\sl HST} were obtained in
January and March of 2018 and in January 2019. 
Using the WFC3 F625W filter, a set of
dithered  4 $\times$ 500 s exposures were taken at nine separate pointings
covering the whole remnant in 2018 and 2019. However, due to a failure on the
on-board data recorder, the center set of 
three pointings for the January 2018
images were not obtained. 

Replacement images were subsequently obtained in March 2018 but with a rotation offset of 90 degrees to the east and west
regions due to the satellite's solar array configure constraints. Similarly, some of the 2019 WFC3 images suffered from shorter (but uncertain) scheduled exposure lengths due to on-board camera shutter failures. 
These issues, along with the lower throughput of the WFC3 vs.\ the ACS/WFC and the limited number of filters used, effectively means the 2004 ACS/WFC multi-filter survey images comprise the deepest and most complete {\sl HST} survey of the remnant.

Lastly, a more recent set of ACS/WFC images of six regions covering the whole remnant and
using the same four filters used in 2004 were obtained in November 2022. Because a few filter images were not usable for a variety of reasons, replacement images were requested and obtained in May 2023. 

\subsection{Optical Spectra}

In order to investigate the velocity dispersion of ejecta knots following their encounter with the
reverse shock, moderate-dispersion spectra of several ejecta knots and filaments 
were obtained in September 2011 with the MDM 2.4m Hiltner telescope. 
These spectra were taken using a Boller \& Chivens
Spectrograph (CCDS) which employs a Loral 1200 $\times$ 800 CCD detector. A
600 grooves mm$^{-1}$ grating blazed 4700 \AA \ was used to yield a wavelength 
coverage of 4500 to 5400 \AA \ with a spectral scale of 0.79 \AA \ per pixel.  Exposure
times varied from 300~s to 1500~s depending filament brightness.

A $1'' \times 5'$ slit was used which resulted in a measured FWHM of 2.5 pixels
corresponding to a measured spectral resolution of 1.95 \AA \ providing an effective R
$\simeq$2500 at the [\ion{O}{3}] $\lambda\lambda$4959,5007 lines and thus
a velocity resolution $\approx$ 120 km s$^{-1}$. The slit was sufficiently long to cover
much of the remnant's large and bright northern emission ring \citep{MF13}.

Standard pipeline data reduction for all MDM image and spectral data was performed using
IRAF/STSDAS\footnote{IRAF is distributed by the National Optical Astronomy
Observatories, which is operated by the Association of Universities for
Research in Astronomy, Inc.\ (AURA) under cooperative agreement with the
National Science Foundation. The Space Telescope Science Data Analysis System
(STSDAS) is distributed by the Space Telescope Science Institute.}.  This
included debiasing, flat-fielding, geometric distortion corrections,
photometric calibrations, and cosmic ray and hot pixel removal.  The STSDAS
{\it drizzle} task was used to combine single exposures in each filter.

\begin{figure*}[t]
\begin{center}
\includegraphics[width=\textwidth]{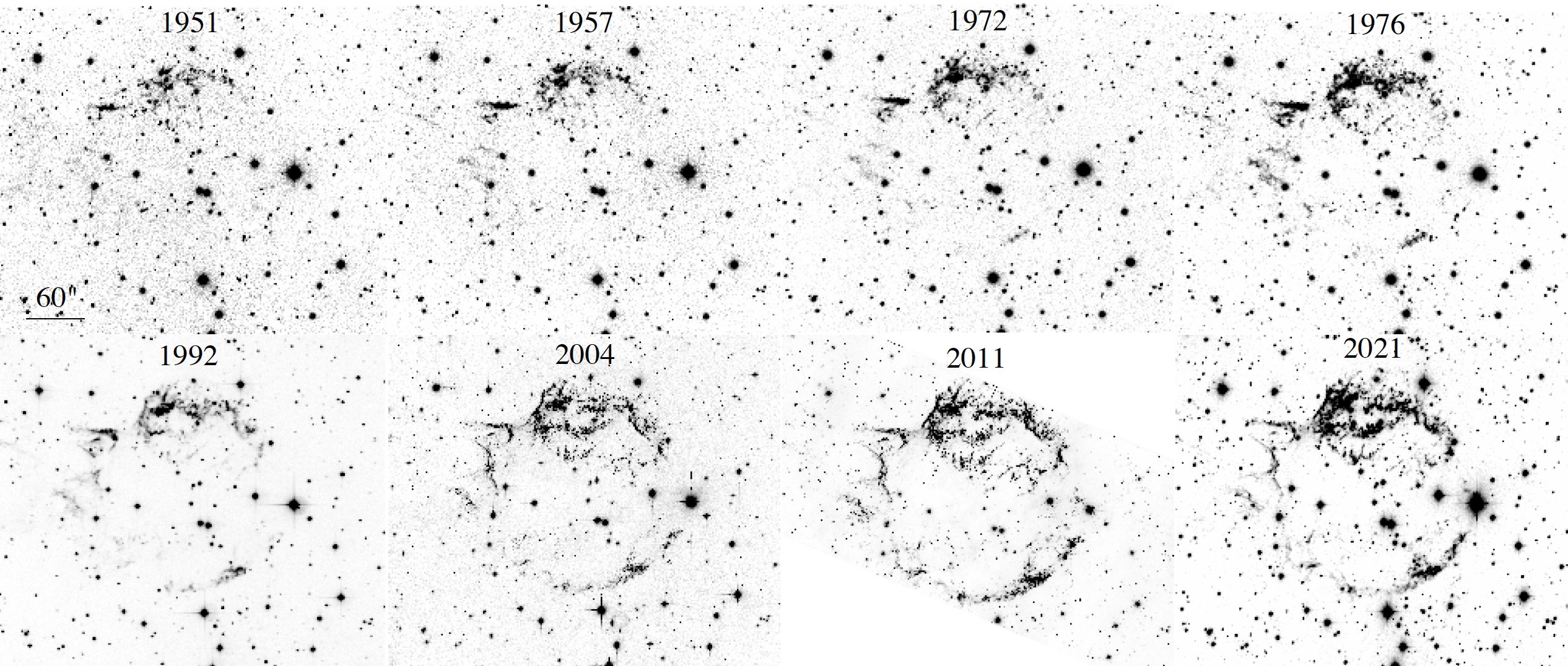}
\caption{Seventy years of evolution of Cas A's red optical emission between 1951 and 2021. 
Images between 1951 and 1976 are digitally scanned broadband red Palomar 5m plates, 1992 and 2021 are CCD MDM 2.4m images, with the 2004 and 2011 images red and near-IR {\sl HST} images (see Table 2 for details).  North is up, East to the left. 
\label{Evolution}
}
\end{center}
\end{figure*}

\begin{figure}[ht!]
 \centering
  \includegraphics[width=\columnwidth]{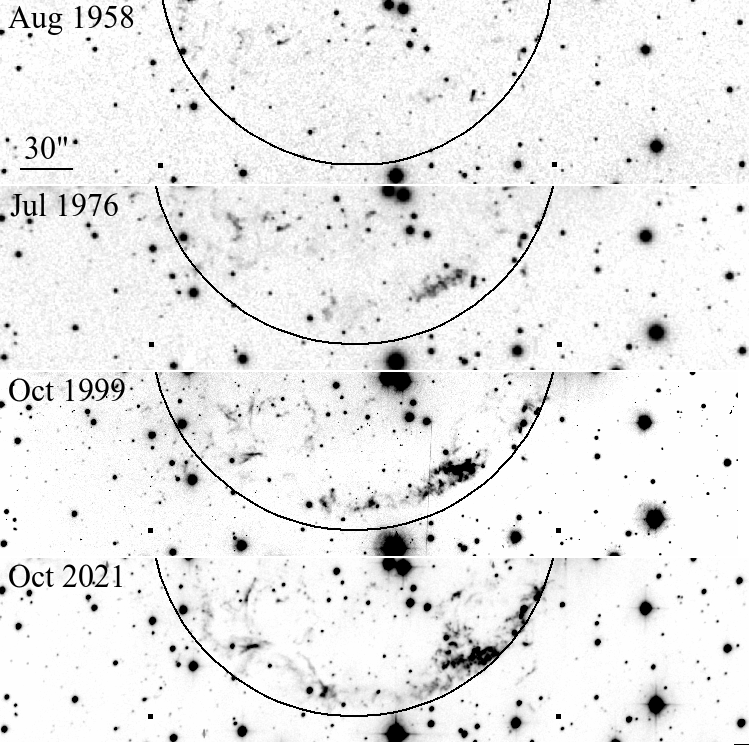}
\caption{Evolution of Cas A's southern optical emission. Circles shown are 115$''$ in radius centered on the 
center of expansion of Cas~A's radio emission knots and shell \citep{Anderson1995}. Note the southern's limb spherical morphology present in 2021.
North is up, East to the left. \label{Southern_Evol} }
\end{figure}

\section{Overview of Cas A's Emissions}


Until the realization of the existence and role of
Cas~A's reverse shock in the mid-1970's, changes in the remnant's optical emission and appearance remained a mystery. 
However with the evolutionary SNR models describing the formation of both forward and
reverse shocks \citep{McKee74, Gull1975}, the rapid appearance of Cas~A's expanding
ejecta knots was finally understood as due to shock heating as the
expanding ejecta collided with a slower reverse shock traveling behind the
faster forward shock, without resorting to interstellar or circumstellar cloud interactions.

Because Cas~A exhibits strong X-ray emission along with optical and IR emissions, parts of the reverse shock must travel through a low density component of the remnant's ejecta cloud ($n_{ej}$ $\sim 1-3$ cm$^{-3}$ \citep{LH03,Patnaude14}. Equating ram pressures, a $\sim$2500 km s$^{-1}$ reverse shock moving through a low density ejecta medium will drive $\sim$100 km s$^{-1}$ shocks into dense ejecta knots ($n \sim 1 - 10$ $\times 10^{3}$ cm$^{-3}$).
Even a 100 km s$^{-1}$ shock in ejecta composed of either pure oxygen or a mixture of O through Ca can produce
ion kinetic temperatures above $10^{6}$ K. However,
the enormous cooling rate of the electrons leads to
T$_{\rm e} \ll$ T$_{\rm i}$ where ions that produce strong optical and UV emissions are formed
\citep{Raymond18}.

Although largely optically invisible initially, portions of an ejecta knot quickly become heated to temperatures around $2 - 8 \times 10^4$ K but where T$_{\rm e}$ and T$_{\rm ion}$ are not equal.
Ejecta which are relatively dense and metal rich become subsequently visible emitting numerous UV, optical and IR line emissions. The strongest optical emissions from metal-rich SN ejecta include: [\ion{O}{1}] $\lambda\lambda$6300,6364, [\ion{O}{2}] $\lambda\lambda$3726,3729 and 
$\lambda\lambda$7319,7330, [\ion{O}{3}] $\lambda\lambda$4959,5007, [\ion{S}{2}] $\lambda\lambda$6716,6731, [\ion{S}{3}] $\lambda\lambda$9069,9531, [\ion{Ar}{3}] $\lambda\lambda$7136, and [\ion{Ca}{2}] $\lambda\lambda$7291,7330. An internal shock will eventually advance through whole of the ejecta knot increasing the mass of heated material and thereby increasing, at least initially, the knot's luminosity.
 
\begin{figure*}[t]
\begin{center}
\includegraphics[width=\textwidth]{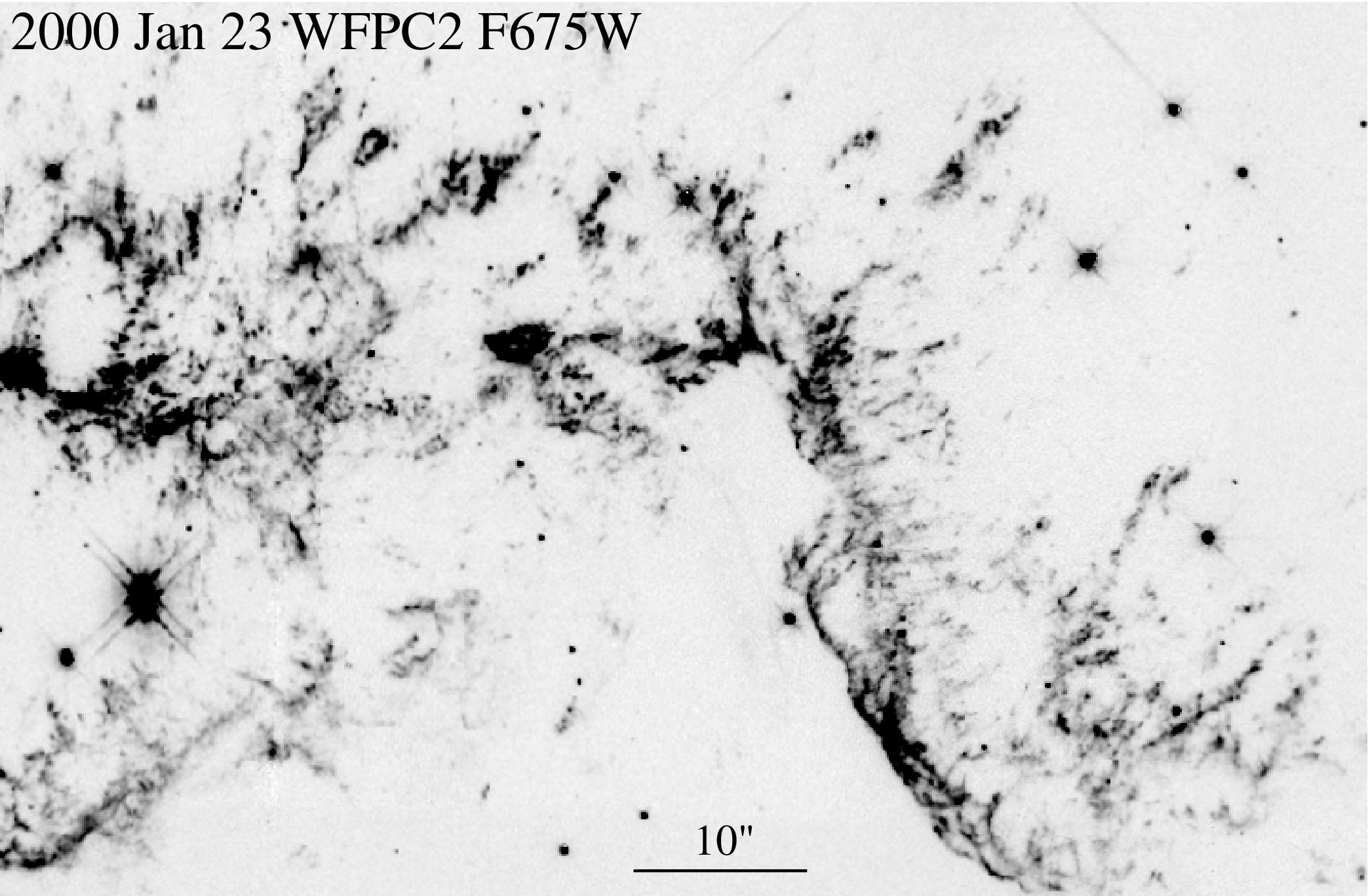}
\includegraphics[width=\textwidth]{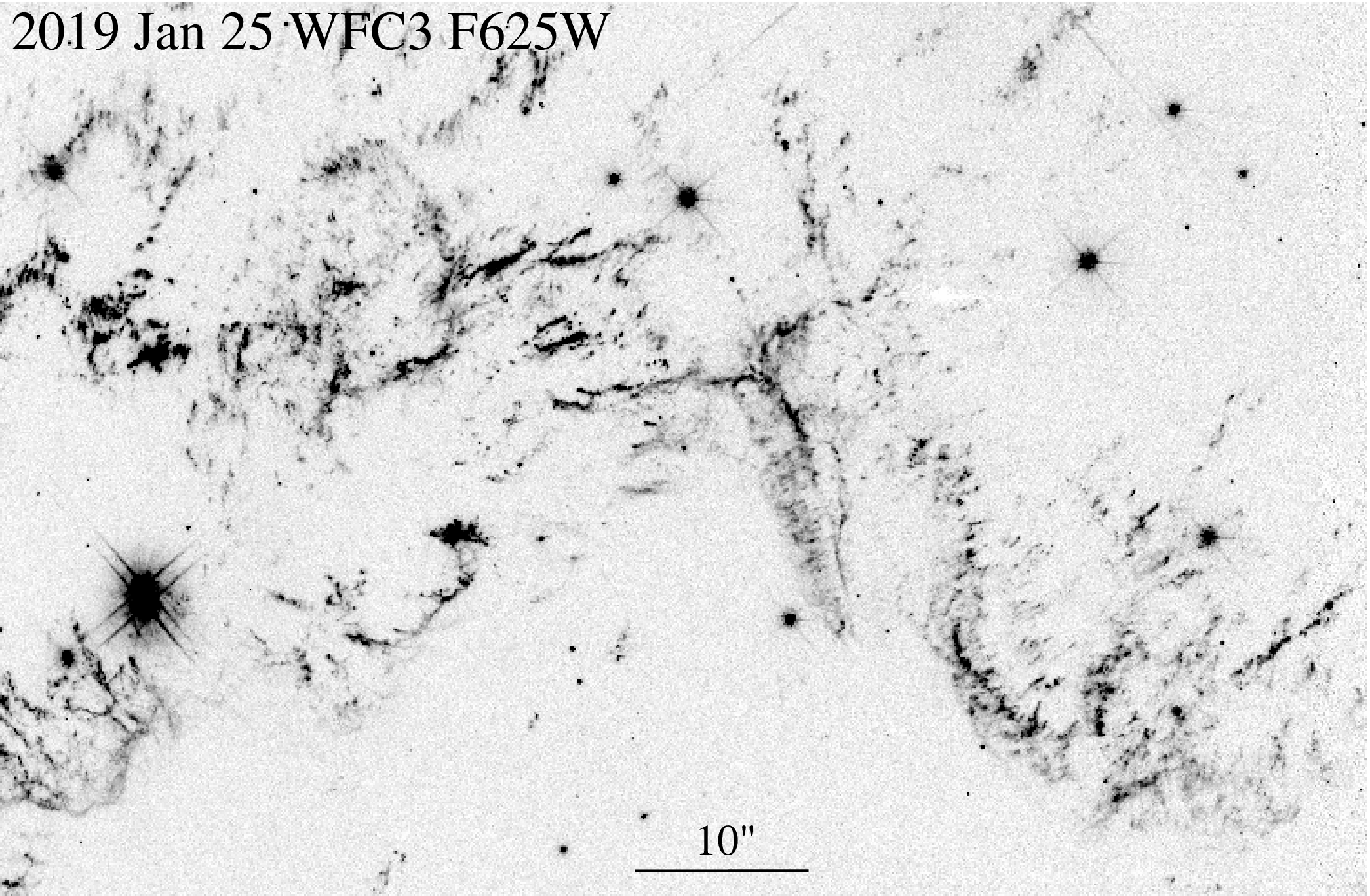}
\caption{Comparison of 2000 and 2019 red {\sl HST} images of exactly the same $80'' \times 50''$ northwest region illustrating large changes in Cas~A's optical morphology over a 19 year period.
North is up and East is left.
\label{NW_comparison}  }
\end{center}
\end{figure*}
 
 \subsection{Evolution of Cas A's Optical Emission }
 
Following the remnant's radio discovery in 1948 by \citet{Ryle1948}
and an accurate location  determined  subsequently by \citet{Smith1951}, associated optical emission was quickly searched for and discovered, with the first deep Palomar 5m images obtained in late 1951 (see Table 3 and \citealt{vdbd70}). From 1951 to the present,
there has been a steady increase of optical emission across the remnant with its structure undergoing substantial changes at both large and small scales. 

This evolution is readily apparent in Figure~\ref{Evolution} where we show a series of broad red passband images covering the period of 1951 to 2021. Palomar plates taken in the early 1950's 
showed only a relatively faint emission 
nebula mainly confined to Cas~A's northern rim and the remnant's NE jet. Starting
from the middle of the 1970's into the early 1980's, optical emission across a much wider area became visible.  Many of the remnant's early optical emission changes have already been discussed in detail by \citet{vdbk85} using Palomar images covering the 32 year time span of 1951 thru 1983.

\begin{figure*}[t]
\begin{center}
\includegraphics[width=\textwidth]{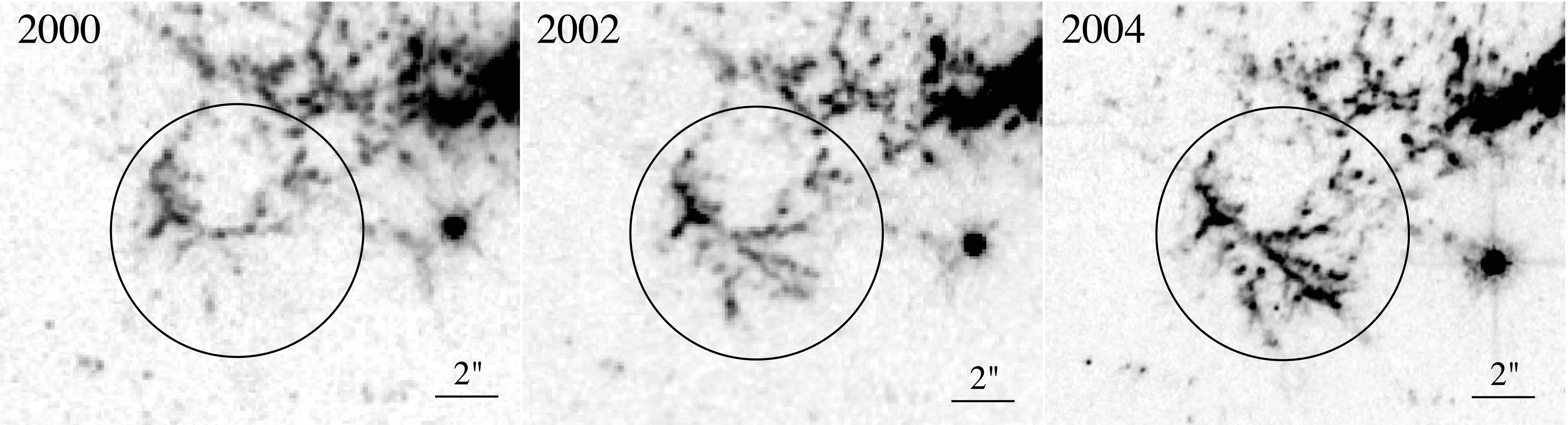}
\includegraphics[width=\textwidth]{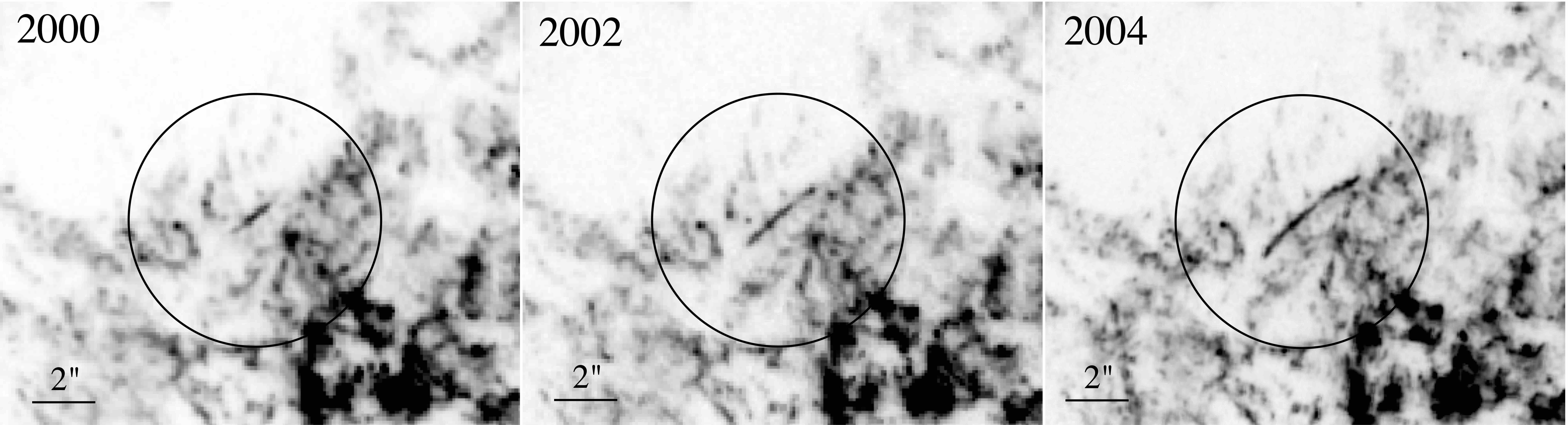}
\includegraphics[width=\textwidth]{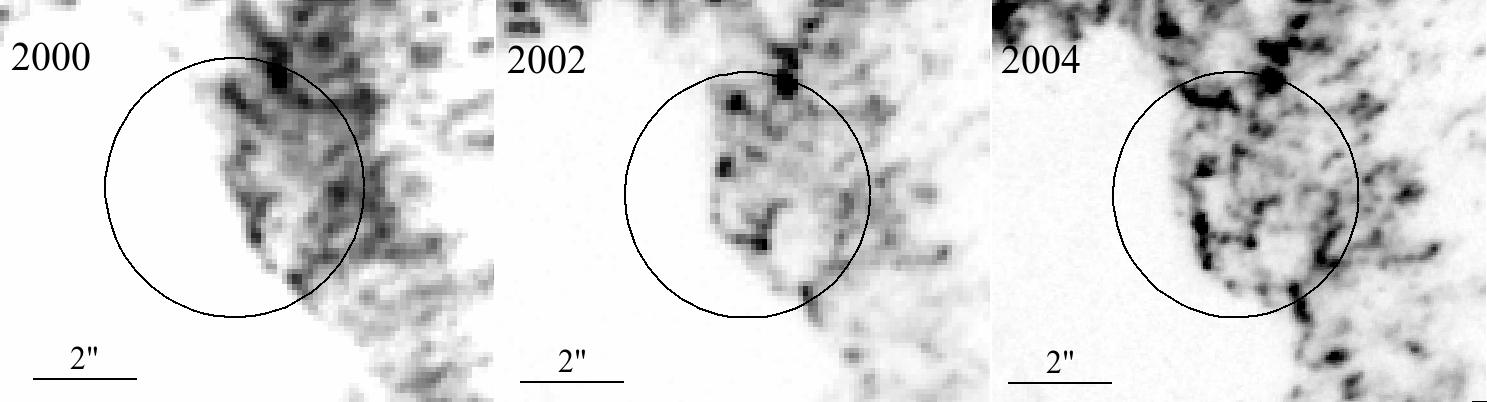}
\caption{January 2000 and January 2002 WFC2 images, and a March 2004 ACS {\sl HST} WFC2 image of ejecta in Cas~A's north regions showing ejecta brightness changes due to the reverse shock. Images are shown centered on the expanding ejecta. Circles in the top two panels are $4"$ in radius and $2.4''$ in the lower panels. \label{fingers}  }
\end{center}
\end{figure*}

However, by the turn of the century, Cas~A's emission grew noticeable in both extent and brightness, so much so, that by 2010 a nearly complete shell of emission became visible along most of its rim.  
This was largely due to emission appearing along its southern limb as shown in 
Figure~\ref{Southern_Evol}.
Whereas only a few pre-SN H-rich mass loss clumps (`QSFs') were visible in the south in the 1950's and 1960's, long and broad emission along the SW rim appeared by the 1970's. This was followed by a steady increase of additional rim emission. By late 2021, some optical emission can be seen nearly all along Cas~A's southern limb (Fig.~\ref{Southern_Evol}) in a fairly well-defined circular arrangement. This striking evolution along the remnant's southern rim will be
discussed further in the Discussion section.

From the earliest optical studies of Cas~A, it was realized that across timescales greater than about a decade, the remnant's emission changes were so substantial that made it hard to follow the motion or evolution of individual knots \citep{vdbd70,vdb71a}.
An example of gross changes in Cas~A's optical structure is presented in Figure~\ref{NW_comparison} which shows a $80'' \times 50''$ section of the remnant's northwest region imaged by {\sl HST} in January 2000 and January 2019. This figure highlights the extensive large-scale as well as sub-arcsecond changes across a large portion of Cas~A's optical emission shell over a time span of just two decades.

While images taken over the last decade can look remarkably similar on large scales, substantial smaller-scale changes (i.e., $0.1' - 0.5'$) have occurred throughout the remnant. Changes are most notably along much of the northwestern and southwestern limbs, and within the large northern ejecta ring with new recent emission extending more southward toward Cas A's center.

\subsection{Factors Affecting Measurements of Cas A's Reverse Shock}

In principle, it should be possible to accurately determine both the location and velocity of the reverse shock through sequential optical brightness changes of the expanding ejecta. However, there are a number issues which make it challenging to accurately measure the reverse shock's velocity even using {\sl HST} images. Below we briefly describe these and how they contribute to measurement uncertainties in both the location of the reverse shock and its velocity via the proper motions of brightening SN ejecta.

\subsubsection{Data Time Coverage}

Given an optical image data set spanning the better 
part of a century (see Table~\ref{data_sources}), one might initially conclude that these images would greatly facilitate the accurate measurement of the 
reverse shock's velocity and location via sequential knot brightening as ejecta light up due reverse shock heating. Unfortunately, in practice, this is generally not the case.

Firstly, the sparse and irregular cadence of both   ground-based and {\sl HST} images of Cas~A lead to an uncertainty stemming simply in determining the precise epoch when an optical knot became bright.  Moreover, as Figure~\ref{NW_comparison} illustrates, large-scale changes in Cas~A's optical emission make it sometimes impossible to follow individual ejecta knots with high confidence even over a time span of just one or two decades.
Consequently, we have mainly focused on relatively shorter time intervals and the high resolution data afforded by {\sl HST} images. 

\subsubsection{The Clumpy Nature of the SN Ejecta}

Cas A's main shell consists of many hundreds of optically bright and discrete knots and filaments of SN ejecta \citep{vdb71a,Reed95,Fesen01etal,Mil15}. This clumpy, inhomogeneous structure can lead to a highly discontinuous emission brightening 
sequence following reverse shock passage.  

Unlike the situation of X-ray emission associated with the roughly spherical forward shock moving through a relatively smoother and more continuous local CSM/ISM medium, 
the clumpy nature of optically bright ejecta leads to significant distortions of the reverse shock front.
This, in turn, can lead to jumps and gaps in the sequence of shocked brightened ejecta, creating a discontinuous 
brightening process at times for even for seemingly adjacent ejecta knots. 
Moreover, the uneven and random time coverage of quality optical images, especially in the archival ground-based data, adds significant uncertainty in determining when individual features became optically bright.

These problems in measuring the remnant's reverse shock due to the ejecta's clumpy nature can be seen in the three sets of images shown in Figure~\ref{fingers}. These show small sections in the remnant's bright northern rings of emission over a roughly four year time span. Although clear emission changes are visible, they are not in a purely radial direction relative to the remnant's center of expansion. 

The top panel illustrate a commonly observed situation where rapid emission changes are seen over a few epochs but which stop abruptly as through the reverse shock transitions out of an area of dense ejecta into a low density region.

Substantial emission changes of a small northern ring region are seen in the first two images taken in January 2000 and January 2002. But little subsequent changes is observed other than basically ejecta brightening over the subsequent 2.3 year time span, namely between January 2002 and March 2004. In fact, no substantial changes in the emission `fingers' are seen from 2004 through 2008.  

The middle three images illustrate a different case where strongly non-radial ejecta brightening is seen. The region shown is a small region in the remnant's northeastern section.
A small SE-NW oriented filament is seen to  
steadily grow in extent from January 2000 thru March 2004 with a curvature parallel to the remnant's NE rim. Such non-radial brightening, although clearly showing the reverse shock's progressive motion in the expanding ejects,  does not lend itself to accurate or meaningful reverse shock proper motion measurements.

Lastly, the lower sequence of images of 
Figure~\ref{fingers} reveal the formation of wavelets as the reverse shock front moves through the clumpy structure of ejecta knots. These images are of one of three obvious parts of the remnant showing this process, and lie in  a section  in the northwestern region of the remnant.

A small bubble or wavelet is visible in the January 2000 image which develops into a larger bubble emerging in between two bright ejecta knots in the 2002 image. By March 2004, this bubble has expanded and has merged with fainter bubbles above and below it, coalescing into a fairly coherent emission shock front seen in the last frame. Such instances 
where the reverse shock forms wavelets are not uncommon and emphasizes the clumpy nature of the ejecta on sub-arcsecond scales affecting and emphasize how the reverse shock does not always proceed as a smooth coherent wave.

\begin{figure*} [t]
\centering
\includegraphics[width=\textwidth]{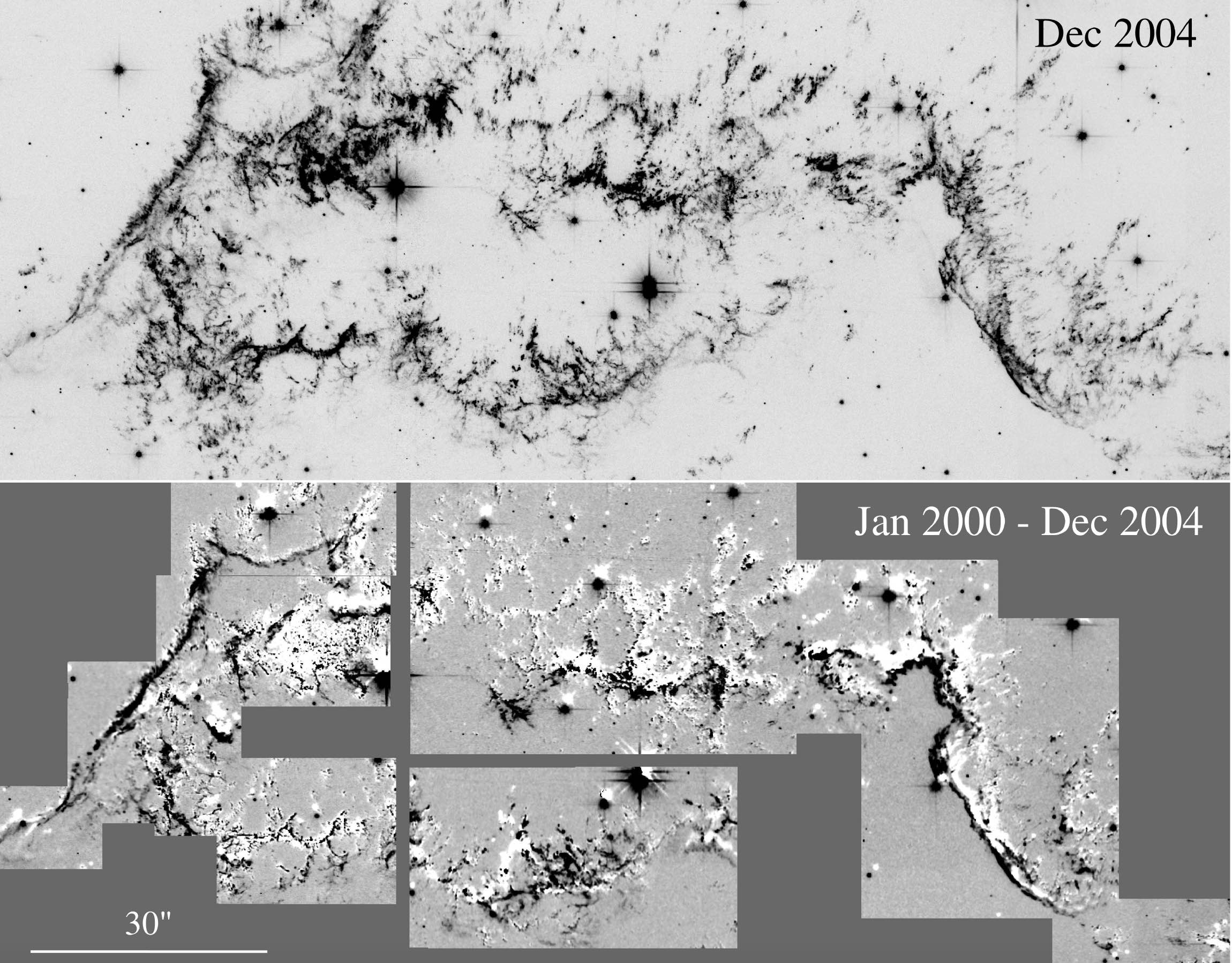}
\caption{Top panel: A December 2004 HST image of the Cas~A's northern portion covering the remnant's two large rings of optically bright ejecta. 
Bottom panel: A mosaic January 2000 and December 2004 difference images made comparing the 2004 image (black) with the 2000 image (white). 
The progression of the reverse shock is seen to be highly fragmented in the central sections, but exhibiting a coherent front seen as long and continuous black filaments along the northeastern and northwestern sections. North is up and East is left. \label{Diff_00_04}  }
\end{figure*}

\subsubsection{Timescale for Post-Shock Ejecta Brightening}

There is also the time delay between when an ejecta knot first encounters the reverse shock and when it becomes sufficiently bright optically to be readily visible.
Such delays place limits on accurately identifying the location of Cas~A's reverse shock at any particular time. Just how quickly Cas~A's dense ejecta become optically bright has not been firmly established observationally but is likely a function of knot density and reverse shock velocity.

\cite{Smith2009} found fairly rapid radiative cooling was likely give the difference between unshocked and shocked ejecta but gave no quantitative details.   
Dozens of the hundreds of the unresolved outer, high-velocity [\ion{N}{2}] emission knots have been observed to brighten in less than nine months \citep{Fesen11}. However, these outer ejecta are quite different from main shell ejecta. They lie outside both the forward and reverse shock fronts and are heated via shocks driven into them due to their high-velocity passage through the local CSM and ISM. Moreover, these knots exhibit [\ion{N}{2}] and H$\alpha$ line emissions and hence are chemically distinct from the remnant's main shell O,S,Ar,Ca-rich ejecta which show virtually no H or N emissions.

The examples presented in the top two panels of Figure~\ref{fingers} show  ejecta can grow substantially in less than two years. But other 
examples seen elsewhere make it clear that ejecta can become luminous within a period of a year or less. A brightening timescale of $\simeq$ 1 yr means that, in principle, it is possible to identify the location of Cas~A's reverse shock with relatively good precision.  A reverse shock even with a velocity as high as 5000 km s$^{-1}$ will not advance more than $0.5''$ in the observers rest frame in the time period when the knot becomes optically visible. This is a more favorable situation than that seen in the X-rays \citep{Patnaude14,Vink2022}.

However, {\sl HST} images have shown that a knot's emission morphology can significantly change within just a few years, thereby adding to measurement uncertainty. Such changes in a knot's emission structure are not unexpected given the time a 100 km s$^{-1}$ shock will taken to move through a typical optical ejecta knot of dimensions $\sim0.1''$ ($5 \times 10^{10}$ km), namely $\sim10 - 20$ yr which is approximately the observed lifetime of Cas~A's optical ejecta knots \citep{vdb76a,vdbk85}. 

\subsection{Methodology}

The various issues described above mean that obtaining accurate reverse shock information can be challenging in certain situations. They contribute to measurement uncertainty in both the location of the reverse shock and its velocity via proper motion of brightening ejecta. As has previously been done for Cas~A's forward shock, measurements of the remnant's ejecta are made in the rest frame of the 
observer (i.e., the sky frame). The development of new emission as seen in the frame of the expanding ejecta is taken as indicating the proper motion of the reverse shock. The difference between this motion and that of the ejecta in the sky frame is then the proper motion of the reverse shock in the observer's frame.

Reverse shock proper motion estimates were made in three
ways. The first was to align ejecta and then subtract images for relatively small regions
which highlighted emission changes of reverse shocked ejecta. Alignment accuracy was typically $0.25''$ for ground-based images and $0.04''$ for {\sl HST} data. Examples of image alignment are shown
in Figure~\ref{Diff_00_04}. Shifts in emission between the two images are viewed as due to the advancement of the reverse shock in the out going ejecta.

Alternatively, proper motion of the reverse shock was determined by comparing the coordinates of newly appearing ejecta in images taken at different epochs.  Knot position was determined through circular aperture
photometry to compute the central moment of flux. Since a majority of ejecta
emission are highly asymmetric, these results were then checked manually. If this and calculated position differed by more then 20\%, the result was discarded and a new ejecta feature was selected and measured.

Lastly, especially for the {\sl HST} images, we aligned small regions of images both in the sky frame or centered on moving ejecta knots to measure the positional shift of sequential regions of brightened ejecta. Sequences of different epoch images made by following ejecta expanding outward lead to the creation of very short movies which proved highly valuable to witnessing and thus measuring the brightening reverse shocked ejecta. 
We note that for ground-based images. all three methods were affected by differences (sometimes significant) in measured stellar FWHM.

\begin{figure*}[t]
\begin{center}
\includegraphics[width=\textwidth]{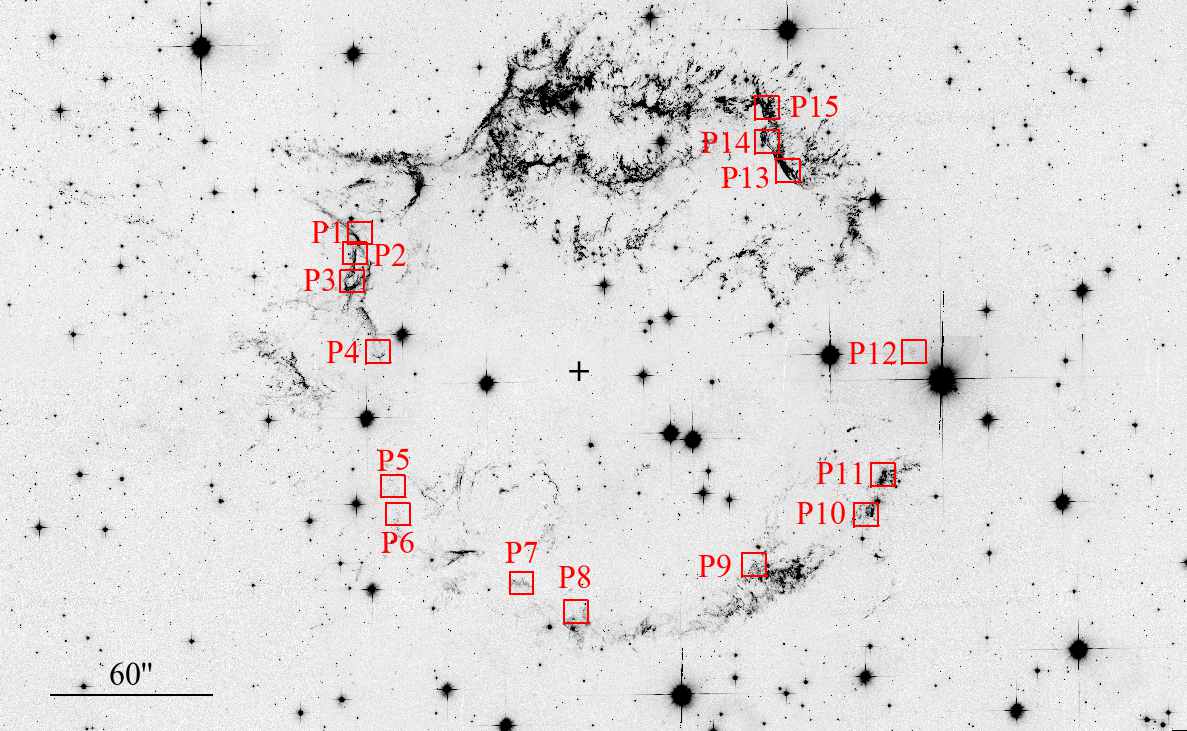}
\caption{Combined March 2004 {\sl HST} ACS F625W and F775W images of Cas~A showing locations where reverse shock velocity measurements were made. The black cross marks the remnant's optical center of expansion (\citealt{Thor01}.) 
\label{RS_map}  }
\end{center}
\end{figure*}


\section{Estimated Reverse Shock Velocities Derived from Optical Images}

Proper motion measurements were made for 15 selected regions in 
Cas~A's optical emission shell where especially large knot and reverse shock transverse motions are expected, namely, along the remnant's outer periphery.
Proper motion measurements  were translated into a transverse velocity assuming a distance of 3.4
kpc (\citealt{Reed95,Alarie14,Neumann2024}).  While ICRS positional coordinates on the World
Coordinate System (WCS) were applied to both Palomar and MDM ground-based
images,
positional accuracy for the Palomar images was limited to $\leq 0.35''$ 
due to the images being taken with the Ross corrector at the prime focus 
of the Palomar 5m telescope which distorted the plate scale by 
some 20\% going from field center to the image edge. 

A majority of Cas A's optical main shell emission ejecta do not lie along the remnant's outer limbs, and hence exhibit non-zero radial velocities. This is especially true for the remnant's two large overlapping rings of ejecta which have substantial radial velocities and thus relatively small proper motions (see \citealt{Mil15}). Because of the large radial velocities and consequently smaller proper motions in these northern ejecta rings, we chose not to make reverse shock measurements for this section of the remnant.

Nonetheless, the progression of the reverse shock within these bright northern rings could be clearly seen. This is shown in
Figure~\ref{Diff_00_04} where a four year progression of the reverse shock is shown in the difference image between the 2000 WFPC2 F675W image (white) to 2004 ACS F625W+F775 image (black). Both the 2000 and 2004 images are sensitive to [\ion{O}{1}] $\lambda\lambda$6300,6364, [\ion{S}{2}] $\lambda\lambda$6716,6731 and [\ion{O}{2}] $\lambda\lambda$7320,7330 emission lines. Due to the remnant's isotropic expansion, only limited sections of the Jan 2000 and Dec 2004 images were able to be accurately compared.

Black appearing features mark ejecta brightening due to the advancement of the reverse shock.
The ejecta in the central portions
of the remnant's northern wreath of ejecta show a jumble of short and separate
brightened ejecta of varying lengths and orientations.
In contrast, the far northeastern and far northwestern portions of  the image reveal a fairly continuous and coherent reverse shock front, seen as long black filaments, representing its four year movement in these areas.

It is important to emphasize that estimates regarding reverse shock velocities 
throughout the remnant based solely on transverse proper motions provide only lower estimates of the shock's true velocity and its radial distance from the center of expansion.  Thus, we have made use of the extensive optical radial velocity measurements described in \citet{MF13} to estimate true space velocities for the reverse shock at measured positions.

\begin{deluxetable*}{crrcccccr}
\centering
\tablecolumns{9}
\tablewidth{0pc}
\tablecaption{Proper Motion Measurements and Implied Transverse Velocities \label{PM} }
\tablehead{\colhead{Position}   &  \colhead{PA}  &  \colhead{d$_{\rm CoE}$} & 
\colhead{$\mu_{\rm ejecta}^{\rm sky~frame}$} &  \colhead{V${_{\rm ejecta}^{\rm sky~frame}}$}  & 
\colhead{V${_{\rm ejecta}^{\rm sky~frame}}$/V${_{\rm T}}$ }  &  
\colhead{$\mu_{\rm RS}^{\rm ejecta~frame}$} & \colhead{V$_{\rm RS}^{\rm ejecta~frame}$} &  
\colhead{V$_{\rm RS}^{\rm sky~frame}$}    \\ 
\colhead{ID}  &  \colhead{($\degr$)} &  \colhead{($''$)}  &  \colhead{($''$ yr$^{-1}$)} &  \colhead{(km s$^{-1}$)} & 
\colhead{m}  & \colhead{($''$ yr$^{-1}$)} &  \colhead{(km s$^{-1}$)}  &  \colhead{(km s$^{-1}$)}    }
\startdata
 P1  &  59  & 96  &  $+0.327\pm0.008$  &  $+5270\pm130$ & 0.99 & $0.249\pm0.026$  & $4010\pm420$ &  $+1260\pm420$  \\
 P2  &  64  & 90  &  $+0.271\pm0.007$  &  $+4370\pm110$ & 0.97 & $0.153\pm0.016$  & $2470\pm260$ &  $+1900\pm280$  \\
 P3  &  70  & 89  &  $+0.271\pm0.008$  &  $+4370\pm130$ & 0.98 & $0.152\pm0.017$  & $2460\pm270$ &  $+1920\pm300$  \\
 P4  &  88  & 73  &  $+0.216\pm0.010$  &  $+3480\pm160$ & 0.97 & $0.092\pm0.023$  & $1480\pm300$ &  $+2000\pm340$  \\
 P5  & 112  & 81  &  $+0.241\pm0.008$  &  $+3880\pm130$ & 0.99 & $0.131\pm0.021$  & $2100\pm330$ &  $+2110\pm360$  \\
 P6  & 128  & 89  &  $+0.261\pm0.008$  &  $+4210\pm130$ & 0.97 & $0.136\pm0.020$  & $2190\pm320$ &  $+2020\pm360$  \\
 P7  & 165  & 82  &  $+0.239\pm0.009$  &  $+3850\pm150$ & 0.97 & $0.148\pm0.022$  & $2390\pm350$ &  $+1460\pm400$  \\
 P8  & 179  & 95  &  $+0.269\pm0.007$  &  $+4340\pm110$ & 0.94 & $0.197\pm0.025$  & $3170\pm400$ &  $+1170\pm410$  \\
 P9  & 218  & 101 &  $+0.291\pm0.007$  &  $+4690\pm110$ & 0.96 & $0.228\pm0.027$  & $3670\pm440$ &  $+1020\pm450$  \\
 P10 & 242  & 118 &  $+0.342\pm0.008$  &  $+5510\pm130$ & 0.97 & $0.346\pm0.018$  & $5580\pm290$ &    $-70\pm320$  \\
 P11 & 253  & 128 &  $+0.368\pm0.009$  &  $+5930\pm150$ & 0.96 & $0.352\pm0.020$  & $5670\pm320$ &   $+260\pm350$  \\
 P12 & 273  & 125 &  $+0.359\pm0.009$  &  $+5790\pm150$ & 0.96 & $0.362\pm0.018$  & $5830\pm290$ &    $-40\pm330$  \\
 P13 & 314  & 107 &  $+0.306\pm0.007$  &  $+4930\pm110$ & 0.95 & $0.181\pm0.010$  & $2920\pm160$ &  $+2010\pm200$  \\
 P14 & 321  & 113 &  $+0.326\pm0.008$  &  $+5250\pm130$ & 0.96 & $0.360\pm0.022$  & $5800\pm350$ &   $-550\pm370$  \\
 P15 & 324  & 116 &  $+0.330\pm0.008$  &  $+5320\pm130$ & 0.95 & $0.269\pm0.016$  & $4340\pm270$ &   $+980\pm300$  \\
\enddata
\tablenotetext{}{Note: For calculating the transverse velocity of the ejecta in the sky frame a distance of 3.4 kpc is assumed. Deceleration values, m, assume an explosion date of 1671.}
\end{deluxetable*}

Below we describe reverse shock proper motion and hence velocity measurements organized by remnant azimuth regions. Our measurements, 
listed in Table 5, are not intended to be exhaustive but instead representative for various regions around the remnant's periphery.

Our ejecta proper motion and reverse shock velocity results are based on optical images are listed in Table~\ref{PM}. 
Because the motion of the
reverse shock is more
apparent when viewed in the reference frame of the ejecta, that is, without the added outward motion of the ejecta, we first
measured the motion of the ejecta in the sky frame.
Then the difference between
the proper motion of the ejecta and 
the motion of the reverse shock relative to the ejecta yields the velocity of the reverse shock in sky frame.

Table column entries are: 1) ejecta position ID, 2) position angle measured from due north counter-clockwise to the east,
3) projected angular distance from the \citet{Thor01} center of expansion (CoE),
4) typical ejecta proper motions for this region,
5) implied transverse velocities (V$_{\rm T}$) assuming d = 3.4 kpc and rounded to the nearest 10 km s$^{-1}$
(i.e., V$_{\rm T}$ = 4.74$\mu$ $\times$ 3400), 6)
the expansion parameter, m, (radius
$\propto$ t$^{m}$)
which here is ratio of estimated ejecta transverse velocity to the transverse velocity from the CoE assuming a remnant age of 333 yr for the 2004 image epoch\footnote{The possible sighting of the Cassiopeia~A SN by Giovanni Cassini in the summer of 1671 \citep{Soria13} is consistent with \citet{Thor01} and \citet{Fesen06a} explosion date estimates.}, 
7) reverse shock (RS) proper motion measured in the ejecta's rest frame 
 (typical RS proper motion uncertainty: 15\% -- 20\%), 
8) the reverse shock velocity experienced by the ejecta
(i.e., V$_{\rm RS}^{\rm ejecta}$  = 4.74$\mu_{\rm RS}^{\rm ejecta}$ $\times$ 3400),
9) the reverse shock in the sky frame which is the difference of ejecta expansion velocities from 
the reverse shock velocity in the ejecta reference frame; i.e., V$_{\rm RS}^{\rm sky}$ = 
V${_{\rm ejecta}^{\rm sky}}$  $-$ V$_{\rm RS}^{\rm ejecta}$.  

Our ejecta proper motion measurement uncertainties of 3-5\%, mostly using {\sl HST} images covering the $\simeq$5 year period January 2000 thru December 2004 images, are similar to earlier optical studies using images taken over longer time spans.
In the proper motion study of 102 of Cas~A's optical ejecta knots by \citet{vdb76a} using Palomar 5m images spanning 3--24 years (1951 thru 1975) quoted uncertainties were 2-4\% for time spans of 5 to 20 years but $\sim$12\% for time spans of 3--5 years. 

Our results for Cas~A's reverse shock are discussed in comparison to estimates made at other wavelengths in $\S6$.
What follows are short descriptions and explanations of our measurements of ejecta and reverse shock motions for the 15 selected regions along the remnant's periphery. These regions are 
marked in  Figure~\ref{RS_map}.

\begin{figure*}
\centering
\includegraphics[width=\textwidth]{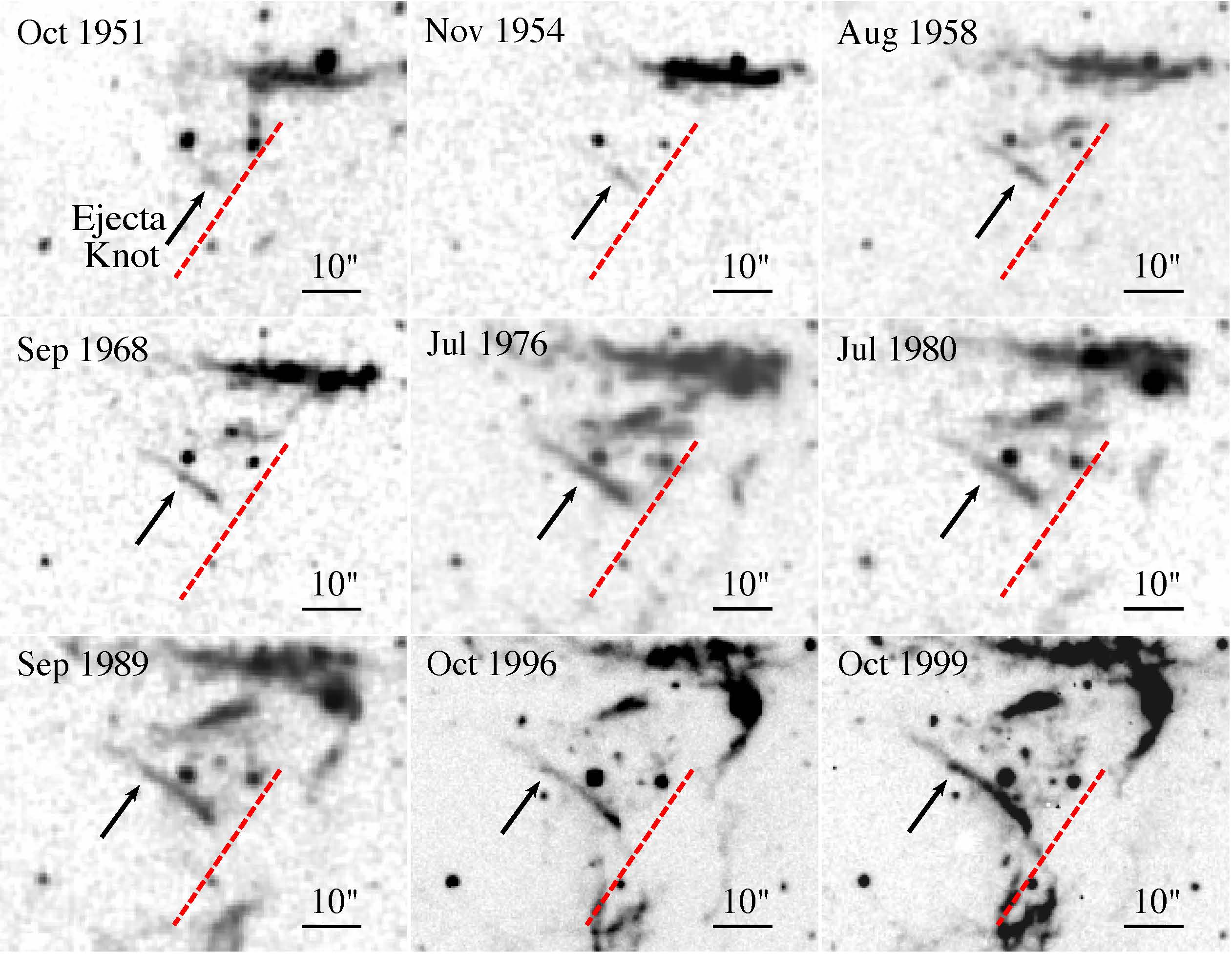}
\caption{P1: A coherent stream of ejecta located in the northeast region of Cas A is shown here as it evolves in ground-based images taken between 1951 and 1999. Ejecta in this stream becomes sequentially visible through interaction with Cas~A's slower expanding reverse shock front whose location in 1951 is seen as the stream's starting point or `base' indicated here by the red dotted line in these images which is fixed in the sky frame. An arrow in each image points to a discrete ejecta knot in the stream which is seen to travel steadily outward, i.e., toward the upper left of these frames. 
The position of the reverse shock, seen here as the start or base of this ejecta feature, also moves outward but far more slowly, measured by the increasing distance of the stream's base away from its 1951 location (the red dotted line) in the observer's rest frame. Note: After 1980, the stream's base begins to shift downward as the ejecta stream loses its dominate radial orientation. North is up, East is to the left in all frames.
\label{NE_streak} }
\end{figure*}

\begin{figure*}
\centering
\includegraphics[width=\textwidth]{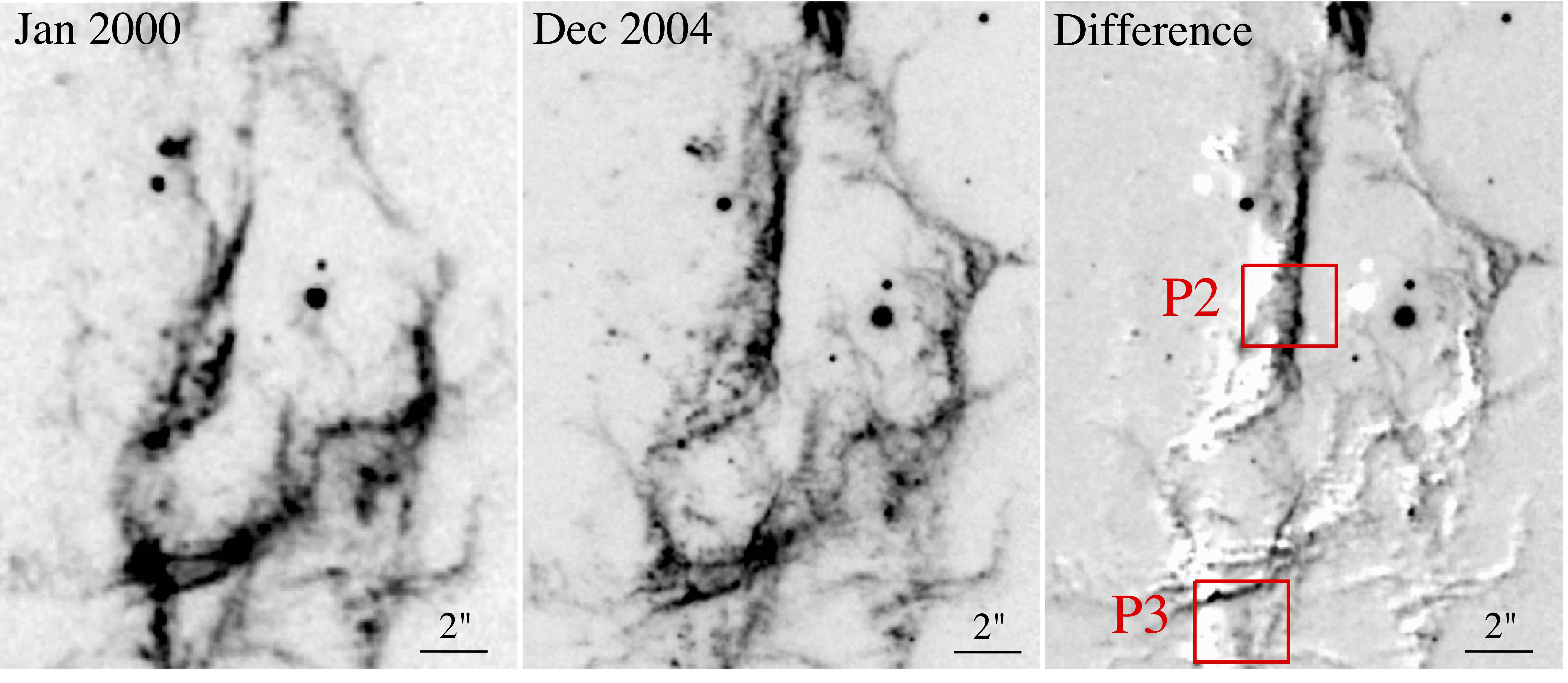}
\includegraphics[width=\textwidth]{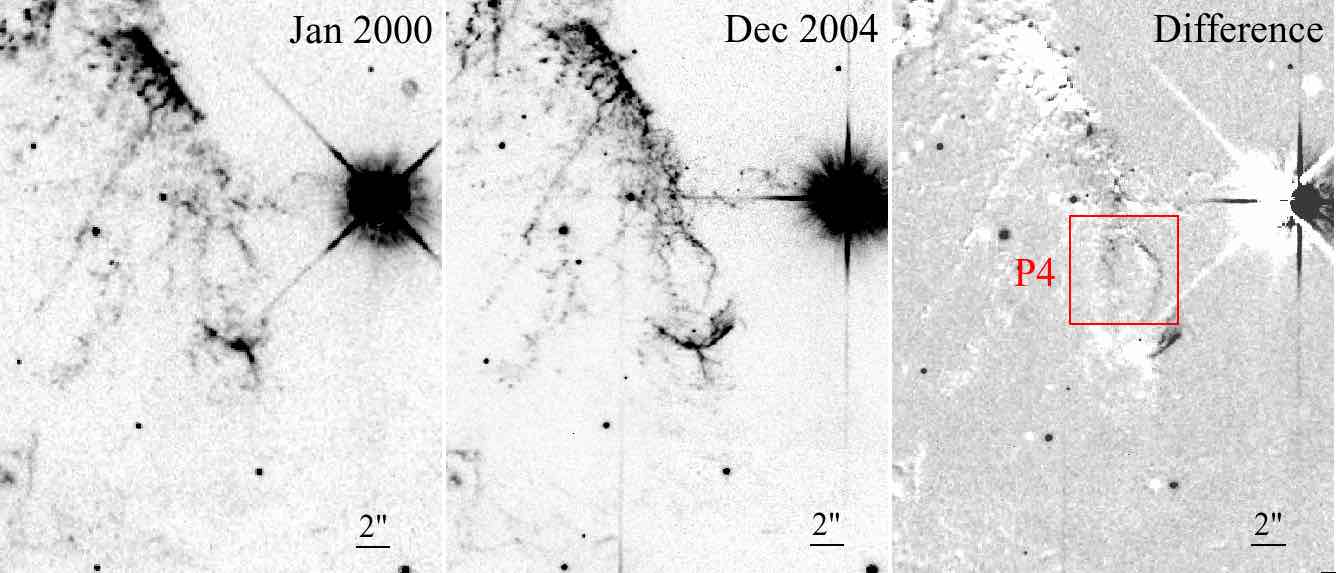}
\caption{Examples of the appearance of new emission features along Cas~A's eastern limb
reflecting the motion of the reverse shock at Positions P2, P3, and P4.
Images as shown are in the rest frame of the ejecta which are moving to the left (eastward); The slower expanding reverse shock leads to the appearance of new emission to the right. North is up, East is to the left.
\label{RS_East} }
\end{figure*}

\begin{figure*}
\centering
\includegraphics[width=\textwidth]{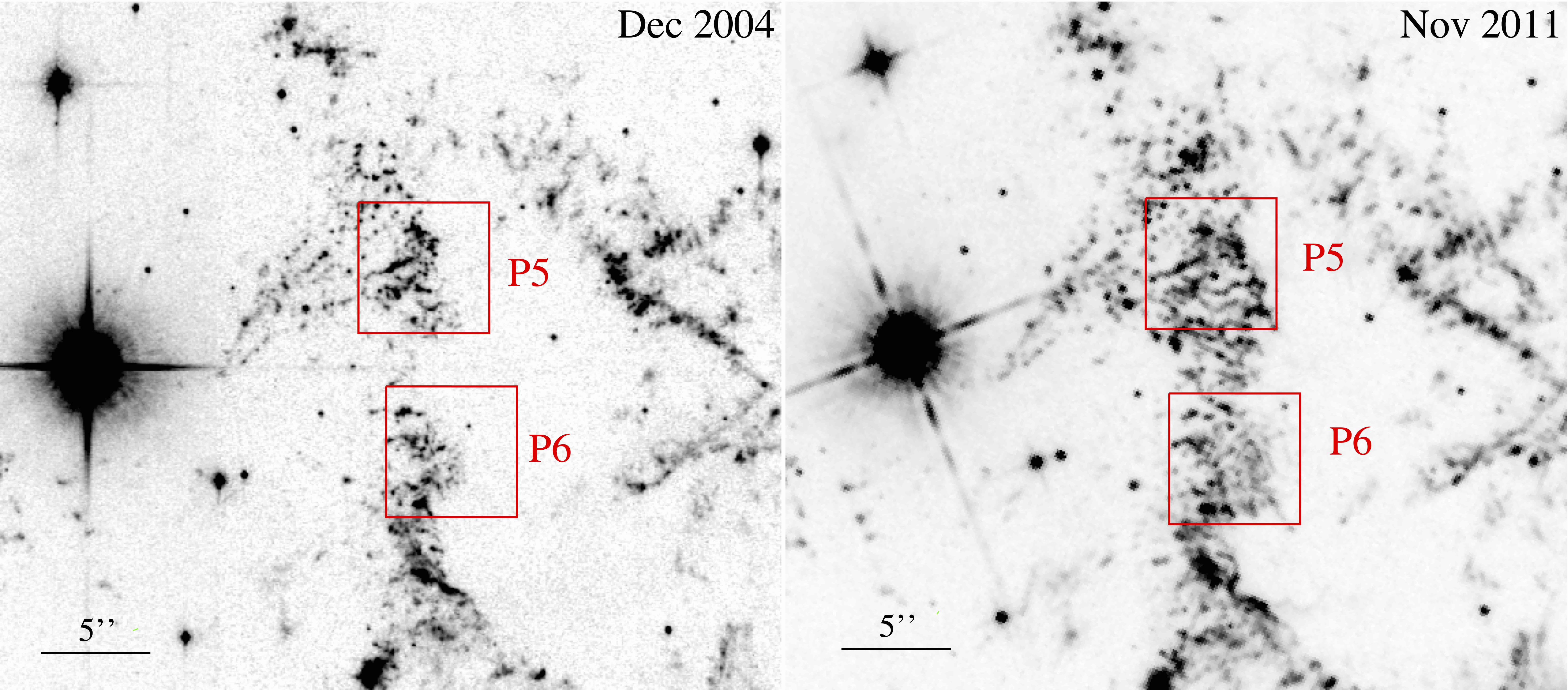}
\caption{December 2004 ACS/WFC F625W+F775W and November WFC3 F980M 2011 images of Cas~A's southeastern limb showing locations of P5 and P6.
in the rest frame of the expanding ejecta. The appearance of new emission within the red boxes  ($6'' \times 6''$) is due to the proper motion of reverse shock relative to the ejecta. North is up, East is to the left.  \label{RS_SE_East} }
\end{figure*}

\begin{figure*}
\centering
\includegraphics[width=\textwidth]{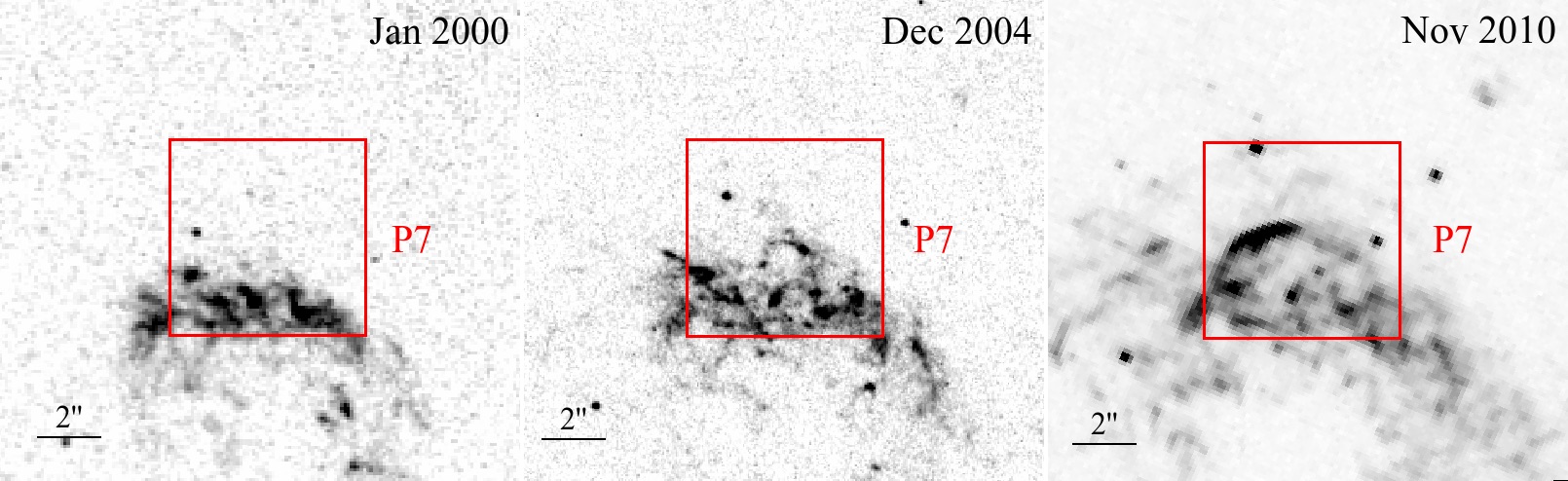}
\includegraphics[width=\textwidth]{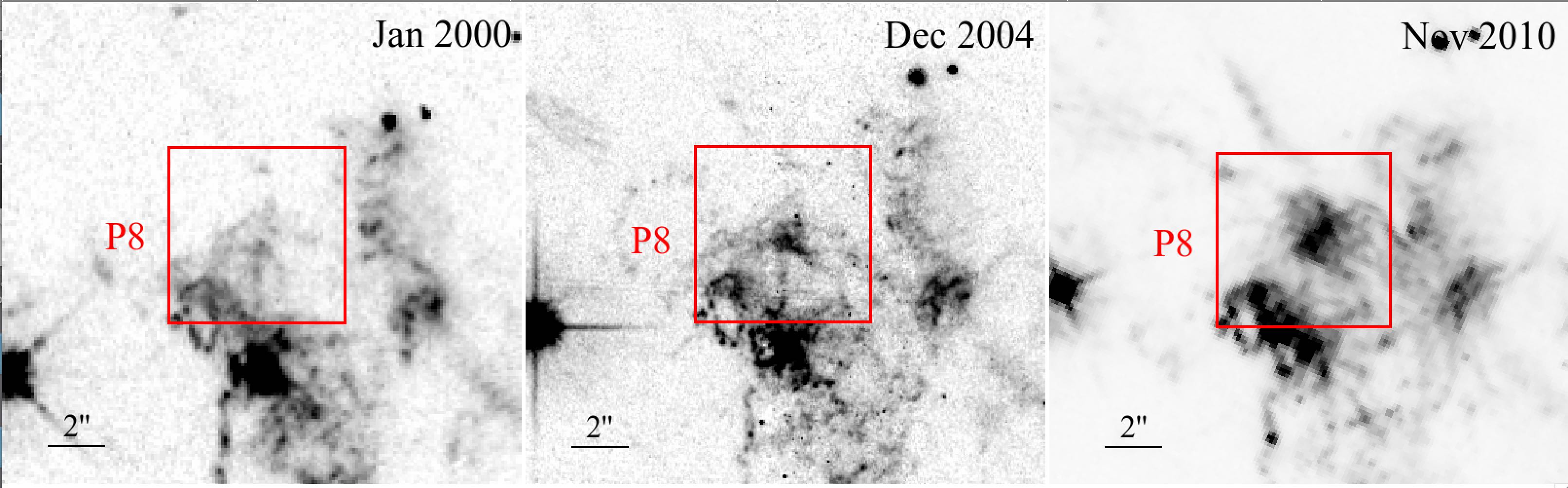}
\includegraphics[width=\textwidth]{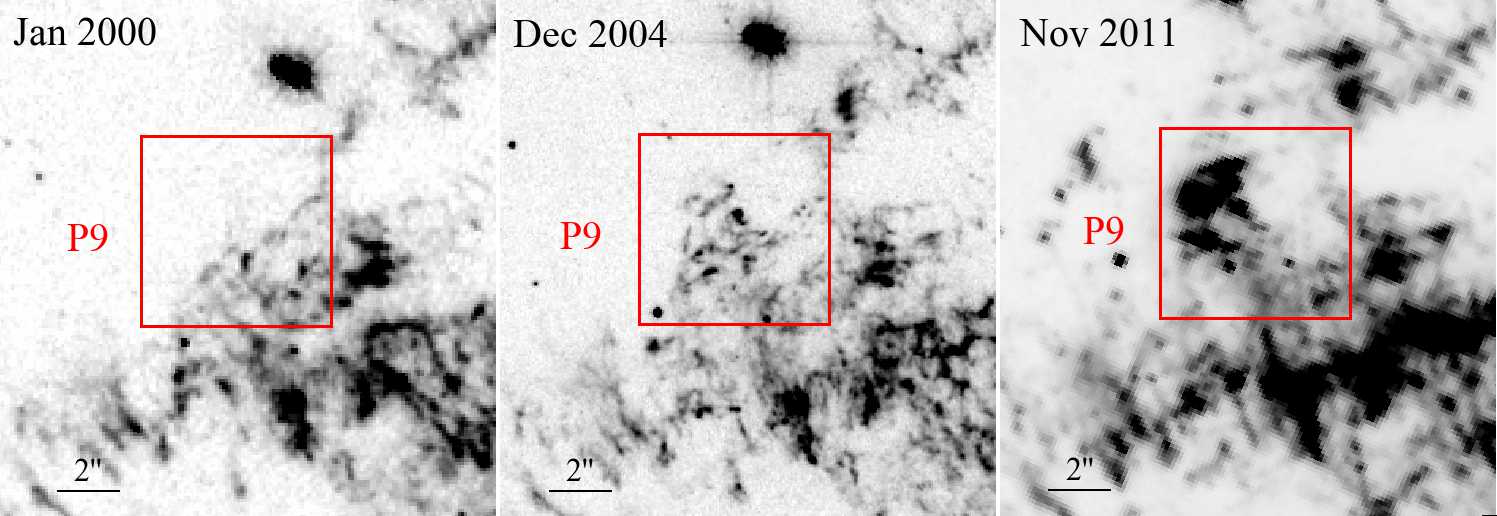}
\caption{January 2000 WFC2 F675W, December 2004 ACS/WFC F6225W+F775W, November 2010 WFC3 F980M 
and November 2011 WFC3 F980M images of Cas~A's southern limb showing the locations of P7, P8, and P9.
Images are shown in the rest frame of the expanding ejecta. The appearance of new emission in the red boxes ($6'' \times 6''$) is due to the proper motion of reverse shock relative to the ejecta. North is up, East is to the left.  \label{RS_P7P8P9} }
\end{figure*}

\begin{figure*}
\centering
\includegraphics[width=\textwidth]{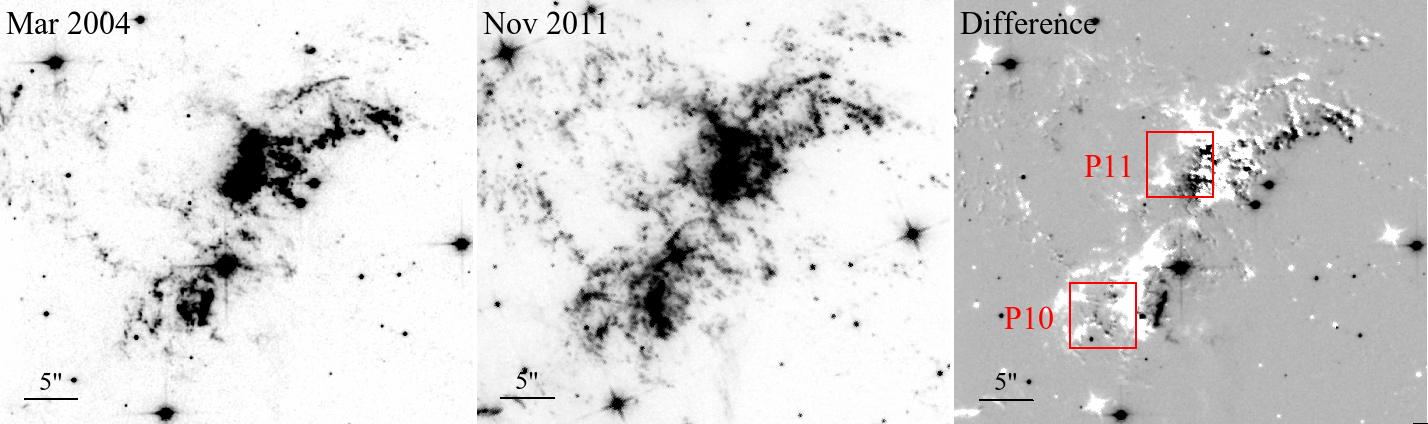}
\caption{March 2004 ACS/WFC F850LP and November 2011 WFC3 F980M images of Cas~A's southwestern limb showing the locations of P10 and P11.
Images are shown in the rest frame of the expanding ejecta. The appearance of new emission features (white) inside the red boxes ($6'' \times 6''$) are due to the proper motion of reverse shock relative to the ejecta. North is up, East is to the left.  \label{P10nP11} }
\end{figure*}


\begin{figure*}
\centering
\includegraphics[width=\textwidth]{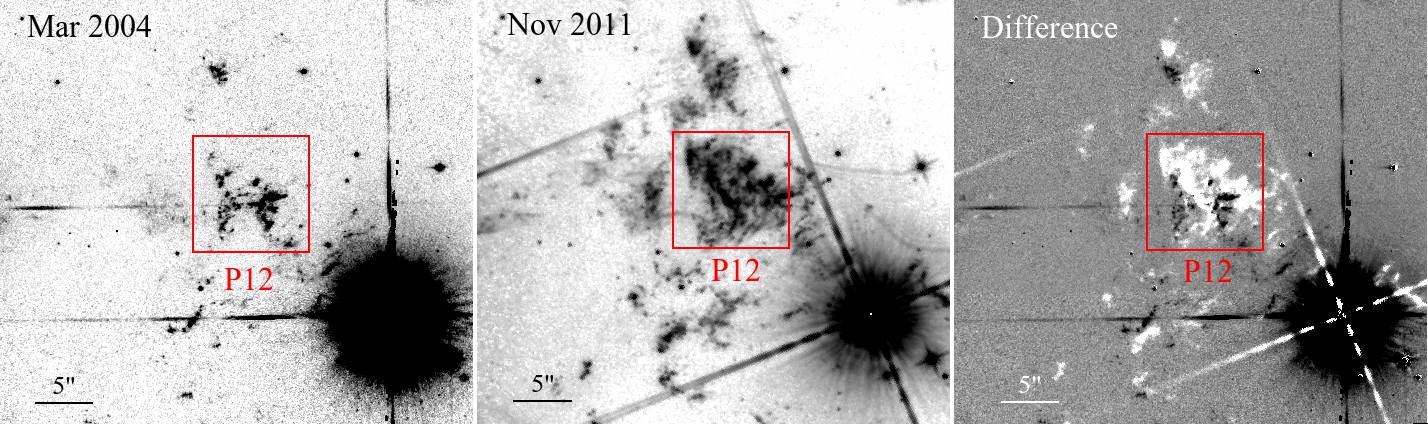}
\caption{March 2004 ACS/WFC F850LP and November 2011 WFC3 F980M images of Cas~A's western limb showing the location of P12
in the sky frame.  The appearance of new emission features (white) inside the red boxes between 2004 and 2011 ($10'' \times 10''$) is due to the proper motion of reverse shock relative to the ejecta. North is up, East is to the left.
\label{P12} }
\end{figure*}

\begin{figure*}
\centering
\includegraphics[width=\textwidth]{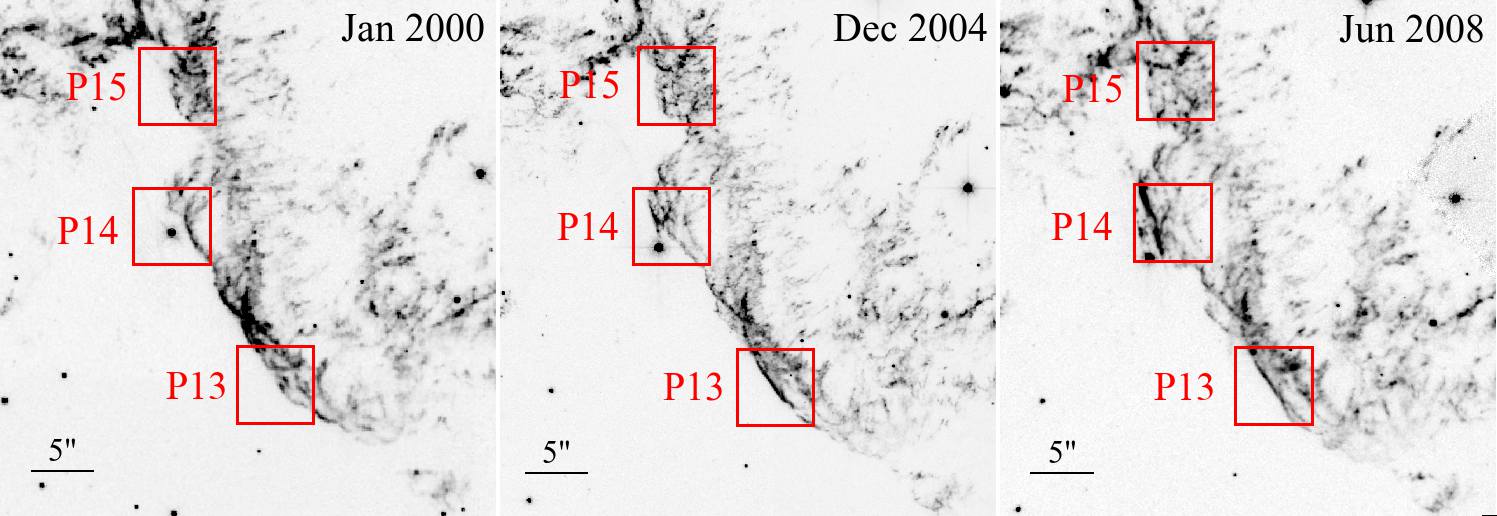}
\caption{January 2000 WFC2 F675W, December 2004 ACS/WFC F6225W+F775W, and June 2008 WFC2 F675W 
of Cas~A's northwestern limb showing the locations of the P13, P14, and P15 regions.
Images are shown in the rest frame of the expanding ejecta. The appearance of new emission in the red boxes ($6'' \times 6''$) is due to the proper motion of reverse shock relative to the ejecta. North is up, East is to the left.  \label{RS_P13_P15} }
\end{figure*}

\subsection{ P1: A Northeast Stream of Ejecta}

We begin with an elongated, streak-like feature located along Cas~A's northeastern limb just south of the large, bright Baade-Minkowski Filament No.\ 1.
This emission feature is a relatively rare instance where an apparent continuous stream of ejecta became and remained visible for more than 50 years.
While \citet{vdbk85} noted that this streak-like feature grew steadily in 
length for more than 30 years, they noted that the position of its starting base moved remarkably very little in the sky frame over those three decades.  

We view the starting point or base of this feature as marking the location of the reverse shock where this largely radially orientated out flowing ejecta stream 
impacts the remnant's reverse shock, thereby making its ejecta optically bright. In this view, the proper motion of the streak's base in the sky frame is then a measure of the reverse shock's velocity.

Figure~\ref{NE_streak} shows the evolution of this ejecta feature as seen in nine images beginning in 1951 up through late 1999. In the 1951 image, the base of the streak is faintly seen just interior to an ejecta knot marked here by an arrow.
A dashed red line oriented roughly perpendicular to the emission streak marks the approximate location of the feature's starting point and hence the reverse shock location as seen in the October 1951 image.
(Note: The ejecta stream is not perfectly radial in 
orientation which leads to the streak's inner section and base to slowly drift southward, most obvious after 1980.)

The discrete ejecta knot first seen in the 1951 image and marked by arrows in the subsequent eight images, exhibits an outward proper motion of $0.327'' \pm0.008''$ corresponding to a transverse velocity of $5270 \pm 130$ km s$^{-1}$ assuming a distance of 3.4 kpc. We view this knot's motion is indicative of the ejecta stream's general outward expansion modulo its greater radial location and hence higher expansion velocity relative to more interior portions of this stream of ejecta. 

The location of the streak's base, which we
attribute as being the location of the reverse shock front due to the appearance of freshly reverse shock heated ejecta, can be seen to move outward like that of the ejecta knot, but much more slowly.
With the assumption that the intensity
of the stream's base location did not significantly vary during this  nearly 50 year time frame, we estimate the base's outward proper motion to be $0.078'' \pm0.016''$ in the sky frame corresponding to v$_{\rm RS}$ = $1260 \pm 260$ km s$^{-1}$ at d=3.4 kpc.

These transverse velocity values indicate a proper motion difference between ejecta and the reverse shock front of $0.249''\pm0.026''$ implying a transverse expansion velocity difference of $\sim$4100 km s$^{-1}$.
A reverse shock outward velocity of only 1260 km s$^{-1}$  means the stream's ejecta moving outward at around 5270 km s$^{-1}$ encounters the reverse shock front here at a velocity $\sim$4000 km s$^{-1}$ in its rest frame. This value is listed in column 8 in Table 5.

We note that after correcting for the stream's observed +800 km s$^{-1}$ radial velocity \citep{MF13}, the
ejecta stream has a space velocity of $\sim5300$ km s$^{-1}$ which is nearly the same as
as that of the $\simeq$5400 km s$^{-1}$ forward shock in this region \citep{Delaney03,Patnaude09,Hwang12,Vink2022}. (A short movie of the evolution of this NE ejecta stream in the sky frame can be found in Appendix B: P1-sky-frame.mp4)

\subsection{Eastern Ejecta Positions: P2 and P3} 

We selected two ejecta filaments along the remnant's northeastern and eastern limbs,
some 10 to 15 arcseconds south of P1 which would provide additional estimates of the motion of the reverse shock
a short distance away from P1 and  farther away from the southern edge of the remnant's NE jet.
Figure~\ref{RS_East} shows Jan 2000 to Dec 2004 red broadband {\sl HST} images
of these filament regions with the two positions where we measured ejecta proper motions
and the motion of the reverse shock front relative to the ejecta rest frame, shown here
as the black features in the 2000-2004 images.

Both P2 and P3 are parts of a largely N-S complex of filaments. Both positions P2 and P3 exhibit similar transverse ejecta proper motions with similar implied transverse velocities $\sim$4400 km s$^{-1}$, and similar reverse shock velocities of 1900 km s$^{-1}$ (see Table 5). Measurement uncertainties on the reverse shock proper motions at P2 and P3 were affected by the WFPC2's resolution ($0.1''$ per pixel) and the weak detection of the reverse shock emission. 

\subsection{Eastern Ejecta Position: P4} 

There are only a few good optical emission features along the eastern portion of Cas~A to observe the reverse shock. One of these is a region roughly 24 arcseconds south of P3. This feature, labeled P4, is shown in the lower panels of Figure~\ref{RS_East}.  This region lies closer to the remnant's expansion center than P2 or P3
with expected lower ejecta transverse velocities $\sim$3500 km s$^{-1}$ compared to the $\sim$4400 km s$^{-1}$ seen at P2 or P3. Due to the motion of the reverse shock moving through this ejecta region creating wavelets, our measurement here is quite uncertain, which is reflected in a larger than usual error bar.

\subsection{Southeastern Ejecta Positions: P5 and P6}

These positions sample relatively faint optical emission features but ones that lend themselves to
relatively good proper motions. P5 covers an unusual optical emission structure consisting of numerous
small ejecta knots, only partially resolved in the 2004 ACS/WFC images. Region P6 lies immediately south of P5 where
few additional discrete ejecta knots are visible. Both regions share similar ejecta transverse and estimated
reverse shock velocities.

\subsection{Southern Ejecta Positions P7, P8, and P9}
Regions P7  and P8 are small emission patches lying almost due south of Cas~A expansion center \citep{Thor01}.
In the ejecta rest frame, images presented in  the upper panels of Figure~\ref{RS_P7P8P9}, show considerable emission for P7 and P8 becomes visible northward in the images taken in Dec 2004 and Nov 2010 relative to that seen in the Jan 2000 image.
Motion of the reverse shock here and for both regions were estimated based on
these ejecta reference frame comparisons. However, due to the highly inhomogeneous nature
of the new ejecta emission following interaction with the reverse shock, our values are seriously subjective and should be taken with caution.

Similarly to Positions P7 and P8 where proper motion measurements of the reverse shock are
uncertain, measuring its motion at Position 9 was particularly challenging (see bottom panels in Fig.\ 11). 
This is reflected in Table 5 also showing an even larger error for P9's reverse shock proper motion.

\subsection{Southwestern Ejecta Positions P10 and P11}

Ground-based images show virtually no emission in this area of the remnant before 1980
(see Fig.1). This together with the lack of {\sl HST} images taken in 2000 or 2002,
forced us to rely on the March and December 2004 images along with the 2010 and 2011 WFC3 near IR images.
In addition, because of the enormous changes in the emission features in this region between 2004 and 2010/2011 making connection between individual emission features wholly unreliable, 
we used the 0.742 yr separation between the 2004 March and December {\sl HST} AFC/WFC images to estimate ejecta proper motions. Fortunately, this gave consistent results across both P10 and P10
regions. As shown in Figure~\ref{P10nP11}, reverse shocked ejecta emission becomes noticeable to the northeast of the emission seen in 2000, in a largely radial direction back toward the remnant center. 

\subsection{Western Limb Position P12}

Although there is only sparse optical emission directly west from the remnant's expansion center, in view of the result of \citet{Vink2022} where they found the reverse shock moving inward with a proper motion suggesting a transverse velocity $\sim$1000, we thought it important to obtain an estimate of the reverse shock here based on optical data.
Figure~\ref{RS_map} shows the location of western limb optical emission labeled P12, with the panels of  Figure~\ref{P12} showing close-ups of the near IR emission seen in March 2004 and November 2011 as well as an image showing the seven year difference. 

As seen in Figure~\ref{P12}, large changes in the region's optical/near infrared emission are seen during the 2004 to 2011 time frame, so much so that it difficult to obtain an accurate RS proper motion using just these 2004 and 2011 images.  
Unfortunately, {\sl HST} 2000 WFC2 images did not go deep enough to be useful and  {\sl HST} 2002 WFC2 images did not cover this region.
Consequently, we were left with only images taken only roughly 1 year apart. These included the March and December 2004 
ACS/WFC F850LP images  ($\Delta$t = 0.759 yr)  and
the November 2010 and 2011 WFC3 F980M images ($\Delta$t = 1.07 yr). 

Changes in the region's relatively bright near IR emission, largely due to [\ion{S}{3}] $\lambda\lambda$9069,9531, seen in the 2004 ACS/WFC images were relatively minor across this short time span. Differences between the March and December 2004 images, although showing advancement of the RS relative to that of the ejecta (white features in the difference image), we were not able
to distinguish between a RS proper motion greater or less than measured for the ejecta (i.e., 0.272$''$ implying a proper motion of 0.359$''$ yr$^{-1}$) from these data (see Table 5).

However, the WFC3/IR images taken in November of 2010 and 2011 showed changes that indicated values around 0.36$''$ yr$^{-1}$ in the rest frame of the ejecta. Although our measurement uncertainties for this rapidly evolving region are not sufficient to firmly reject a value as large as $\approx$0.41$''$ as proposed \citet{Vink2022}, our result suggests that the reverse shock here is nearly stationary or has only a small inward motion ($-50$ km s$^{-1}$) relative to the observer's sky frame. This is in contrast to that reported by \citet{Vink2022} who finds a large inward motion of around $-1000$ km s$^{-1}$.

As a check on our findings, we note that eastern edge of the emission shifts eastward less than 0.2$''$ between March 2004 and November 2011. This small shift suggests a nearly stationary reverse shock front in the sky frame over the 7.74 yr period, consistent with little or only a small inward proper motion.

\begin{figure*}
\centering
\includegraphics{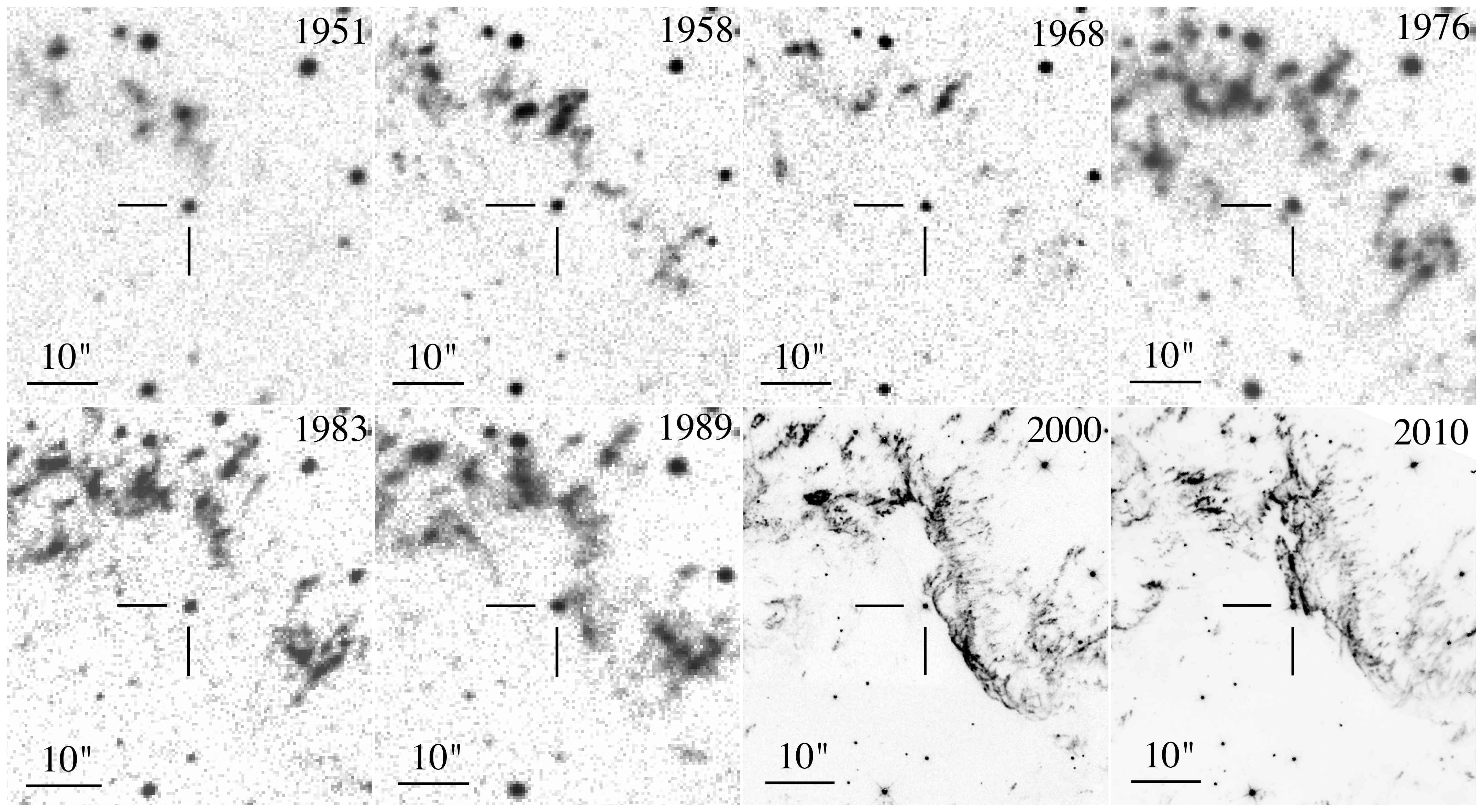}
\caption{Evolution of Cas A's optical emission along its northwestern limb. }
\label{NW_history}
\end{figure*}

\subsection{NW Limb: Positions P13 - P15}

The remnant's NW region exhibits a long, continuous line of filaments, exhibiting a concave curvature, opposite from a convex morphology expected along the rim. Despite an organized appearance, both radial velocities and proper motions indicate the ejecta here comprise several distinct expansion groupings with varying reverse shock motions. Consequently, we have divided this region's emission into three closely spaced regions; P13 thru P15 (see Figure~\ref{RS_map}). 

While all three regions showed similar ejecta expansion proper motions ($0.31'' - 0.33''$ yr$^{-1}$), the motion of the reverse shock was seen to vary widely here. This can be seen in Figure~\ref{RS_P13_P15} where the motion of the reverse shock viewed in the reference frame of the expanding ejects and hence reflected in the appearance of new interior emission varied from relatively limited new emission in the sharp overlapping filaments seen in P13 
of $0.18''$ yr $^{-1}$, to strongly non-radial motions seen in region P15 of order $0.27''$ yr$^{-1}$, to the very large changes $\sim0.36''$ yr$^{-1}$ for the P14 region.  Our proper motion measurements for P14 indicate a small inward proper motion ($\sim$500 km s$^{-1}$) of the ejecta here leading to a nearly stationary appearance in the sky frame. 

In order to explore the development of such a variety of reverse shock proper motion, we looked at the evolution of the remnant's optical broadband red emission in this NW region over the nearly 60 year period from 1951 thru 2010. This is shown in Figure~\ref{NW_history}. A relatively bright star is marked to help observe the optical emission changes over this period.  From 1951 to the early 1980's, emission that appeared in this region would move steadily outward and away from the marked star. 

However, this changed when a 1989 image showed a dramatic increase in emission features along this portion of Cas~A's NW rim from that seen just six years earlier in 1983. New emission appeared projected near the star noted above. Emission grew in extent, whereby in 2010 emission  was seen immediately adjacent to the star and filled in the region immediately due north and southwest of it. The evolution after 2010 is shown in Figure 4 where the distribution of emission close to and north of this star seen in 2000 is largely unchanged in 2019 in terms of general location. 
Interestingly, the bulk of this emission never crosses over the star, appearing largely north or west of it.

\begin{figure*}[t]
\begin{center}
\includegraphics[angle=0,width=16.0cm]{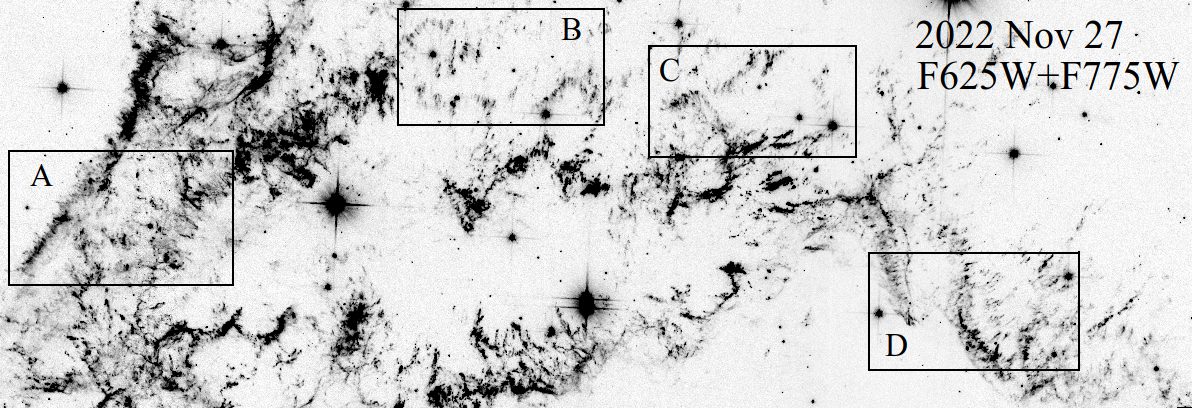}
\includegraphics[angle=0,width=7.98cm]{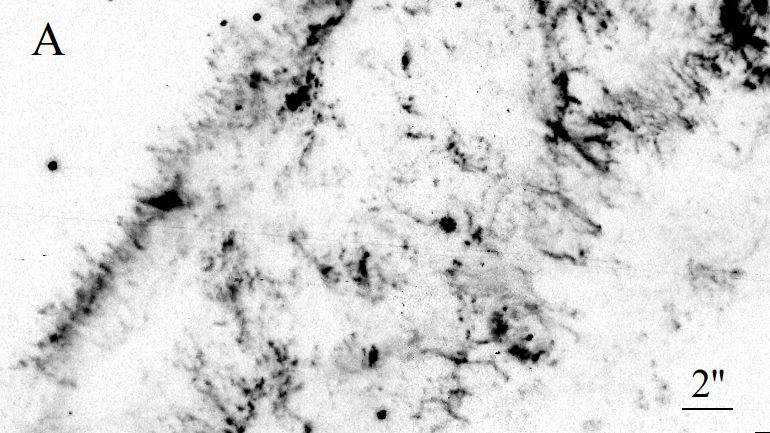}
\includegraphics[angle=0,width=7.98cm]
{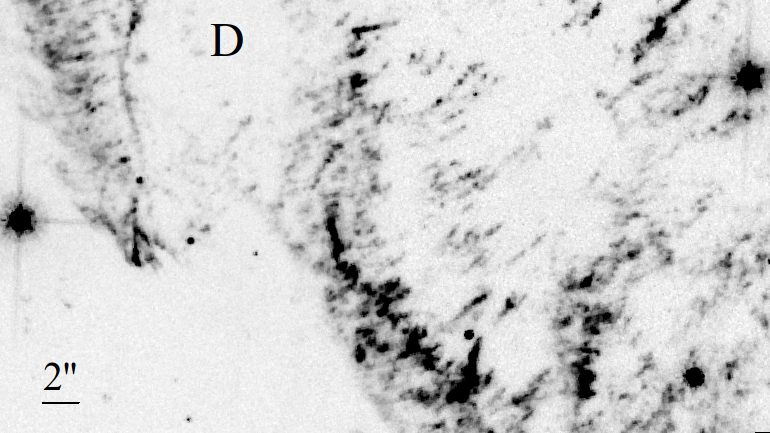}
\includegraphics[angle=0,width=7.98cm]
{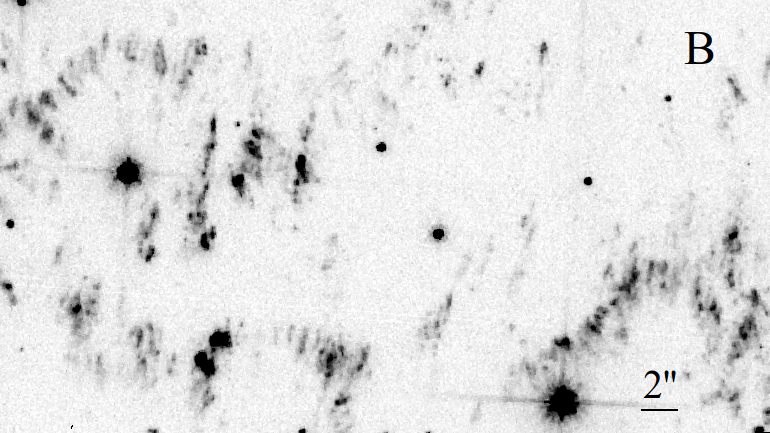} 
\includegraphics[angle=0,width=7.98cm]{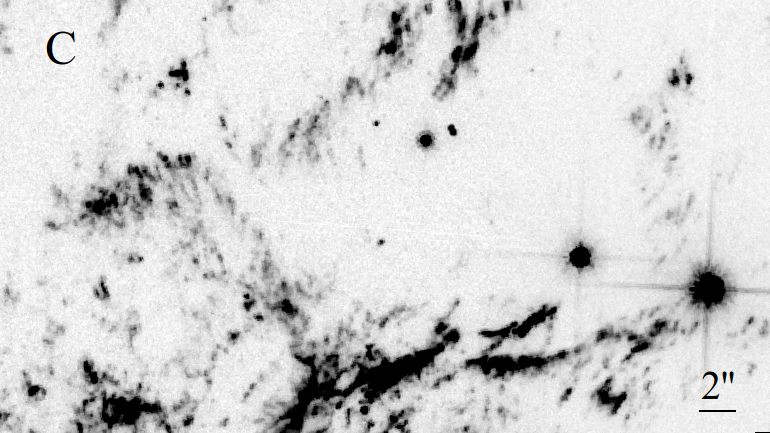} \\
\caption{ 
Combined {\sl HST} ACS/WFC 2022 November 27 F625W + F775W filter images of four northern regions exhibiting radially distorted ejecta morphology due to interaction with the reverse shock.
Top panel shows the location of the four enlarged regions. Middle panels: East and West limb regions (A \& D respectively). Lower panels: Neighboring regions (B \& C) along Cas~A's northern shell boundary.
North is up, East is to the left. 
\label{ablation_North}
}
\end{center}
\end{figure*}

\begin{figure*}[t]
\begin{center}
\includegraphics[angle=0,width=16.0cm]{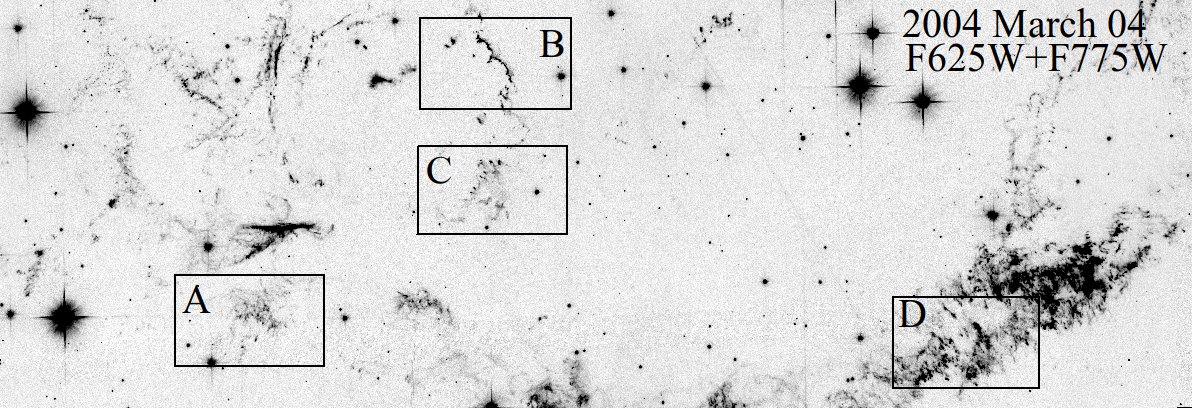}
\includegraphics[angle=0,width=7.95cm]{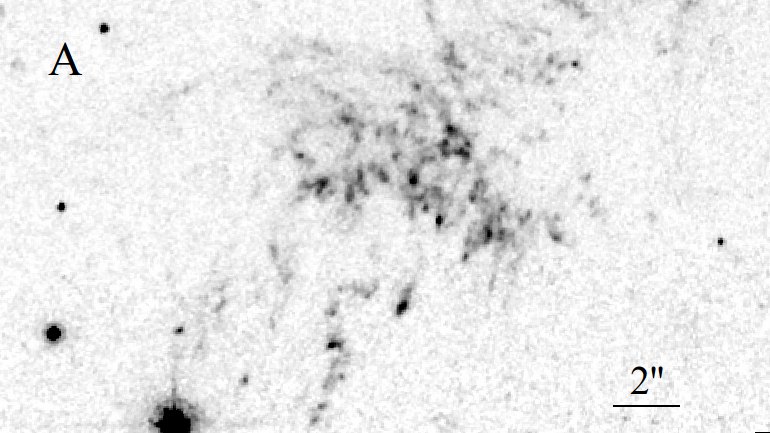}
\includegraphics[angle=0,width=7.95cm]{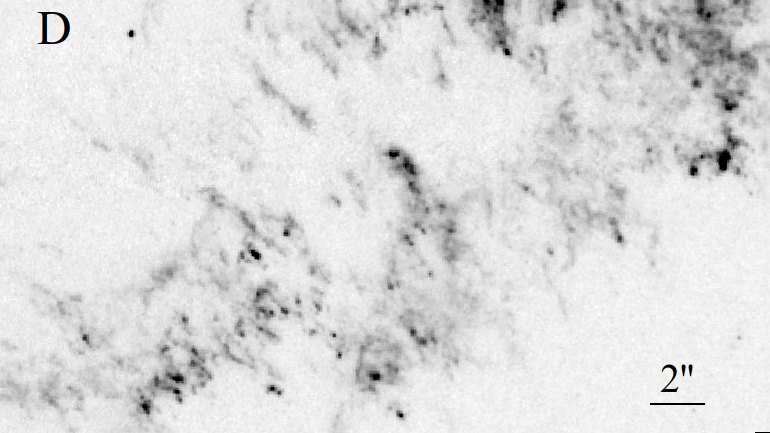} \\
\includegraphics[angle=0,width=7.95cm]{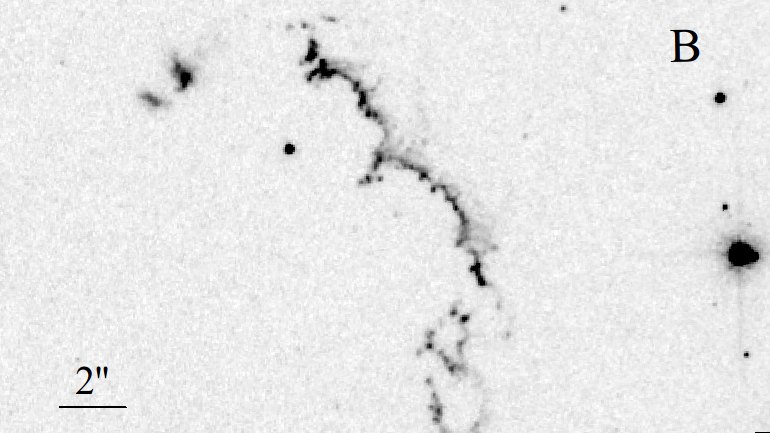} 
\includegraphics[angle=0,width=7.95cm]{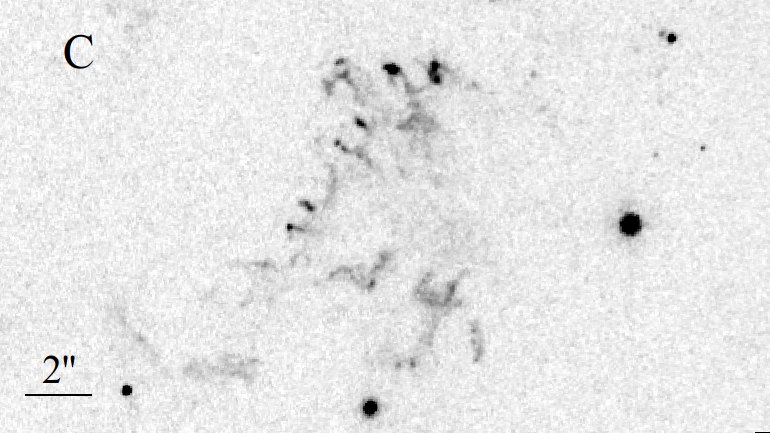} \\
\caption{ 
Combined {\sl HST} ACS/WFC 2004 March 04 F625W + F775W filter images of four southern regions exhibiting distorted ejecta morphologies following interaction with the reverse shock.  
Top panel shows the location of the four enlarged regions. Middle panels: southeast and southwest limb regions (A \& D respectively). Lower panels: Neighboring parts (B \& C) of the so-called parentheses feature near Cas~A's center.
North is up, East is to the left. 
\label{ablation_south}
}
\end{center}
\end{figure*}


\section{Effects of the Reverse Shock on Cas~A's Expanding Ejecta} \label{sec:effects}
The heating of Cas~A's ejecta via impact with its reverse shock leads to the generation of a range of optical and infrared emission lines. Strong emission lines of low to moderate ionization states ($\sim$ 10 - 60 eV) such as
[\ion{O}{1}], [\ion{O}{3}], 
and [\ion{Ar}{3}] indicate postshock electron temperatures in the range of $1 - 5 \times 10^{4}$ K. 
Denser portions of an ejecta clump will experience a slower internal shock and hence will predominately show strong low-ionization
lines such as [\ion{O}{1}] and [\ion{Fe}{2}], whereas much lower
density material like that around the outer parts of an ejecta knot will lead
to brighter lines of higher ionization such as [\ion{O}{3}]
and [\ion{S}{3}] due to higher shock velocities, as we will show in Section 5.2.

Because SN ejecta knots are neither sharply bounded clumps nor uniform in density, the presence of ejecta core and envelope density differences will lead to deceleration differences
which can be seen in high resolution images where recently reverse shocked ejecta show  elongated shapes and trailing ablation tails.
Observable structural and emission changes are in addition due to 
hydrodynamic effects including Rayleigh-Taylor (R-T), Kelvin-Helmholtz, and Richtmyer-Meshkov instabilities.  These include changes in knot size and emission morphology, mass loss due to mass ablation, and fragmentation leading to eventual knot disintegration
\citep{Fryxell1991,Klein94,MacLow94,Pol04a,Pol04b,Silvia2010,Hammer2010,Pittard09,Pittard2016}. Given that the velocity difference of the outgoing ejecta with the remnant's more slowly expanding reverse shock can range from 1000 to 5000 km s$^{-1}$ (\citealt{Vink2022}), this interaction has significant effects on the structure of reverse shock heated ejecta.



Although typical ground-based images do not have 
sufficient resolution to detect knot deceleration
caused by the impact of the reverse shock 
\citep{vdbk83}, higher resolution {\sl HST} images reveal
extended knot emission consistent with both knot
mass ablation and deceleration to due to internal
knot density variations.
Below we present high resolution images of Cas~A taken using both {\sl HST}  and 
JWST along with ground-based optical spectra that reveal
structural changes and mass ablation off Cas~A's optical ejecta knots following contact with the reverse shock. 


\subsection{ {\sl HST} ACS/WFC Images }

In their final paper on Cas~A's optical emission
where they presented some of their best Palomar 5m images, \citet{vdbk85} remarked that ejecta knots in the remnant's northern limb region  showed either faint 
``tails'' pointing approximately toward the center of expansion, or the knots themselves displayed elongated radial shapes.

Subsequent higher resolution {\sl HST} images supported this knot description and  
morphology of
Cas~A's 
ejecta.
Using {\sl HST} WFCP2 images ($0.1''$ pixels), \citet{Fesen01etal} found much of the remnant's
ejecta to be resolved into clouds of small knots ($0.2'' - 0.5''$) with a R-T `head-tail' like morphology with a bright knot at the tip of a thin stem
appearing much like that of model of a
shock running over a dense ejecta cloudlet or knot 
\citep{Klein94}.

In Figures 16 and 17, we present  {\sl HST} ACS/WFC images ($0.05''$ pixels) which reveal some of the
dynamical effects that the reverse shock imprints on Cas~A's ejecta knots. We have selected eight regions  that illustrate ejecta knot shapes; four regions in the remnant's northern emission feature and four in the south/central emission regions.  
The {\sl HST} images for these eight regions were taken in 2004 and 2022.

Figure 16 shows images of regions along Cas~A's northern rim 
taken using the 2022 F625W and F775W filter images which are sensitive to 
[\ion{O}{1}] $\lambda$6300,
[\ion{S}{2}] $\lambda\lambda$6716,6731, and
[\ion{O}{2}] $\lambda\lambda$7319,7330.
These images
reveal dozens of elongated knots with extended emission tails. In all four northern regions, extended streaks of emission are seen  
($\simeq0.5''$)
originating from small, bright knots, while others are much longer ($\sim1'' - 2''$). At a distance of 3.4 kpc, these angular sizes 
correspond to projected dimensions of $2.5$ to $ 10 \times 10^{11}$ km.

Streaked and elongated emission features are seen along
both northeast and northwest regions (panels A and D).
Similarly distorted knot emission morphology is present
along the remnant's northernmost limb (panels B and C).
Unlike the sharp point like appearance of the (few) stars in these images,
the fine scale structure of the remnant's ejecta almost universally display a
streaked or elongated appearance.

Figure 17 shows similarly enlarged 2004 ACS/WFC images  but now for four central and southern remnant regions. Images of regions A and D
show ejecta along the remnant's southeastern and southwestern sections, respectively.
Region B shows a central emission 
south of remnant 
center that appears as a line of ejecta knots, all showing
faint extended emission giving the impression of a shock front moving lower left to upper right
encountering a line of ejecta knots
which have undergone
mass ablation seen as faint trailing emission.

We note that mass stripping has been reported before for Cas~A's especially high-velocity (8,000 to 10,500 km s$^{-1}$), outlying N-rich ejecta knots \citep{Fesen06b,Fesen06a, Fesen11} due to these ejecta knots' catching up and then
passing through the slower forward shock followed by their interaction with local CSM and ISM. While somewhat similar to what is found for main shell ejecta knots, the degree of mass loss and the duration of mass loss appears substantially different likely due
to main shell ejecta being both physically larger and more massive 
as well as much higher shock velocities for these outer ejecta knots.

\begin{figure*} [t]
\centering
\begin{tabular}{cc}
\includegraphics[scale=0.45]{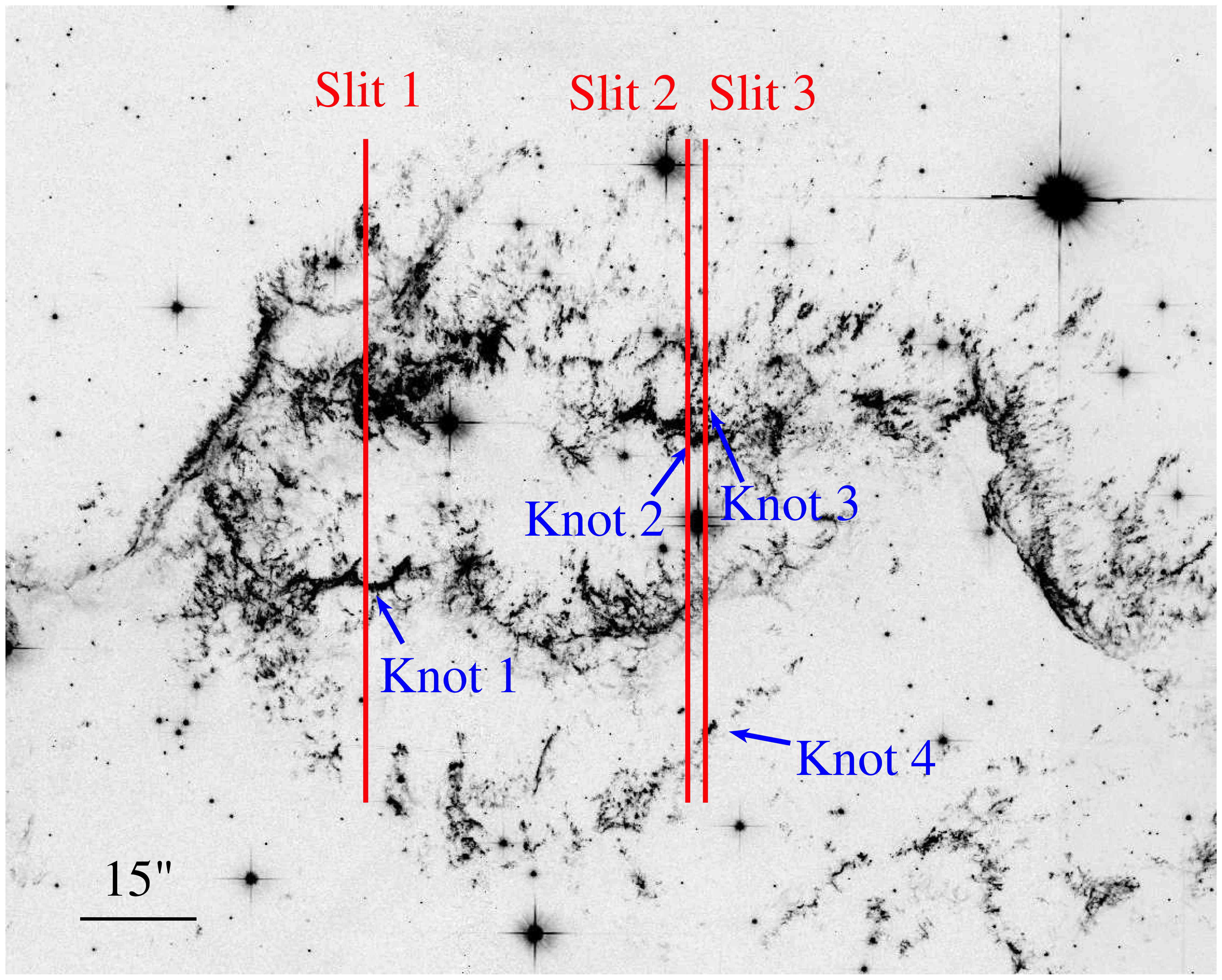}  & \includegraphics[scale=0.45]{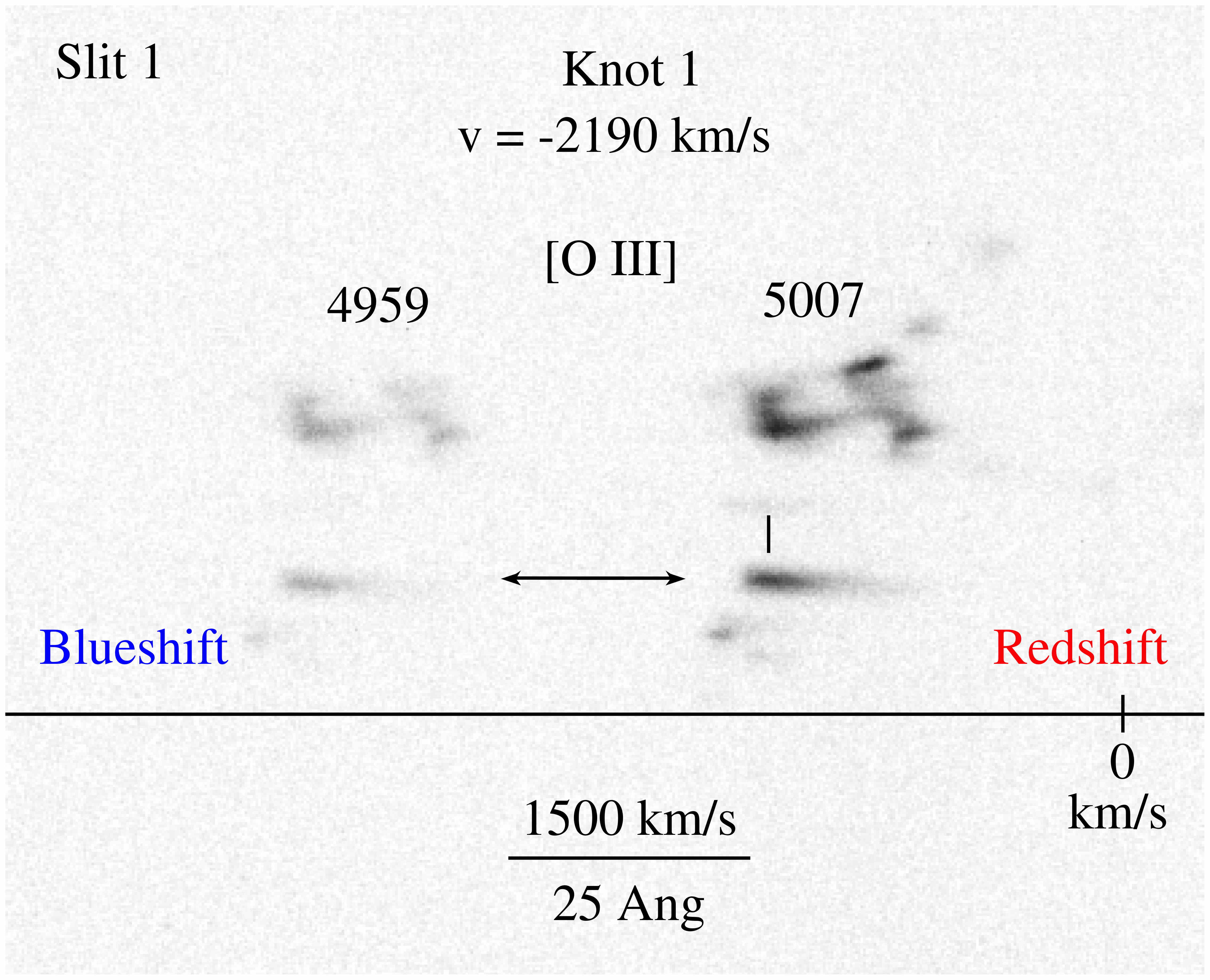} \\
\includegraphics[scale=0.45]{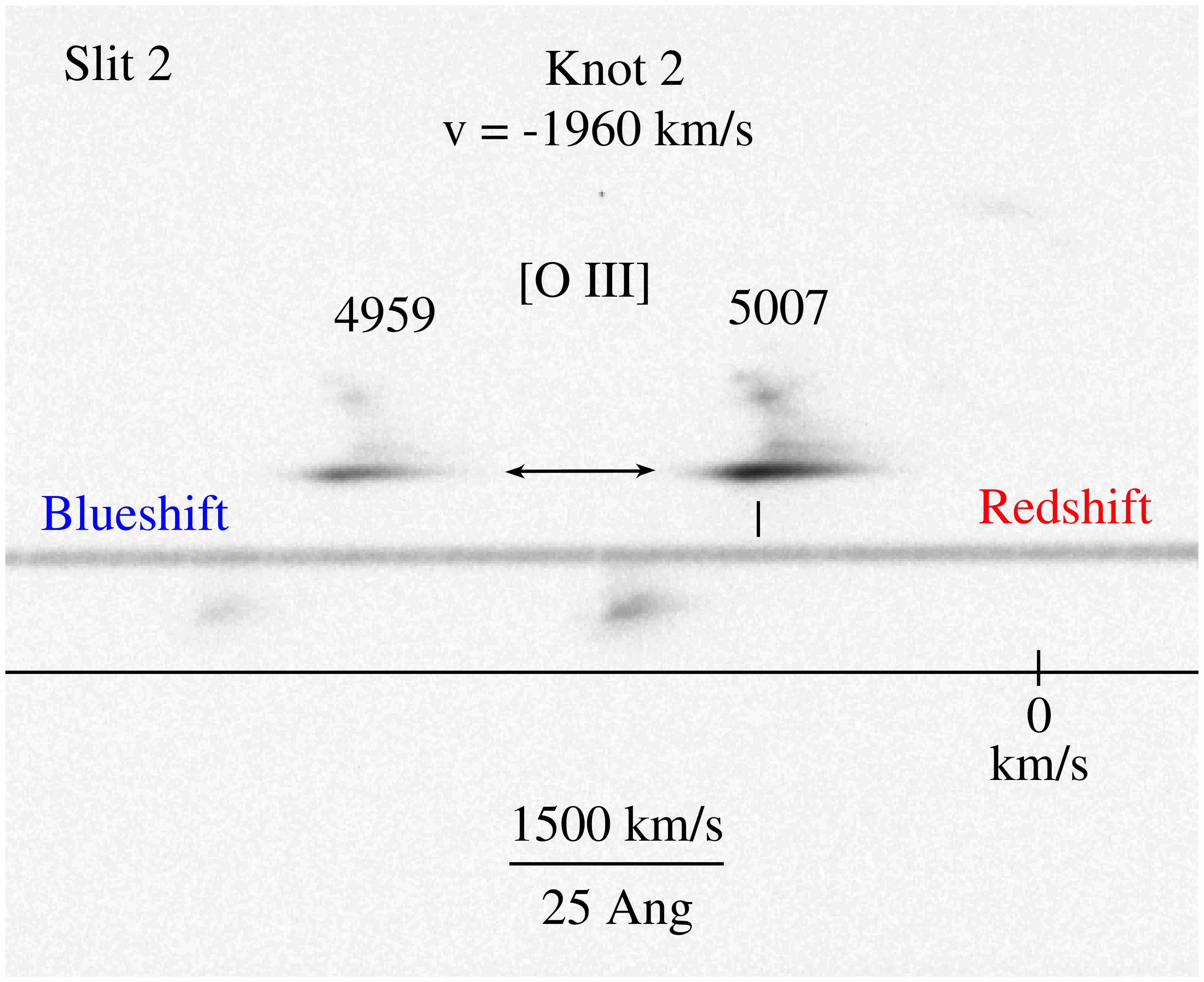} & \includegraphics[scale=0.45]{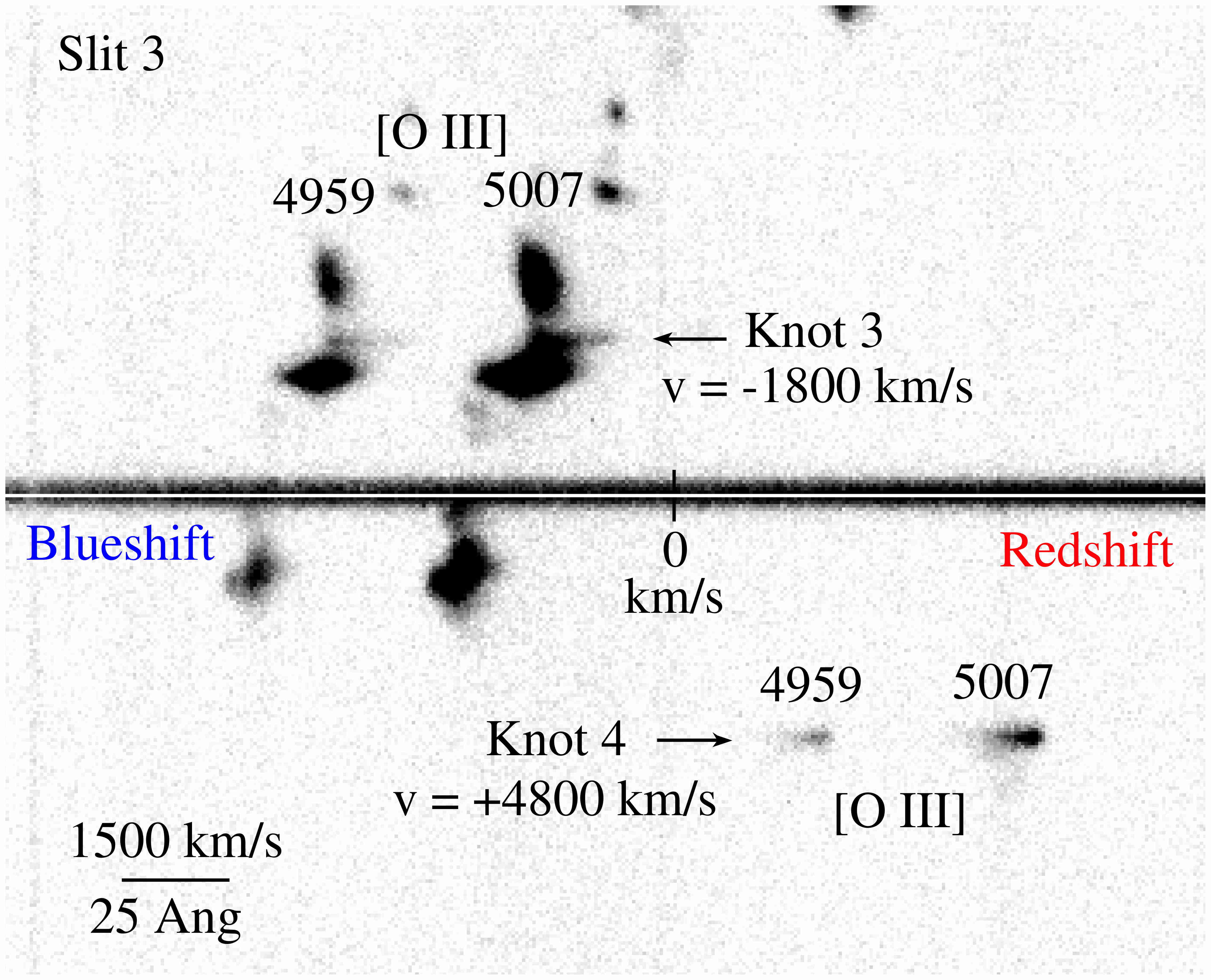} \\
\end{tabular}
\caption{Top left panel - Reference image where the locations of the three
slits used and the knots studied are indicated. Top right panel - The spectrum from
Slit 1's spectrum shows Knot 1's [O III] $\lambda\lambda$ 4959,5007 emissions. Marked by
the black arrow, Knot 1 has a radial velocity of $-2190$ km s$^{-1}$, with a faint decelerated ablation tail extending  
$\sim$ 900 km s$^{-1}$ from the knot. Bottom left panel - The spectrum for Slit 2. Knot 2, marked by the black arrow, has
a radial velocity of $-1950$ km s$^{-1}$ and an ablation tail that shows
deceleration of $\sim$ 500 km s$^{-1}$. Bottom right panel - The spectrum for Slit 3, showing [O III] emissions from Knots 3 and 4. Knot 3
has a radial velocity of $-1800$ km s$^{-1}$ and a decelerated ablation tail extending
$\sim$ 700 km s$^{-1}$. Knot 4 has a radial velocity of +4800 km s$^{-1}$ and
has an ablation tail extending  $\sim$ 500 km s$^{-1}$ in radial velocity. \label{spectra}  }
\end{figure*}

\subsection{Velocity Dispersions Seen in Ejecta Spectra}

In a limited spectroscopic survey of Cas~A, 
\citet{vdb71b} noted that although most ejecta knots exhibited a half-width Doppler velocity dispersion around 200 km s$^{-1}$, a few filaments were observed to have much larger velocity dispersions. \citet{Fesen01etal}
reported spectra showing much larger
$\simeq$1000 km s$^{-1}$
velocity dispersion in a western region which they attributed to the formation of mass-loss, ablation/stripping follow passage of the reverse shock.
The velocity shear was most prominent in high-ionization emission lines such as [\ion{S}{3}] 
$\lambda\lambda$9069,9531 and [\ion{Cl}{2}] $\lambda$8579 but was
nearly absence in lower ionization lines such as
[\ion{S}{2}] $\lambda\lambda$6716,6731 and
[\ion{Fe}{2}] 
$\lambda$8617.  Below, we present additional spectroscopic  data which reveal clear evidence for 
mass-loss stripped off Cas~A's ejecta as a
consequence of encountering the reverse shock.

Figure~\ref{spectra} shows the results of three long slit spectra
of the remnant's bright northern ring of ejecta
for their [\ion{O}{3}] line emissions. The slit locations were chosen to intersect bright emission filaments and knots in the remnant's northern region.
This section of Cas~A contains a large and mainly redshifted ring of ejecta along with a smaller largely blueshifted ring (see figures and animations in \citealt{MF13}). Because these ejecta knots have different projected line of sight radial velocities,  both red or blue radial velocities were seen.

In the upper right panel of Figure~\ref{spectra}, we show the [\ion{O}{3}] line emissions from a cluster of bright knots, all showing blueshifted velocities.  Several of these  display ablation emission 'tails'  extending to smaller blueshifted velocities, i.e., to the right in the figure.
Such extended velocity emission is most clear in the lowermost knot, with a velocity of $-2190 \pm 75$ km s$^{-1}$  for its brighter part.
It has a tail of decelerated 
[\ion{O}{3}] emission extending down  to $-1300$ km s$^{-1}$ amounting to a velocity shear or deceleration of nearly 900 km s$^{-1}$.
Smaller extended emission tails can also be seen for other ejecta knots in this panel.

A similar velocity dispersion is seen for Knot 2 (lower left panel)
where a velocity spread of $\sim 900 - 1000$ km s$^{-1}$
is observed. 
Shown in the lower right hand panel
are spectra of two other knots,
namely Knots 3 and 4 where
smaller blue and red shifted velocity shears can be seen.

The fact that one sees such significant
velocity dispersions in ejecta knots
in the spectra of three slit positions that were simply exploratory and not chosen to sample any particular ejecta knots or regions suggests that
ejecta ablation due to contact with the remnant's
reverse shock is not rare but common.
These results are in line with the 
$\sim$1000 km s$^{-1}$ velocity shear reported by \citet{Fesen01etal} and 
consistent with the elongated shapes of ejecta knots in {\sl HST} images shown in 
Figure~\ref{ablation_North} and 
Figure~\ref{ablation_south}.
The degree of velocity deceleration due to mass ablation, limited by the visibility of the trailing emission, is expected to vary considerably based
on the mass of the ejecta knot, the knot's density profile (halo to core),
and the reverse shock velocity experienced by the knot in its rest frame.

Most of the knots shown in 
Figure~\ref{spectra}, with the exception of Knot 4, have radial velocities close to $-2000$ km s$^{-1}$ with velocity dispersions of 
between 500 and 1000 km
s$^{-1}$.  In some cases, this approaches half the knot's observed radial velocity.
Due to uncertain project effects and initial knot densities, it is not
easy to correlate knot head-tail sizes with spectroscopic radial velocity dispersions. 


In summary, it is clear that many if not most of Cas~A's ejecta knots undergo some structural shear due to impact with the reverse shock which can lead to trailing optical emission. Such ablated material is most visible in higher ionization emission lines 
e.g., [\ion{O}{3}] and [\ion{S}{3}] suggesting a relatively low densities and higher shock temperatures, presumably  arising from lower density regions of ejecta knots.

\begin{figure*}[ht]
\begin{center}
\includegraphics[angle=0,width=7.95cm]{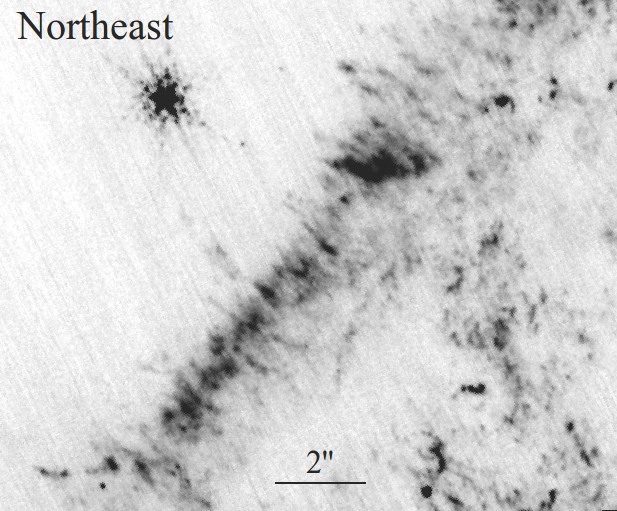}
\includegraphics[angle=0,width=7.95cm]{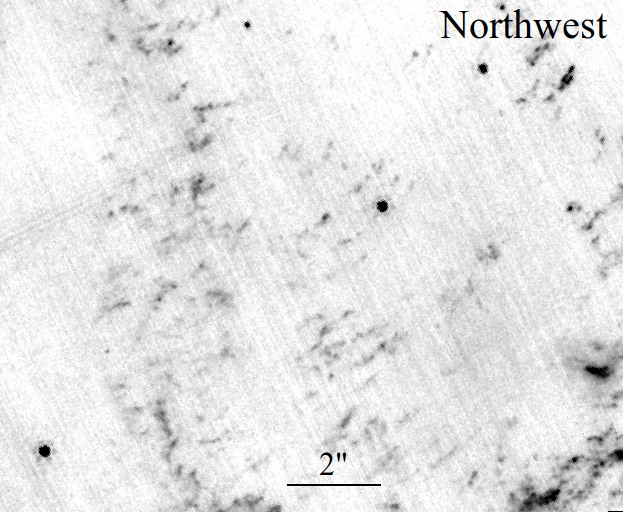} \\
\includegraphics[angle=0,width=16.0cm]{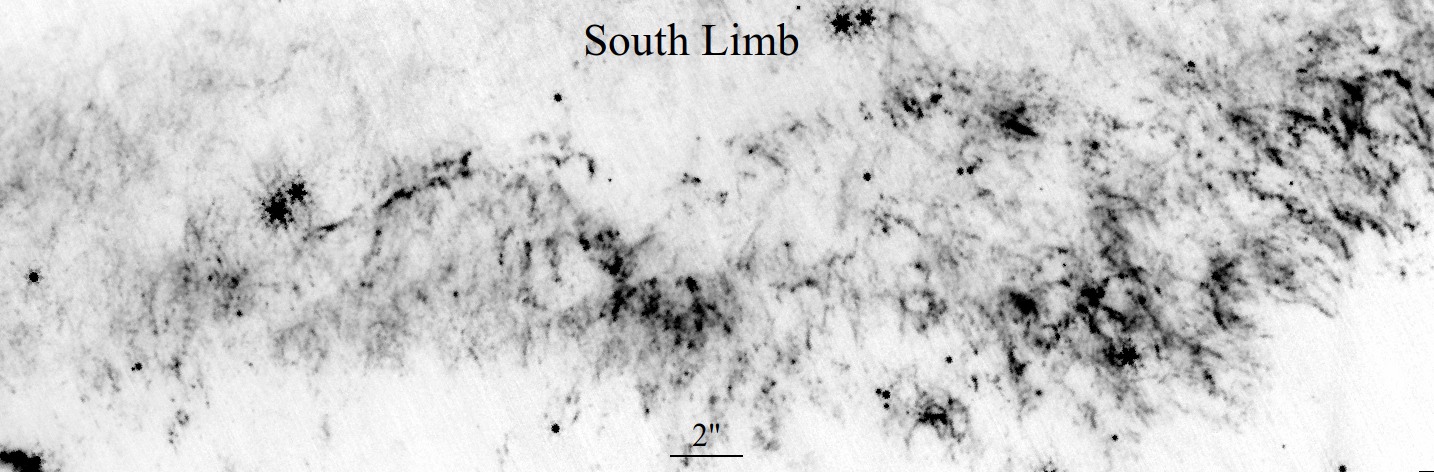} \\
\caption{High resolution JWST [\ion{Fe}{2}] F162M images of ejecta 
showing fine scale ejecta knot
morphologies 
along Cas~A's northeast and northwest regions (top panels) and its southern limb (bottom panel).
North is up, East is to the left. 
\label{JWST}
}
\end{center}
\end{figure*}

\subsection{JWST NIRCam [\ion{Fe}{2}] Images}

Since Cas~A's optical ejecta can remain visible for years and in a few cases even one or two decades 
following the reverse shock front passage 
\citep{vdb71a}, density variations in a knot's density structure
will naturally lead to differences in the decelerated velocity of
parts of a knot thereby leading over time to an elongation morphology of reverse shocked ejecta knots.
Unlike high ionization line emissions which best show low density ablated material, low ionization lines 
are expected to 
be more concentrated in the denser portions
of shocked ejecta. Consequently, 
emissions from low ionization species
allow us to trace the morphological evolution of the denser
portions of ejecta knots.

A low ionization emission line that typically shows little extended trailing emission 
in {\sl HST images}
is [\ion{Fe}{2}].
Because JWST NIRCam images provide even higher resolution
that {\sl HST} images (specifically, JWST NIRCAM F162M PSF FWHM: $0.055''$ vs.\ {\sl HST} ACS/WFC F625W PSF FWHM: 
$0.095'' - 0.120''$) we have investigated the fine-scale morphology of  shocked ejecta in low ionization line of [Fe~II].
With an ionization potential of 16.2 eV, recent JWST 1.644 $\mu$m images of Cas~A sensitive to 
[\ion{Fe}{2}] emission offers a better 
view of cool, dense ejecta thus a good tool for investigating late-time structural evolution of Cas~A's ejecta knots years after being impacted by the reverse shock.

Figure~\ref{JWST} shows enlargements of JWST NIRCam  [\ion{Fe}{2}] 1.644 $\mu$m line emission for several regions in Cas~A's main shell.
These images highlight the elongated and shredded 
appearance of the ejecta  in the
along the NE corner of the remnant (top left panel),
short filamentary fragments in the
remnant's northwest region (top right panel), and whole banks of R-T shaped
ejecta knots along the remnant's southern main shell edge.
The scale of knot extensions and distortions is of order $0.3'' - 1.0''$.
We note that such extended and distorted ejecta morphology is wide spread throughout the JWST image of Cas~A's main shell.

The elongated morphology of ejecta knots is likely due to internal density
variations of individual knots.
The electron density of Cas~A's  main shell ejecta has been reported to cover a wide range
of values; N$_{\rm e}$ = 2000 to 50,000 cm$^{-3}$ \citep{Peimbert71,ck79, HF1996}.
Internal knot density variations of just 20\% to 30\%  will lead to
velocity dispersion of $\sim$ 500 km s$^{-1}$ or more.
Assuming equal ram pressures 
($\rho_{1}$v$_1^{2}$ = 
$\rho_{2}$v$_{2}^{2}$), such velocity differences inside an ejecta knot mean that
lower density regions will experience higher speed shocks and thus
be decelerated more causing them to lag behind denser portions of a shocked knot. 
This will lead to a knot's cooler and denser emission regions to be less extended down stream.
A spread in internal knot
velocities of 500 to 1000 km s$^{-1}$ can lead to angular elongation of 
$\sim1.0''$
over a time span of 10 to 20 years after which it cools sufficiently to become optical faint.
Thus, variations in the internal  density structure
of an ejecta knot can naturally lead to extended ejecta emissions over the time span of a few decades.
Figure~\ref{JWST} shows a sample
of such post reverse shocked knot morphologies.

It is important to note, however, that this
analysis does not apply to the low density regions inside the large ejecta bubbles and rings that make up the greater volume of the Cas~A remnant. The reverse shock will proceed faster inside the less dense bubbles
than in the surrounding rings of denser ejecta and this difference will be important to understand
differences in reverse shock velocity measurements made in X-rays vs.\ the optical that are
discussed below.

\begin{figure*}[ht]
\begin{center}
\includegraphics[angle=0,width=7.8cm]{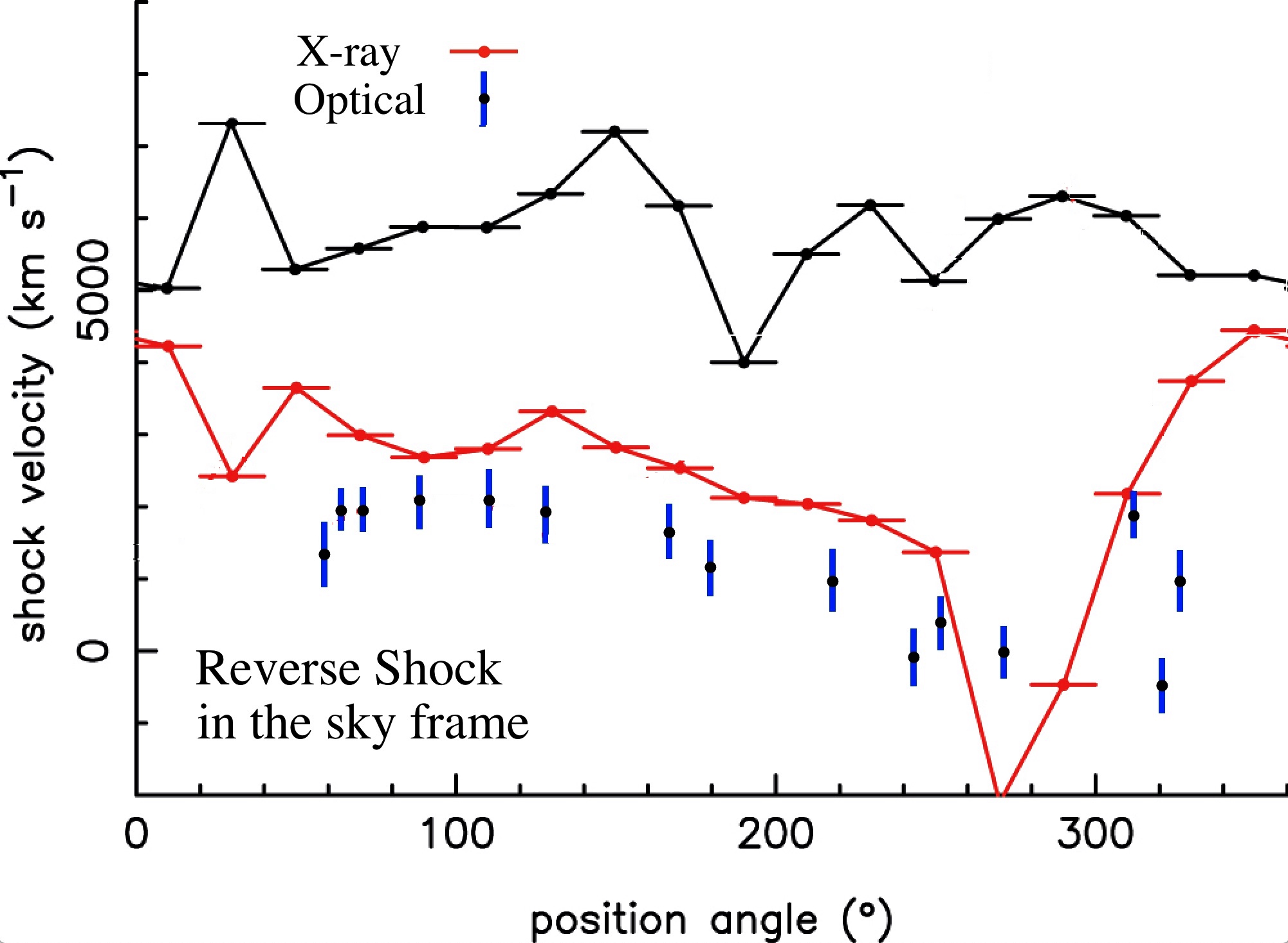} 
\includegraphics[angle=0,width=7.8cm]{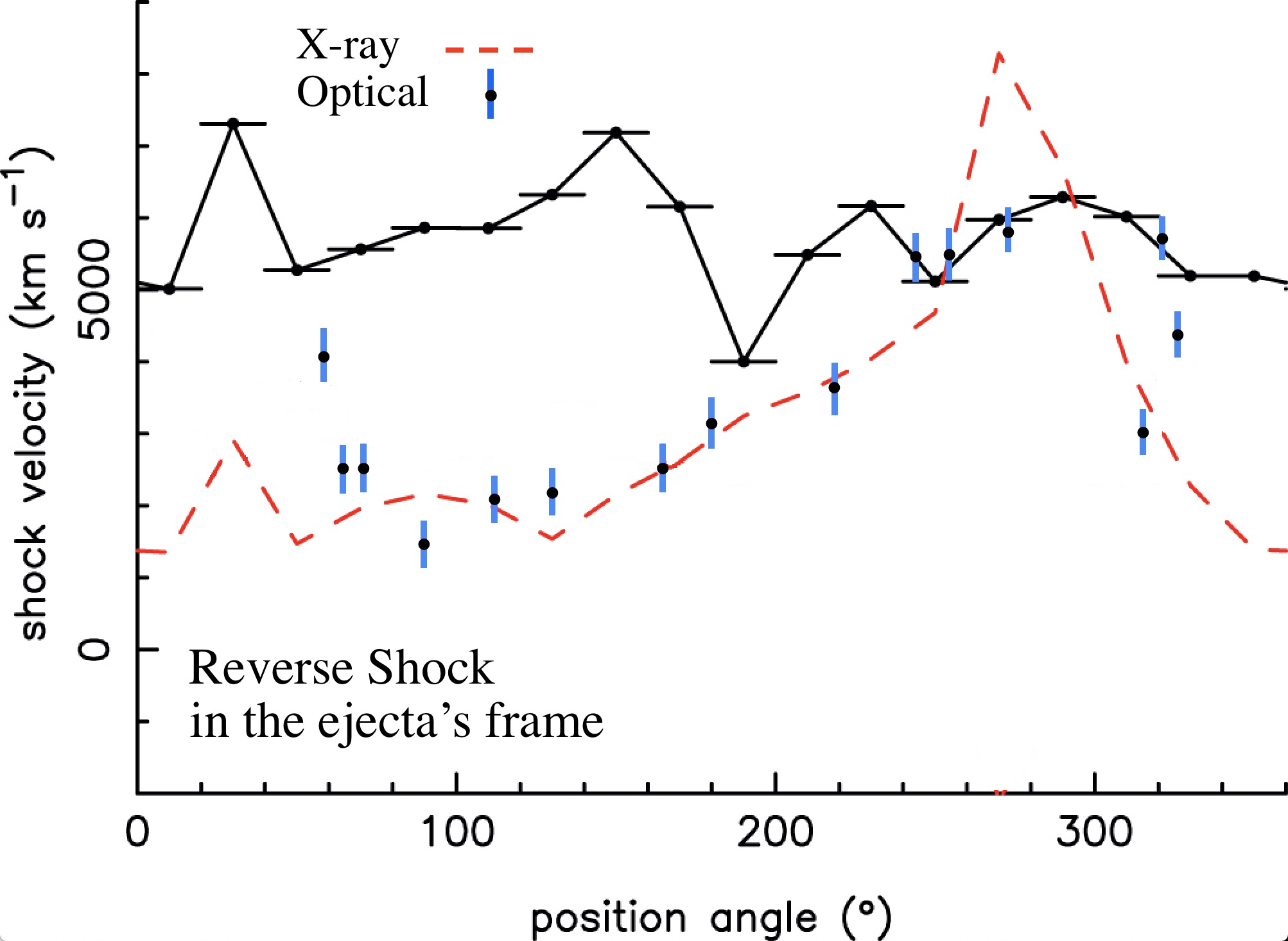}  \\
\caption{Comparison of our optically estimated vs. X-ray inferred reverse shock velocities made by \citet{Vink2022} using their velocity plots for values in the sky frame (left panel) and in the ejecta's reference frame (right panel). The black line and black points are the same in both plots and shows X-ray inferred forward shock velocities.
Red solid and dashed lines are X-ray derived reverse shock velocities
relative to sky/observer (left panel) and ejecta reference frames (right panel). The blue points are our corresponding values measured from optical images and should be compared to the solid red line in the left plot and the
dashed red line in the right plot.
(Adapted from \citet{Vink2022} with permission.)
\label{Vink_n_us}
}
\end{center}
\end{figure*}

\section{Discussion}

The young and nearby Cas~A remnant offers us valuable insights into the elemental and kinematic
properties of a supernova from a high mass 
progenitor. 
However, details regarding some of it properties
such as its reverse shock are
areas where our understanding is still poor or incomplete. 
For example, as can be seen in Tables 2 and 3, 
despite both model and observational investigations, there is not yet a clear consensus reached regarding the range of reverse shock velocities 
or the shock's exact location in some parts of remnant's main ejecta shell.
Although investigations of Cas~A
began more than 70 years ago in the early 1950's, our picture of many of its basic properties  have evolved slowly, becoming clear only in the last few decades.

Comparison of photographic plates taken of Cas~A with the Palomar 5m telescope 
from 1951 through the 1970's revealed high-velocity ejecta knots 
(the so-called `Fast Moving Knots' or FMKs) 
continually appearing while old ones slowly faded away
\citep{vdb71a,vdb71c}.  Because new knots seemed to appear mainly in regions
where optical ejecta knots were already present, little significant large-scale
emission structure was apparent during this nearly two decade time frame. 

This situation initially led observers to view Cas~A's high-velocity ejecta as
``ploughing through stationary interstellar cloud banks at hypersonic
velocities'' \citep{vdbd70,vdb71c} thereby giving rise to temporary optical emission through
collisional ionization via shocks \citep{Pem_vdb71}. However, they were puzzled  
by the complete absence of any hydrogen emission lines in 
any Cas~A knot spectrum that would indicate an interaction of 
the ejecta with H-rich CSM or ISM material \citep{Baade1954a,Minkowski1954, Peimbert71, Pem_vdb71}.

The eventual realization of the presence of an inner ``reverse shock'' 
behind the outer forward shock front
\citep{McKee74, Gull1975} 
resolved these issues. It is now recognized that the interaction of 
Cas~A's expanding main shell of ejecta with 
the remnant's reverse shock is the main process that makes the remnant's 
dense ejecta knots visible in the optical and infrared \citep{Chevalier1982a}. 



Our understanding of the 
emission structure of Cas~A advanced again with subsequent discovery that the majority of the remnant's optical ejecta are arranged in
a dozen or so well-defined and nearly circular ring-like structures with diameters between approximately
$30''$ and $2'$ (0.5 pc - 2 pc) 
\citep{Reed95,Law95,Delaney10,MF13}.
Unlike that of a smooth, filled expansion sphere of ejecta as some early models assumed, 
these rings of reverse shocked ejecta, visible in both infrared and optical surveys of Cas~A
\citep{Delaney10,MF13}, are structures 
related to the large scale structure of the remnant’s expanding debris field, independent of influences of the remnant’s local CSM/ISM environment.

These rings of dense, optically bright ejecta represent cross-sections of expanding ``bubbles'' likely generated by energy released by ejecta rich in radioactive elements
such as $^{56}$Ni and $^{44}$Ti.
The rings mark the areas of
intersection of a largely spherical reverse shock front with these structures of dense ejecta clumps.
The presence of these ejecta rings and their on-sky projections  
affects our ability to accurately
measure the motion of the remnant's reverse shock via optically bright ejecta due both to
the highly clumped nature of the expanding ejecta
forming these rings and the absence of emission in the ring centers.
Because lower density ejecta inside the
rings are more prominent in X-rays,  
studies of the motion of the reverse shock
in X-rays are less affected by ejecta clumping and projection effects.

\subsection{Our Estimated Reverse Shock Velocities}

Our estimates of Cas~A's reverse shock in the sky reference frame listed in Table 5 show a wide range of values at different points around the remnant, not unlike those of previous measurements (cf. Table 2 and 3).
Although velocities are consistently at around +2000 km s$^{-1}$ in the sky frame across much of its eastern and sections (PA = $60\degr - 130\degr$), values drop down
to +1000 to +1500 in the southeast and southwest 
(PA = $130\degr - 220\degr$), and become quite variable along the western and northwestern rim
areas (PA = $240\degr - 325\degr$), with values ranging from +1000 to +2000 km s$^{-1}$ to near zero. 

We find the range of
velocity of the reverse shock experienced by the outwardly expanding ejecta is 
between 2000 and 4000 km s$^{-1}$ for most regions
with the notable exception of the southwest and
west where the reverse shock velocity the ejecta sees is between 5500 and
6000 km s$^{-1}$. 

The only previously reported reverse shock study of Cas~A based on optical images is that of
\citet{Morse04} who used two years of 
{\sl HST} images to measure
ejecta and reverse shock velocities.
They reported finding 
proper motions of the 
emitting features 
of $\sim 0.31'' - 0.34''$ yr$^{-1}$ corresponding to
transverse velocities of 
$5000 - 5500$ km s$^{-1}$.
While these values closely match our P13-P15 ejecta velocities of $4950 - 5320$
km s$^{-1}$ covering this same region, their 
reverse shock velocity
estimate
of a +3000 km s$^{-1}$
differs greatly from our
$-500$ to +2000 km s$^{-1}$
estimates.

This discrepancy most likely arises from differences in measurement
technique. \cite{Morse04} measured proper motions by aligning shock fronts in
small regions of the remnant and creating differences images. Aligning the
images on the shock fronts increases uncertainty because of the irregular
nature of the shock. Additionally, aligning images in this way over just a two year
time period resulted in extremely small changes in reverse shock position that
have a high measurement uncertainty. Our measurements were done in reference to
ejecta knots just outside the shock front, which gave us a clearer reference
point from which we measured shock front progression.

\subsection{Comparison of Our Optical Derived Reverse Shock Velocities with those Derived from Other Data}

A recent in-depth proper motion survey of Cas~A's X-ray emissions 
by \citet{Vink2022} estimated both its forward and reverse shocks
based on 19 years of data from the
Chandra X-ray Observatory.
They found an  outward motion of the remnant's reverse shock velocity of between +2000 and +4200 km s$^{-1}$ in the sky frame for many eastern and southern regions of Cas~A, but a relatively low velocity ranging from $\sim$ +1000 km s$^{-1}$ to an actual inward motion 
reaching as much as $\simeq$ $-1900$ km s$^{-1}$ along parts of the remnant's western regions.
If correct, this would mean that the ejecta in the remnant's
western region effectively experiences a much higher reverse shock velocity than elsewhere, as high as 8000 km s$^{-1}$. We find it is between 2000 and 4500 km s$^{-1}$ in most regions except for in the west where it can be as large as $\sim$5800 km s$^{-1}$ (see Table 5).
\citet{Vink2022} viewed this very high effective
reverse shock velocity
they estimated the ejecta experienced along the remnant's western regions as a way of  
explaining the strong X-ray synchrotron emitting filaments in western portions of Cas~A \citep{Helder08}. 
However, if outward moving western ejecta are really experiencing a shock  as high as a 8000 km s$^{-1}$, one should see unusually high ionization levels in these ejecta. Yet, unusual optical line emissions have not been reported for western ejecta to date
\citep{Delaney10,MF13}. 

Figure~\ref{Vink_n_us} shows comparisons of our reverse shock velocity estimates relative to those of \citet{Vink2022}, both in the observer's sky frame (left plot) and as experienced by the ejecta in its reference frame (right plot).
In both graphs, the solid black line
shows \citet{Vink2022} measurements for the forward shock
expansion velocity as a function of position angle where zero is due north and is measured eastward.
We note that although their forward shock velocities, which range from 5000 to 7500 km s$^{-1}$, are on the high side of previous velocity estimates (Table 1), \citet{Sakai2024}
using the same {\sl Chandra} data found similar forward shock velocities.


Our reverse shock velocity estimates in the sky reference frame shown as blue symbols in Fig.\ 20
are consistently lower by
roughly 1000 km s$^{-1}$ than that reported
by \citet{Vink2022}
for the region between
PA of 60$\degr  - 240\degr$. This might indicate
differences associated with X-ray and optical ejecta densities \citep{Patnaude14}.
However, this trend disappears
in west and northwest regions where the X-ray derived
values plunge sharply to a inward value of $-1880 \pm 22$ km s$^{-1}$ around PA = $260\degr - 280\degr$. In contrast, our optical proper motion
measurements show less of a  decrease in reverse shock velocities down to small values near zero km s$^{-1}$, indicating a near stationary shock front in the sky frame.
Our most negative velocity
of $-550 \pm 370$ km s$^{-1}$ occurs at a PA of 321, which is much farther along to the northwest where \citet{Vink2022}
report velocities near +4000 km s$^{-1}$.

The notion of a near stationary or fast inward moving reverse shock front in such a young SNR as Cas~A is unexpected.
Models of SN shocks expanding into a uniform progenitor $\rho$ $\propto$ 1/r$^{2}$ wind
do not predict an inward motion of
the reverse shock until very late times when a remnant has swept up a significant amount of
material behind the forward shock
\citep{McKee74, True1999} and recent 
hydrodynamical simulations of Cas~A
show an outward moving reverse shock with always outward velocities around 2000 to 4000  km s$^{-1}$
\citep{Orlando21}.
\citet{Vink2022} proposed possible scenarios for explaining this unexpected western
reverse shock motion including the
forward shock interaction with a particular dense CSM/ISM cloud generating a strong
reflected shock like that considered by
\citet{Orlando22}, or a brief Wolf-Rayet phase of the progenitor which could have
created a highly aspherical, low density cavity (n$_{\rm e}$ $< 1$ cm$^{-3}$; \citealt{LPHS14}) along the remnant's western boundary through which the forward shock expanded into leading to a more evolved forward shock.


\subsection{The Nature of the Asymmetry of Cas~A's Reverse Shocked 
Ejecta Relative to Its Expansion Center}

Cas A's center of expansion, determined by the ballistic motions of ejecta knots \citep{Thor01}, is very nearly at the center of the remnant's nearly spherical forward shock as defined by the  thin outer nonthermal X-ray filaments 
\citep{Gotthelf01,Patnaude09}. But this is clearly not true  for the remnant's reverse shock as indicated by the shell of reverse shocked ejecta being clearly
displaced to the northwest across all wavelengths: radio, infrared, optical, and X-ray emissions (Fig.\ 1). The question is why should this be.

An asymmetry in Cas~A were seen 
from the very start, with a
striking difference in the remnant's northern and southern optical emissions.
When first studied back in the 1950s and 1960s, Cas~A's optical emission was largely limited to just its northern limb leading to a  strongly lopsided
appearing remnant (Figs.\ 2 \& 3). 

This was not due to an absence of the forward shock in the south, since 
several shocked, pre-SN mass loss knots,
the so-called QSFs, were prominently visible along Cas~A's SW rim. 
Although optical emission along the remnant's southern region has gradually filled in over the last 50 years, its north, NE, and NW areas still dominate.
While this asymmetry is also seen in X-rays (see \citealt{Patnaude14} for a 1979 Einstein HRI and 2004 {\sl Chandra} X-ray images), it is not the case in the radio and infrared (Fig.\ 1).

Part of the cause for the North--South  asymmetry in the optical and X-rays may be due to 
slower velocity ejecta in the south. We find ejecta proper motions for southern locations P5 to P8
to be $0.22''$ to $0.28''$ yr$^{-1}$, which is significantly lower than $0.33''$ to $0.37''$  yr$^{-1}$ seen for P1, and P13-15 in the NE and NW. Indeed, the highest transverse proper motions of it main shell ejecta $\sim$$0.33''$ to $0.36''$  yr$^{-1}$ are found along its northern rim.
However, it is not clear if such
ejecta velocity difference explains the displacement of the remnant's shell of reverse shocked ejecta to the northwest. 

One explanation is for there to be a strong asymmetry in Cas~A's expanding debris toward the north and west, and there are
several lines of evidence regarding gross asymmetries in ejected mass in Cas~A related to its explosion dynamics (see \citealt{Wheeler2008} for an
in-depth discussion). A  displacement of ejecta to the north
and west would be consistent with a bipolar SE-NW expansion asymmetry suggested by the presence of high-velocity, Fe-rich
ejecta seen in X-rays along the remnant's SE rim and in the opposite direction in the remnant's NW.

The remnant's compact X-ray point source's transverse motion of 350 km s$^{-1}$ \citep{Fesen06b} to the southeast
(PA = 169$\degr \pm 8.4\degr$)
would also be consistent with a greater mass ejected to the north to balance the momentum of the
$\sim$1.5 M$_{\odot}$ mass in the
presumed neutron star 
\citep{Wongwath2017, Katsuda2018, Holland2020, Picquenot2021}.
Greater mass toward the rear of remnant's projected northern and western regions might also
help explain the remnant's
smaller blue shifted radial expansion velocity of only $\sim$4000 km s$^{-1}$ versus a redshift of 
$\sim$6000 km s$^{-1}$
\citep{Delaney10, MF13},
as well as its evolution over the past 70 years.

\subsection{Knot Ablation Tails and Distortions Following Reverse Shock Encounter}

Only a handful of previous papers have remarked on the morphology of Cas~A's reverse shocked main shell ejecta 
\citep{vdb76a,vdbk85,Fesen01etal}.
While a variety of instabilities in post-shock radiative knots have
been discussed and modeled \citep{Gull1975,Mell02,Kif03,Raga07,Silvia2010, Silvia2012}, only limited observational data has been presented and discussed 
regarding the morphological and kinematic changes to SN ejecta knots following their collision with a reverse shock. 

High resolution {\sl HST} and JWST images plus optical spectra presented above
show clear evidence for ejecta distortion
and shredding by the reverse shock plus significant mass ablation off
Cas~A's ejecta knots in the form of long emission tails. These extended emission tails are best seen in higher ionization emission lines, presumably because greater mass loss off the lower density envelopes of ejecta knots imply decelerated velocities
as high as $\sim$1000 km s$^{-1}$. JWST 
[\ion{Fe}{2}] images, being sensitive to dense
ejecta regions,
show that over time scales of decades Cas~A's optical emission ejecta knots will
become extended due to internal density variations.
Both immediate and long term effects mark stages of ejecta
knot disruption and destruction that eventually lead to the metal enrichment of the surrounding CSM/ISM. 

However, as shown in Table 5, bright and presumably denser 
portions of ejecta knots with
significant mass show surprisingly little deceleration, with a deceleration parameter typically between 0.95 and 0.99 where 1.00 is free expansion.
This means that the shocks driven into the ejecta knots
are a few percent of the 4500 to 5500 km s$^{-1}$ ejecta velocities or roughly 50 to 200 km s$^{-1}$.

We can compare these observations with a 
series of recent papers that
specifically investigated the
effect on Cas~A's ejecta clumps by its reverse shock \citep{Kirch2019, Kirsh2023, Kirch2024}.
With the goal of exploring dust destruction rates via sputtering and grain–grain collisions, 
they modeled the evolution and disruption of 
metal rich ejecta clumps lumps 
at different remnant ages
through magneto-hydrodynamical simulations. 
They show clump destruction maps in density and temperatures as a function of clump/ambient density contrast ratios, $\chi$,  ranging from 10 to 1000 and magnetic field strengths.
These simulations, like many other previous shocked clump models, start with a sharply bounded spherical clump of uniform electron density and temperatures with radius of $\sim 10^{16}$ cm 
matching the observed size of 
main shell ejecta clumps 
($0.1''$ to $0.5''$; \citealt{Fesen11})
at a distance of 3.4 kpc. Unlike real ejecta
clumps which are unlikely to be sharply bounded and have a range of internal
densities and temperatures, these simulations
result in post shock clump distortions that
do not closely match the reverse shocked clumps seen in Cas~A at an age of 350 yr in that they show highly filamentary and turbulent downstream features which are not observed in {\sl HST} or JWST images.

Such model-observation differences for shocked ejecta clumps are 
understandable since these seemingly `tiny'
ejecta knots seen in {\sl HST} and JWST images
are not physically small gaseous clouds, but rather 
are roughly 300 to 1600 AU in diameter. To put this in perspective, they are many times larger than the $\sim80$ AU
diameter of the Solar System if defined by Pluto's orbit.
Consequently, the density, temperature and ionization state of internal regions of a single ejecta clump can be expected to vary considerably
across their 100's AU dimensions given
the high cooling rates for metal-rich plasma
\citep{Raymond2018}. The authors themselves realize this by not assuming a homologous expansion of Cas~A's ejecta which would lead to
very low densities and thus much higher gas temperatures at younger ages than observed.

Reverse shocked ejecta in Cas~A
may be particularly sensitive to other effects brought on by the enhanced cooling rates due to
its undiluted H,He-poor but metal-rich 
ejecta which might explain differences between observed spectra of SN
ejecta and steady-flow, planar shock emission models \citep{Raymond18}.
Significant mass stripping of ejecta knot material following reverse shock passage may also
result in 
downstream flow of material of varying lengths and densities 
depending on the initial
properties of the shocked knot and the reverse shock's velocity in the knot's reference frame.

A study of Cas~A's radio emission properties led \citet{Anderson1995} 
to conclude that there was 
a nearly constant transfer of clumpy ejecta into a more diffuse, lower density state following reverse shock passage leading to
knot deceleration, a scenario consistent with knot structural changes and mass stripping.
These effects have been modeled
for near solar composition plasmas
\citep{Klein94,MacLow94,Hansen17} but the very high cooling rates for SN ejecta make such model simulations challenging \citep{Raymond18}.


\section{Conclusions}

Our analysis of both ground- and space-based
optical images of Cas~A reveals a rather 
complicated and highly asymmetrical expanding remnant.
Although the available optical data set spans a period of over 70 years,
the remnant's optical emission features change so
rapidly that it is hard (if not impossible) to follow the progression of individual optical knots and filaments over the course of just  one or two decades. Nonetheless, our analysis reveals several important properties of the remnant's main shell of reverse shocked ejecta. These include:

1) Judged by the appearance of new, bright optical emission
generated by the impact of high velocity, outward moving ejecta with a more slowing moving reverse shock,
we find reverse shock velocity in the observer/sky frame of reference to be typically $1000 - 2000$ km s$^{-1}$ over most of the remnant. This outward velocity range is notably less than
that predicted by hydrodynamical models of the remnant and is
generally lower by around 1000 km s$^{-1}$ than estimates from X-ray analyses.

2) We find the remnant's reverse shock velocity along the remnant's western regions to be unusually low and nearly stationary in the sky frame within measurement uncertainties. While this
is roughly consistent with that suggested from a recent X-ray study, we do not find evidence for high velocity inward reverse shock motions.

3) Tangential proper motions indicate velocities
of 3500 to 5000 km s$^{-1}$ along the remnant's
eastern and southern sections, but higher
values, between 5500 and 6000 km s$^{-1}$ in the west and northwestern limb regions. Bright newly reverse shocked main shell ejecta show little deceleration, with a deceleration parameter typically between 0.95 and 0.99 indicating nearly free expansion. Combined with the slower outward moving reverse 
shock in the remnant's western regions, this
suggests a much higher effective reverse shock
velocity for western limb ejecta, around 
5500 km s$^{-1}$ versus
1500 to 3500 km s$^{-1}$ elsewhere in the main shell.

4) Optical emitting ejecta appear highly clumpy
in the form of small distinct `knots'
with typical knots dimensions of $0.2'' - 1.0''$
($\sim 1 - 5 \times 10^{16}$ cm). Following interaction with the reverse shock, ejecta knots often exhibit
extended mass ablated tails up to  $\sim1''$ in length. Optical spectra reveal faint, extended emission indicating reverse shock induced decelerated velocities as large as $\simeq$1000 km s$^{-1}$. These emission tails are most prominently seen in high ionization emission lines such as [\ion{O}{3}] and [\ion{S}{3}].
High resolution JWST images of the low ionization emission lines such as [\ion{Fe}{2}] show elongated knot structures which may reflect internal density variations leading a spread of knot material due to differential post-shock decelerations.

5) The Cas~A remnant appears as an asymmetric cloud of reverse shocked SN debris significantly displaced to the northwest from the point of explosion. Some of this asymmetry is likely due to the progenitor's explosion dynamics
where more mass was ejected rearward to the northwest while its X-ray point source
and presumed neutron star was ejected toward us and to the southeast.
The remnant's asymmetrical radial velocity range of $-4000$ to +6000 km s$^{-1}$ may be the most obvious
signature of this asymmetry.
However, the nature of NE and SW  `jets' of ejecta which have transverse velocities  up to 14,500 km s$^{-1}$, nearly three times that of the remnant's main shell, is presently unclear in regard to the overall explosion dynamics.

The recent discovery of a cloud of CSM
on the remnant's near side (the `Green Monster'; \citealt{Mil24, DeLooze2024})
and the lopsided evolution of its optical emission shown in Figs. 2 and 3 may be signs 
of important CSM/ISM interactions
\citep{Orlando22}.
Consequently, it is perhaps not too surprising that the motion of the reverse 
shock varies somewhat around the remnant, despite
the seemingly uniform expansion of the forward blast wave in X-rays.
However, further work could re-examine the presence and nature of the cause
for the reverse shock's apparent
stationary and/or inward motion in the remnant's western regions.

A more in-depth study could also look into the 
morphological effects of the reverse shock on Cas~A's ejecta using higher dispersion spectra yielding both density and temperature data along with 
hydrodynamic modeling. Since ejecta knots are known to display varying S, O, and Ar abundances, investigations of elemental abundances effects on knot ablation and disruption
might prove interesting. Along these lines,
JWST's NIRCam images are capable of providing better
resolutions than even {\sl HST} and
thus might reveal additional details regarding ejecta evolution as a function time since reverse shock passage.   
\acknowledgements

We have benefited from 
discussions with many colleagues, especially Jacco Vink, Bon-Chu Koo, and Salvatore Orlando.  
We thank Bob Barr, Eric Galayda, and Justin Rupert and the entire MDM Observatory staff for assistance that made the ground-based spectral observations possible. We are especially grateful to Sidney van den Bergh for providing the unique archive of Palomar 5m prime focus plate collection of Cas~A images, and to Josh Grindlay and the DASCH plate scanning team at CfA/Harvard for their expert help and 
assistance in scanning the priceless Palomar 5m photographic plates.
RAF acknowledges support
from the National Science Foundation under grant AST-0908237 and from NASA through Guest Observer grants 8281, 9238, 9890, 10286, 12300, 12674, 15337, 15515 and 17210 from the Space Telescope Science Institute, which is operated by the Association of Universities for Research in Astronomy, Inc.
DJP acknowledges
support from the Chandra X-ray Center through GO8-9065A
and from the Space Telescope Science Institute through grant
GO-11337.01-A. DJP also acknowledges support through
NASA contract NAS8-03060. DM acknowledges NSF support from grants PHY-2209451 and AST-2206532.
This work is dedicated to Rudolph Minkowski, Karl Kamper, and Sidney van den Bergh 
who's decades of research on Cas~A formed the foundation for much of the subsequent work on this important and interesting young Galactic supernova remnant.

\facilities{Palomar Hale 5m telescope at Palomar Observatory, Hiltner 2.4m telescope at MDM Observatory, Hubble Space Telescope ({\sl HST}), Chandra X-ray Observatory (CXO), Very Large Array (VLA), James Webb Space Telescope (JWST) }
\software{NOAO/NOIRLab IRAF, DS9 fits viewer \citet{Joye2003}, WCSTools \citet{Laycock2010} }

\section{Appendix A: \\
Ejecta and Reverse Shock Proper Motion Animations   }

As mentioned in the results section ($\S4$), we have constructed several short animation videos which are especially helpful to clarify the proper motions of the reverse shock through the appearance of new optical emission knots and filaments as detected in images spanning several years. Most make use of {\sl HST} images taken between January 2000 and June 2008. 

The titles of the animation files are meant to be self-explanatory as to what regions are being shown and whether the file shows proper motions in the reference frame of the sky or the outwardly expanding ejecta. In all cases, the observed motions are repeated (looped) either 5 to 7 times to make the motions easier to observe. Each animation begins with a descriptive frame.  All movies are in a MPEG-4 container, specifically encoded with the H264 codec.\\

\noindent
List of animation files: \\
NW-ejecta-reference-frame.mp4 \\
NW-sky-reference-frame.mp4 \\
P1-sky-reference-frame.mp4 \\
P2nP3-ejecta-reference-frame.mp4 \\
P2nP3-sky-reference-frame.mp4 \\
P5nP6-ejecta-reference-frame.mp4 \\
P13-ejecta-reference-frame.mp4 \\
P14-ejecta-reference-frame.mp4 \\
P15-ejecta-reference-frame.mp4 \\

\section{Appendix B: On-Line Data \\ Digitized Palomar Hale 5m Prime Focus Photographic Plate Images of Cassiopeia A from 1951 to 1989}

For nearly forty years, starting from September 1951 through September 1989, a series of mainly broadband blue and red prime focus plates were taken of Cas~A using the Hale 5m (200-inch) telescope, what was then the world's large optical telescope. These plates, more than four dozen blue and red images, comprise the earliest optical images of Cas~A, obtained soon after the remnant's discovery in 1948 in the radio \citep{Ryle1948} and therefore the first $\sim300$ yr phase of young remnant evolution's not available for any other Galactic or extragalactic core-collapse SN. As such they constitute an unique and priceless record of Cas~A's optical evolution worth archiving for future researchers. 

Most of these plates were obtained by Walter Baade, Rudolph Minkowski and Sidney van den Bergh, with a few addition ones taken by Halton Arp, Joseph Miller and Donald Shane at the request of Baade and/or Minkowski. These Palomar 5m images formed the basis of the first investigations of the Cas~A's optical emission structure and expansion properties 
\citep{Min59,Min64,Min66,Min68,vdb71a,vdb71b,vdb74,vdb71c,vdb76a,vdbd70,vdb76b,vdbk83,vdbk85,vdbp86}.

These 5m prime focus plates, $5 \times 7$ inches in size, were often obtained  
with the prime focus Ross corrector and have a plate scale of $0.1''$ per millimeter. 
Only the central $7.7' \times 5.9'$ region of these plates were scanned and
then digitized  
at a pixel scale of 0.42$''$ pixel$^{-1}$
using the high-speed digitizing scanning machine DASCH \citep{Simcoe2006,Laycock2010}  at the Harvard–Smithsonian Center for Astrophysics. Individual image files are $\simeq$3.7 MB in size. 

The image quality of these plates vary considerably and only the best 42 plates were selected to be scanned and digitized (see  Table 6). 
These digitized images are available
at https://doi.org/10.7910/DVN/Q0HDOW.
However, three of the best images were separately scanned at a higher resolution ($0.141''$ pixel$^{-1}$) using the Yale Astronomy Department's PDS microdensitometer (see \cite{Thor01} for details. These and a few other high quality images deemed to be the best Palomar images are available separately at
https://doi.org/10.7910/DVN/OBHZB8.
Note: The entire set of original glass plates has been archived at the Carnegie Observatories' astronomical plate collection located in Pasadena, California.

Approximate world coordinates (WCS) was applied to the scanned data and were
determined using astrometric solutions done with
an automated procedure using software routines in WCSTools \citep{Laycock2010}.
Although more accurate WCS coordinates
were later applied using USNO UCAC3 catalog stars \citep{Zach2010},
the distortions due to the fast Ross corrector often resulted in lower than expected coordinate accuracy.  Consequently, we have re-done WCS coordinates using the online nova.astrometry.net
software.

Only a fraction of these plates have been reproduced in published papers, due in part to the large number of plates as well as the large variations in image quality due in part to varying sky and seeing conditions and exposure length.  Many plates also suffer emulsion and photographic emulsion and developer defects.

Table~\ref{plate_list} lists the 42 best Palomar plates taken between 1951 and 1989. The table lists file number, exposure date and exposure time, type of plate emulsion and bandpass filter used, image and plate quality grades, along with the name of the Palomar 5m observer. Because plate emulsions are no longer commonly used we have included a short table which gives information on emulsions. Addition plate emulsion information can be found in \citet{Hoag1969},  \citet{Latham1974}, at
https://gsss.stsci.edu/SkySurveys/Surveys.htm, 
and in the
`Kodak Plates and Films for Science and Industry', Kodak Publication No. P-9. Eastman Kodak Co. (1967). Descriptions of Schott bandpass filters used at Palomar can be found at the Schott website or in various papers published before 2000.

\begin{deluxetable*}{cclcccl}
\tablecolumns{7}
\tablewidth{0pc}
\tablecaption{Selected Palomar Hale 5m Prime Focus Plates Taken of Cassiopeia A Between 1951 and 1989 \label{plate_list} }
\tablehead{\colhead{No.} & \colhead{Obs. Date} &
\colhead{Plate ID ~ ~~} & \colhead{Exposure Time}  &  
\colhead{color: plate+filter} & \colhead{Image/Plate Quality} & \colhead{Observer ~ ~~ ~~~ ~ ~} }
\startdata
 1 & 1951-09-09 & PH520B       & 1800 s      & blue: 103aO+GG1  & good/fair ~ ~   & Baade  \\
 2 & 1951-10-31 & PH553B       & 7200 s      & ~red: 103aE+RG2  & v. good/poor ~ ~ ~ & Baade  \\
 3 & 1951-11-03 & PH563B       & 7200 s      & ~red: 103aE+RG2  & v. good/fair ~ ~ ~ ~ & Baade   \\
 4 & 1953-08-11 & PH778B       & 1500 s      & blue: 103aO+GG1  & good/good & Baade   \\
 5 & 1953-08-13 & PH793B       & 6120 s      & ~red: 103aE+RG2  & good/good & Baade   \\
 6 & 1954-11-25 & PH232M       & 5400 s      & blue: 103aJ+GG11 & fair/good & Minkowski  \\
 7 & 1954-11-26 & PH236M       & 7200 s      & ~red: 103aE+OR1  & fair/good & Minkowski  \\
 8 & 1954-12-23 & PH1159B      & 1800 s      & blue: 103aO+GG1  &   good/good  & Baade \\
 9 & 1957-09-21 & PH1732B      & 5400 s      & ~red: 103aF+RG2  & v. good/good ~ ~ ~ & Baade    \\
10 & 1957-09-22 & PH1736B      & 2100 s      & blue: 103aO+GG1  & fair/fair & Baade   \\
11 & 1958-08-11 & PH3033S      & 5400 s      & ~red: 103aF+RG2  & excellent/good & Shane \\
12 & 1965-08-30 & PH4815       & 5400 s      & ~red: 103aE+0R1  & fair/poor  & Miller  \\
13 & 1967-10-04 & PH5107A      & 4200 s      & ~red: 103aE+RG2  & good/fair  & Arp    \\
14 & 1968-09-26 & PH5254vB     & 5400 s      & ~red: 103aF+RG2  & good/good  & van den Bergh \\
15 & 1968-09-26 & PH5255vB     & 1500 s      & blue: 103aD+GG11 & fair/poor  & van den Bergh \\
16 & 1969-10-13 & PH5405vB     & 5400 s      & ~red: 103aF+RG2  & fair/poor  & van den Bergh \\
17 & 1970-09-01 & PH5643vB     & 7200 s      & blue: 103aJ+GG475& fair/good  & van den Bergh  \\
18 & 1970-09-03 & PH5648vB      & 7200 s     & blue: 103aJ+GG475& fair/good  & van den Bergh \\
19 & 1970-09-04 & PH5659vB      & 7200 s     & ~red: 103aE+RG630& good/good  & van den Bergh  \\
20 & 1971-08-29 & PH5946A       & 5400 s     & ~red: 103aF+RG630& v. good/good ~ ~ ~  & Arp  \\
21 & 1972-09-09 & PH6236vB      & 7200 s     & blue: IIIaJ+GG7  & fair/poor  & van den Bergh \\
22 & 1972-09-10 & PH6249vB      & 7200 s     & ~red: 103aF+RG2  & v. good/good ~ ~ ~& van den Bergh \\
23 & 1973-07-31 & PH6555vB      & 3600 s     & ~red: 098-04+RG2 & good/good  & van den Bergh \\
24 & 1973-08-01 & PH6562vB      & 7200 s     & blue: 103aJ+GG7  & good/good  & van den Bergh \\
25 & 1973-08-04 & PH6573vB      & 12000 s    & [S II]+098 (FWHM=166 A) &excellent/good& van den Bergh \\
26 & 1973-08-13 & PH6891vB      & 4200 s     & ~red: 098-04+RG2 & v. good/fair ~ ~ ~ & van den Bergh \\
27 & 1974-08-14 & PH6902vB      & 7200 s     & blue: IIIaJ+GG7  & good/good & van den Bergh \\
28 & 1974-08-15 & PH6914vB      & 7200 s     & blue: 127-02+GG7 & good/poor & van den Bergh \\
29 & 1975-07-16 & PH7064vB      & 3600 s     & ~red: 098-02+RG645 & good/fair & van den Bergh \\
30 & 1975-07-18 & PH7071vB      & 7200 s     & ~red: 098-02+RG645 & good/fair & van den Bergh  \\
31 & 1975-07-18 & PH7077A       & 6000 s     & ~red: 098-02+RG645 & good/fair & Arp            \\
32 & 1976-06-30 & PH7231vB      & 7200 s     & blue: 127-04+GG7 &  good/fair  & van den Bergh \\
33 & 1976-07-02 & PH7252vB      & 7200 s     & ~red: 098-04+RG645& excellent/good & van den Bergh \\
34 & 1977-10-08 & PH7433vB      & 4500 s     & ~red: 098-04+RG2  & fair/fair     & van den Bergh \\
35 & 1977-10-10 & PH7445vB      & 7200 s     & blue: 103aJ+GG7   & good/good     & van den Bergh \\
36 & 1980-07-13 & PH7766vB      & 6000 s     & ~red: 098-04+RG645& fair/fair     & van den Bergh \\
37 & 1980-07-14 & PH7771vB      & 7200 s     & blue: IIIaJ+GG7   & fair/good     & van den Bergh \\
38 & 1983-07-12 & PH8192vB      & 6000 s     & ~red: 098+RG645   &   good/good     & van den Bergh \\
39 & 1983-07-14 & PH8201vB      & 10800 s    & [S II]+098 (FWHM=166 A) & v. good/good ~ ~  & van den Bergh \\
40& 1989-09-28 & PH8202vB      & 7200 s     & ~red: 098-04+RG645 & good/poor    & van den Bergh \\
41 & 1989-09-28 & PH8204vB      & 7200 s     & [S II]+098(FWHM=166 A) & v. good/fair ~ ~ ~ & van den Bergh \\
42 & 1989-09-29 & PH8206vB      & 7200 s     & ~red: 098-04+RG645 & excellent/good  & van den Bergh \\
\enddata
\end{deluxetable*}
\clearpage
\newpage
\bibliography{casa.bib}
\end{document}